\documentclass[twocolumn]{aastex631}

\usepackage{CJK}
\usepackage{times}
\usepackage{amsmath}
\usepackage{graphicx}
\usepackage{hyperref}
\usepackage{gensymb}
\usepackage{upgreek}
\usepackage{natbib}
\usepackage{subfigure}
\usepackage{multirow}
\usepackage{comment}
\usepackage{mathtools}
\usepackage{mathrsfs}
\usepackage{fontenc}
\usepackage{color}
\usepackage{url}
\usepackage{pifont}

\newcommand{\Ha}{H$\alpha$}
\newcommand{\Hb}{H$\beta$}

\newcommand{\OII}{[O\,\textsc{ii}]}
\newcommand{\OIII}{[O\,\textsc{iii}]}

\newcommand{\sersic}{S\'{e}rsic}
\newcommand{\msersic}{$m_{\rm S\acute{e}rsic}$}

\submitjournal{ApJ}

\shorttitle{Main Sequence of Low-redshift Quasars}
\shortauthors{Zhuang \& Ho}

\begin{document}
\begin{CJK*}{UTF8}{gbsn}

\title{The Star-forming Main Sequence of the Host Galaxies of Low-redshift Quasars}

\author[0000-0001-5105-2837]{Ming-Yang Zhuang (庄明阳)}
\email{mingyangzhuang@pku.edu.cn}
\affil{Kavli Institute for Astronomy and Astrophysics, Peking University,
Beijing 100871, China}
\affil{Department of Astronomy, School of Physics, Peking University,
Beijing 100871, China}

\author[0000-0001-6947-5846]{Luis C. Ho}
\affil{Kavli Institute for Astronomy and Astrophysics, Peking University,
Beijing 100871, China}
\affil{Department of Astronomy, School of Physics, Peking University,
Beijing 100871, China}

\begin{abstract}
We investigate the star-forming main sequence of the host galaxies of a large, well-defined sample of 453 redshift $\sim$\,0.3 quasars with previously available star formation rates by deriving stellar masses from modeling their broad-band ($grizy$) spectral energy distribution.  We perform two-dimensional, simultaneous, multi-filter decomposition of Pan-STARRS1 3$\pi$ Steradian Survey images to disentangle the active galactic nucleus (AGN) from its host galaxy, by explicitly considering, for the first time, the wavelength variation of galaxy structures.  We quantify the S\'ersic profiles and sizes of the host galaxies from mock AGNs generated from both real and idealized galaxies. Detailed morphological classifications of the calibration galaxy sample with Hubble Space Telescope images enable us to estimate crude morphological types of the quasars. Although the majority ($\sim$\,60\%) of the quasars are hosted by bulge-dominated, early-type galaxies, a substantial fraction ($\sim$\,40\%) reside in disk-dominated, late-type galaxies, suggesting that at least in these systems major mergers have not played a significant role in regulating their AGN activity, in agreement with recent simulations and observations of nearby quasars. The vast majority ($\sim$\,90\%) of the quasars have star formation rates that place them on or above the galaxy star-forming main sequence, with more rapidly accreting AGNs displaced further above the main sequence.  Quasar host galaxies generally follow the stellar mass-size relation defined by inactive galaxies, both for late-type and early-type systems, but roughly 1/3 of the population has smaller sizes at a given stellar mass, reminiscent of compact star-forming galaxies at higher redshift. 
\end{abstract}


\section{Introduction} \label{sec1}

The discovery of tight correlations between the masses of supermassive black holes (BHs) and the properties of their host galaxies suggests that BHs coevolve with galaxies \citep[e.g.,][]{Magorrian+1998AJ, Richstone+1998Natur, Gebhardt+2000ApJ, Ferrarese2000, Ho2004coa..book, Kormendy&Ho2013ARA&A, Heckman&Best2014ARA&A}. What dictates their joint evolution? How were the correlations between BH mass ($M_{\rm BH}$) and host galaxy properties established?  Popular scenarios suggest that active galactic nuclei (AGNs) play a significant role in this narrative by injecting energy and momentum into their environment. The feedback produced by AGNs in the form of fast outflows or jets acts as a double-edged sword for their host galaxies. On the one hand, they are expected to prevent cooling of intergalactic gas by relativistic jets that inflate large-scale cavities \citep[radio/kinetic mode; e.g.,][]{Bruggen&Kaiser2002Natur, McNamara&Nulsen2007ARA&A} and expel gas in the interstellar medium by radiatively driven outflows \citep[quasar/radiative mode; e.g.,][]{Di_Matteo+2005Natur, Hopkins+2008ApJS, Feruglio+2010A&A} and hence suppress star formation (negative feedback). On the other hand, outflows can compress ambient gas and directly induce in situ star formation in the outflows themselves \citep[positive feedback; e.g.,][]{Zubovas+2013MNRAS, Maiolino+2017Natur, Gallagher+2019MNRAS, Qiu+2020NatAs}. Positive and negative feedback are not necessarily mutually exclusive; they have been found to occur within the same galaxy at the same time \citep[e.g.,][]{Cresci+2015ApJ, Shin+2019ApJ, Mandal+2021MNRAS}.

Star-forming galaxies, both in the local and distant Universe, follow a main sequence (MS), a tight correlation between star formation rate (SFR) and stellar mass \citep[e.g.,][]{Brinchmann+2004MNRAS, Elbaz+2011A&A, Speagle+2014ApJS, Renzini&Peng2015ApJ}. In principle, the galaxy star-forming MS offers an effective framework to advance our understanding of the impact of AGNs on galaxy evolution.  By comparing galaxies at fixed stellar mass, especially if other conditions such as gas fraction and morphology can be specified, the position of AGN hosts relative to the MS informs us of the degree and manner in which BH accretion-related processes influence star formation.  However, the current observational landscape is quite complex. Nearby, low-luminosity AGNs mainly lie in the green valley \citep[e.g.,][]{Salim+2007ApJS, Leslie+2016MNRAS}, the sparsely populated region with reduced SFR below the MS, or are altogether quiescent \citep{Ho2003}. By contrast, their more luminous counterparts, the quasars, can be found below \citep[e.g.,][]{Shimizu+2015MNRAS, Stemo+2020ApJ}, on \citep[e.g.,][]{Husemann+2014MNRAS, Woo+2017ApJ, Smirnova-Pinchukova+2021arXiv}, or even above \citep[e.g.,][]{Jarvis+2020MNRAS, Shangguan+2020bApJ, Xie+2021ApJ, Koutoulidis+2022A&A} the star-forming galaxy MS. Several factors contribute to this confusing status of affairs, including the wavelength used for sample selection (X-ray, optical, infrared, or radio), the methodology and uncertainties associated with SFR and stellar mass determination, and even the very shape and normalization of the MS used for reference comparison. Recent works suggest that the position of an AGN relative to the MS is correlated with Eddington ratio \citep[e.g.,][]{Shimizu+2015MNRAS, Ellison+2016MNRAS, Woo+2020ApJ, Torbaniuk+2021MNRAS}, with higher SFRs found in hosts of AGNs with higher Eddington ratio. 

A practical difficulty is that the emission from rapidly accreting BHs (Eddington ratio $\lambda_{\rm E}\gtrsim 10^{-3}$)\footnote{We define the Eddington ratio as $\lambda_\mathrm{E} \equiv L_\mathrm{bol}/L_\mathrm{E}$, with $L_\mathrm{bol}$ the bolometric luminosity and $L_\mathrm{E}=1.26 \times 10^{38}\,(M_\mathrm{BH}/M_\odot)$ the Eddington luminosity.} in type~1 AGNs can easily dominate the observed spectral energy distribution (SED) of the host galaxy starlight at ultraviolet, optical, and near-infrared wavelengths, posing a major obstacle to obtaining reliable information on the stellar mass and SFR of the galaxy. To address these challenges, much attention has been devoted to deriving accurate SFRs for AGN host galaxies.  Efforts include the development of comprehensive AGN dust emission models \citep[e.g.,][]{Honig&Kishimoto2017ApJL, Stalevski+2019MNRAS}, new or improved emission-line diagnostics based on empirical methods and photoionization modeling \citep[e.g.,][]{Ho2005, Kim2006, Thomas+2018ApJ, Zhuang&Ho2019ApJ, Zhuang+2019ApJ}, mid-infrared polycyclic aromatic hydrocarbon features and the far-infrared continuum \citep{Xie+2021ApJ}, and modern SED fitting codes incorporating sophisticated models \citep[e.g.,][]{Calistro_Rivera+2016ApJ, Zhuang+2018ApJ, Yang+2020MNRAS}. 

This paper focuses on the equally pressing issue of how to derive stellar masses for type~1 AGNs.  In inactive galaxies, stellar mass can be inferred by comparing galaxy spectra or multi-band integrated photometry with stellar population and dust attenuation models \citep[e.g.,][]{Bell+2003ApJS, Conroy2013}. Active galaxies require the additional complicated step of deblending the AGN from the host.  Several strategies have been pursued, including direct broadband SED fitting using AGN and stellar templates \citep[e.g.,][]{Merloni+2010ApJ, Ciesla+2015A&A, Suh+2019ApJ}, detailed spectral modeling of the spectrum using stellar absorption features \citep[e.g.,][]{Vanden_Berk+2006AJ, Matsuoka+2015ApJ}, and photometric analysis after AGN-host image decomposition \citep[e.g.,][]{Matsuoka+2014ApJ, Bentz+Manne-Nicholas2018, Yue+2018ApJ, Ishino+2020PASJ, Bennert+2021ApJ, Li_Junyao+2021ApJ, Zhao+2021ApJ}. A practical difficulty, however, is that in the most powerful quasars the integrated spectrum will be completely overwhelmed by the nucleus, barely leaving room for the starlight to be detected in the residuals. In this regard, image decomposition offers a distinct advantage by exploiting the inherently different spatial distribution of the physical components: whereas the host galaxy spans a wide range of physical scales and can comprise multiple subcomponents, the active nucleus is uniquely localized in (or near) the galaxy center, and its light distribution follows the point-spread function (PSF) of the image.  These priors can be used to one's advantage to decouple the AGN from its host.

Measuring the stellar mass, nevertheless, is more challenging than simply detecting the light. To estimate the stellar mass requires photometry in at least two bands with sufficient wavelength separation to yield, at the bare minimum, a rudimentary color, which can be used to constrain the age and hence mass-to-light ratio of the stellar population. Previous studies of bright AGNs derive mainly from relatively small samples of high-resolution imaging with the Hubble Space Telescope (HST), usually acquired in only one filter, or at most a very small number of filters \citep{McLure1999, Dunlop2003, Greene2008,Kim+2008ApJS,Jiang2011,Zhao+2021ApJ}.  Consequently, most investigations have focused on the relation between BH mass and host galaxy stellar luminosity, not stellar mass. To compare with the local scaling relation, one needs to assume a certain galaxy evolution model to obtain the equivalent galaxy luminosity at redshift $0$ \citep[e.g.,][]{Peng+2006aApJ}, which unavoidably introduces more uncertainties. In the largest HST sample assembled to date, \citet{Kim+2017ApJS} studied 235 low-redshift ($<0.35$) type~1 AGNs of heterogeneous origin from the data archives, for which they have only single-band photometry.  This necessarily limits the physical interpretation of the sample \citep{Kim&Ho2019ApJ}.  Large ground-based imaging surveys can furnish multi-band information more readily, but at the expense of lower spatial resolution.  Previous works using images from Stripe 82 \citep{Annis+2014ApJ} of the Sloan Digital Sky Survey \citep[SDSS;][]{York+2000AJ} have shown that it is possible to decompose AGN host galaxies up to redshift $\sim$\,0.6 \citep[e.g.,][]{Falomo+2014MNRAS, Matsuoka+2014ApJ}. With the advent of deeper, higher resolution data from the Hyper Suprime-Cam Subaru Strategic Program \citep{Aihara+2018PASJ}, AGN image decomposition from ground-based surveys has been pushed to redshift $\sim$\,1 \citep{Ishino+2020PASJ, Li_Junyao+2021ApJ}.  

Host galaxy parameters can be extracted from either one-dimensional parametric fits of the azimuthally averaged radial surface brightness profile or decomposition of the two-dimensional image itself. The popular two-dimensional image decomposition code \texttt{GALFIT} \citep{Peng+2002AJ, Peng+2010AJ} can model simultaneously a variety of galaxy substructures besides the traditional bulge and disk, including bars, lenses, rings, and spiral arms, as well as close companions.  More recent softwares (e.g., \texttt{Imfit}: \citealt{Erwin2015ApJ}; \texttt{ProFit}: \citealt{Robotham+2017MNRAS}; \texttt{lenstronomy}: \citealt{Birrer&Amara2018PDU}) offer a Bayesian framework to enable more reliable error estimates.  When more than one band is available, it is common practice, for the purposes of minimizing parameter degeneracy, to fix the structure of the galaxy to that determined from the band with best resolution, highest signal-to-noise ratio, and most optimal host-to-AGN contrast ratio \citep[e.g.,][]{Matsuoka+2014ApJ, Yue+2018ApJ, Ding+2020ApJ, Li_Junyao+2021ApJ, Li_Jennifer+2021ApJ, Zhao+2021ApJ}. However, the morphology and structure of a galaxy can vary significantly with wavelength owing to variations in stellar population, metallicity gradient, or dust attenuation. For example, a spiral galaxy tends to have brighter disk emission at shorter wavelengths and a more prominent bulge component at longer wavelengths, such that the galaxy looks more extended and flatter toward the ultraviolet and more compact and steeper in the near-infrared. In their decomposition of multiwavelength images of nearby galaxies, \citet{Kelvin+2012MNRAS} find that the \cite{Sersic} index increases while the half-light radius decreases toward longer wavelengths. To our knowledge, no work has taken the wavelength dependence of galaxy structure into consideration when performing AGN host galaxy decomposition, even though this capability is possible (e.g., \texttt{GALFITM}; \citealt{Haussler+2013MNRAS, Vika+2013MNRAS}). Neglecting this phenomenon introduces unknown systematic biases.  

\cite{Zhuang&Ho2020ApJ} constructed a well-defined sample of 453 redshift $0.3-0.35$ unobscured AGNs (hereinafter, the redshift $\sim$\,0.3 quasar sample), based on the presence of broad \Ha\ emission in SDSS spectra, primarily for the purposes of investigating their star formation properties.  The $3\pi$ Steradian Survey of Pan-STARRS1 \citep[PS1;][]{Chambers+2016arXiv} conveniently provides five-band\footnote{The five filters employed by PS1 officially are designated $g_{\rm P1}$, $r_{\rm P1}$, $i_{\rm P1}$, $z_{\rm P1}$, and $y_{\rm P1}$. For brevity, hereinafter we simply refer to them without the ``P1'' subscript.} ($grizy$) images for the entire sample, with better red sensitivity and improved spatial resolution compared to SDSS. The large sample size and rich ancillary information, which includes SFR, BH mass, and molecular gas mass estimated using dust attenuation \citep{Zhuang+2021ApJ} make this sample ideal to study the coevolution between supermassive BHs and their host galaxies. In this paper, we perform two-dimensional decomposition of the PS1 images of the redshift $\sim$\,0.3 quasar sample, with the primary goal of deriving their structural properties and stellar masses. Section~\ref{sec2} introduces the sample and data, and performs an extensive investigation of the PSF models needed for the analysis. In Section~\ref{sec3}, we present the two-dimensional multiwavelength simultaneous decomposition and verify our results using mock data. We fit the AGN-decomposed galaxy SED to derive stellar masses and use structural parameters to obtain morphological classifications (Section~\ref{sec4}).  Section~\ref{sec5} discusses the physical properties of AGN host galaxies in the context of the galaxy star-forming MS, and the triggering mechanism of AGNs.  Our main conclusions are summarized in Section~\ref{sec6}. This work adopts a cosmology with $H_0=70$ km~s$^{-1}$~Mpc$^{-1}$, $\Omega_m=0.3$, and $\Omega_{\Lambda}=0.7$. Stellar masses and SFRs assume the initial mass function of \citet{Kroupa2001MNRAS}.

\begin{figure}[t]
\centering
\includegraphics[width=0.48\textwidth]{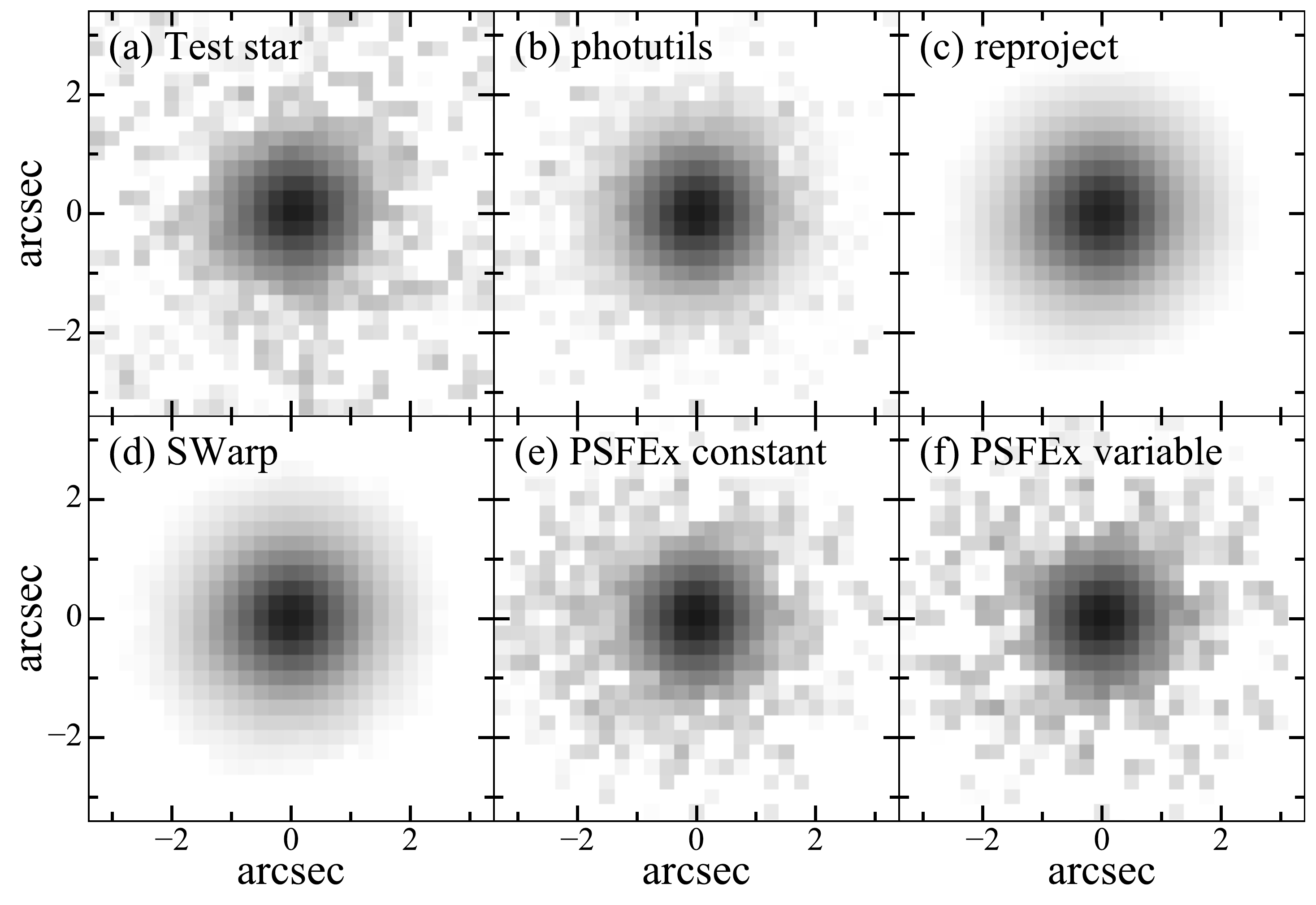}
\caption{An example of a test star ($m=18.8$~mag; panel a) and five PSF models (panels b--f) generated using all other stars in the $i$-band image of SDSS~J021652.47$-$002335.3 ($m_{\rm PSF}=18.2$~mag). Test stars are selected to be within 1~mag of the PSF magnitude of the AGN. PSF models are generated using \texttt{photutils}, \texttt{reproject}, \texttt{SWarp}, and \texttt{PSFEx} are shown in panels b--f. \texttt{PSFEx} provides two versions of PSF models: position-constant (\texttt{PSFEx constant}) and position-variable (\texttt{PSFEx variable}); for the latter, we consider a second-order polynomial variation on pixel coordinates.}
\label{fig1}
\end{figure}

\begin{figure*}[t]
\centering
\includegraphics[width=1.0\textwidth]{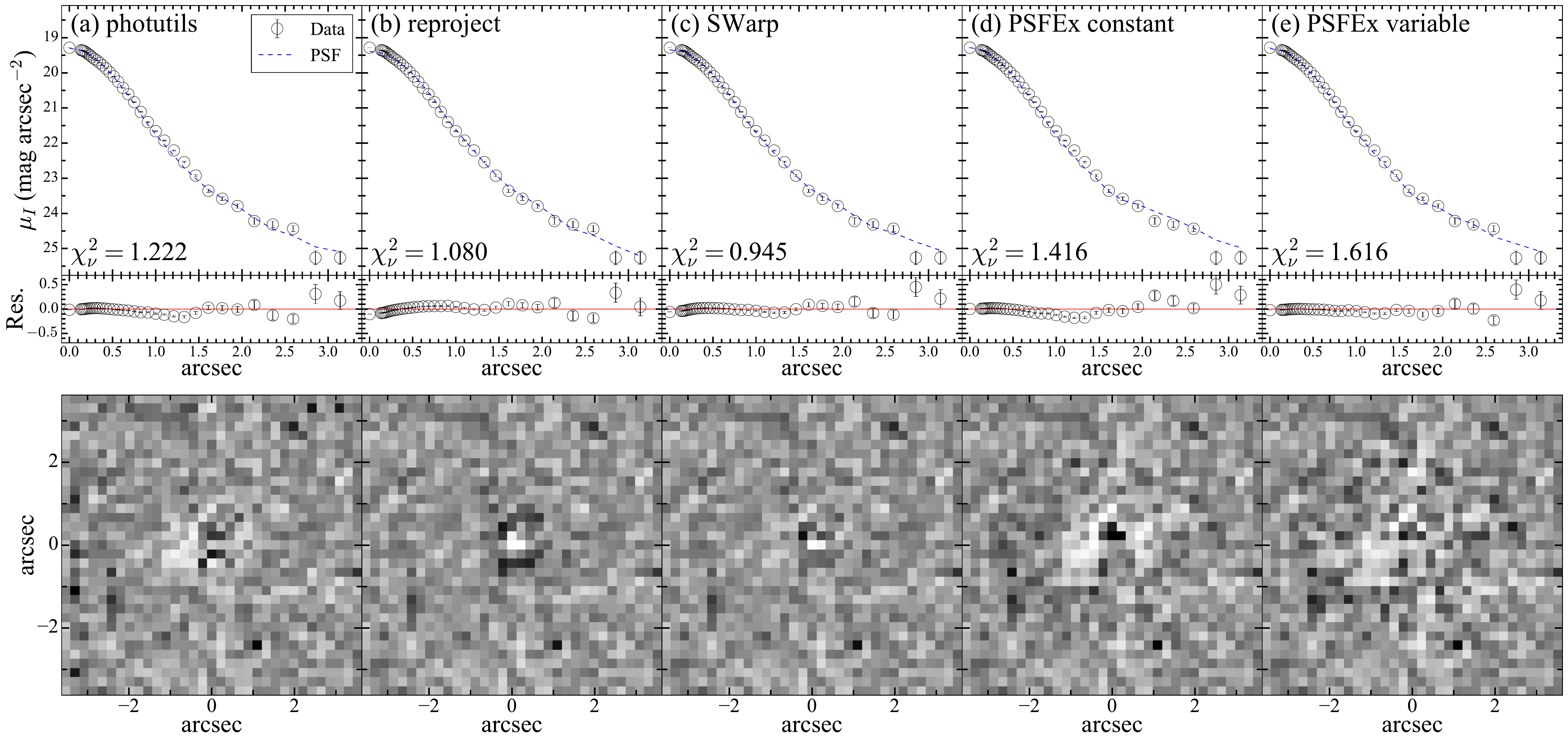}
\caption{Top row: Surface brightness profiles of the test star (data points with error bars) shown in Figure~\ref{fig1} and best-fit \texttt{GALFIT} models of PSFs generated from different methods: (a) \texttt{photutils}, (b) \texttt{reproject}, (c) \texttt{SWarp}, (d) \texttt{PSFEx constant}, and (e) \texttt{PSFEx variable}. The radial profile of the residuals between the star and the PSF model is shown in the lower subpanels. Bottom row: Residual maps (data $-$ PSF) for the different PSF models.}
\label{fig2}
\end{figure*}

\begin{figure*}[t]
\centering
\includegraphics[width=0.9\textwidth]{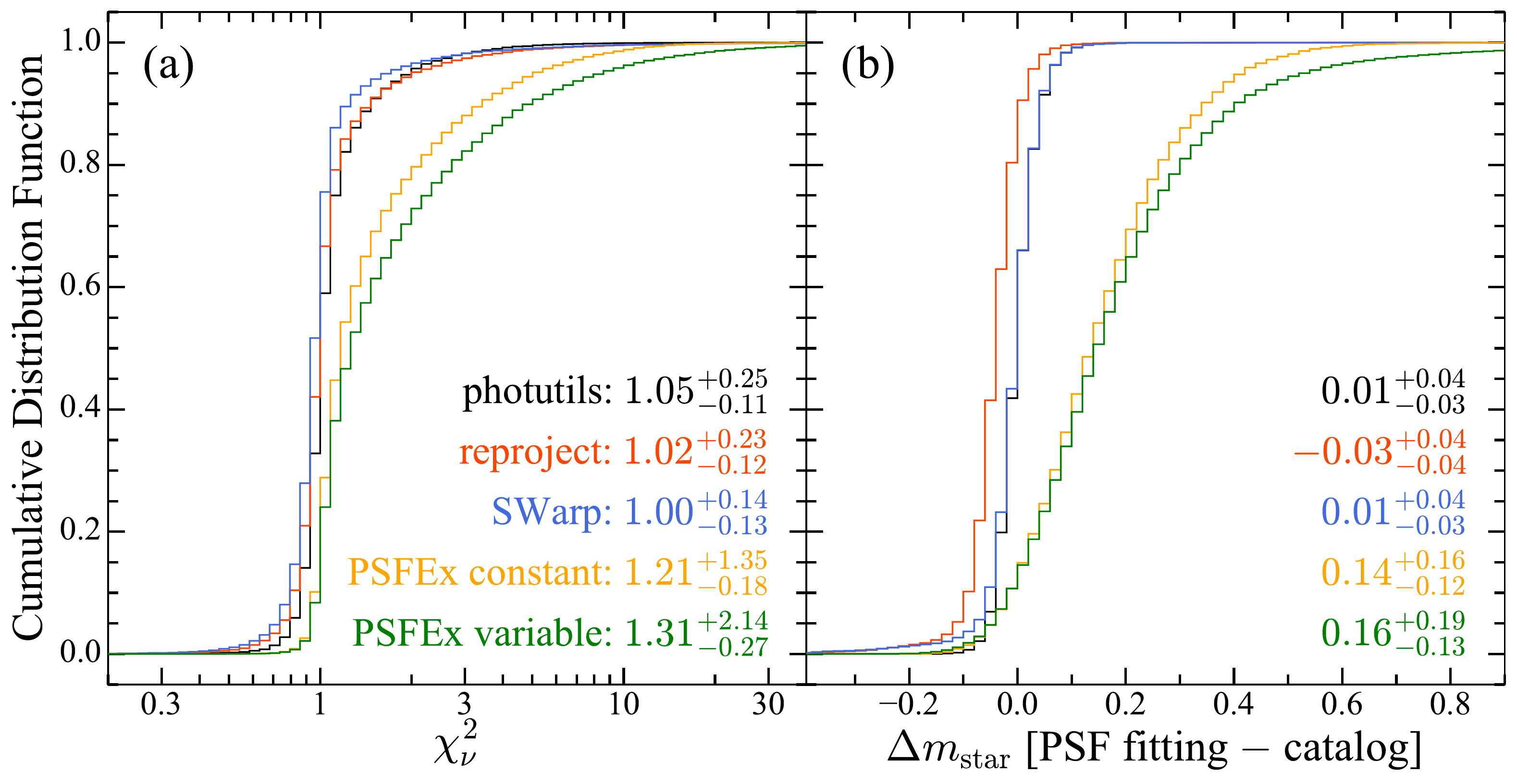}
\caption{Cumulative distribution function of (a) $\chi_{\nu}^2$ of the fits to test stars with PSF models using \texttt{GALFIT} and (b) the differences between the magnitudes of stars ($m_{\rm star}$) from the PS1 DR2 catalog and those derived from PSF fitting using different PSF models. Various colors represent PSF models generated from different methods. The median and upper and lower $1\,\sigma$ (84th and 16th percentiles $-$ median) values of the distributions are shown in the lower-right corner of each panel.}
\label{fig3}
\end{figure*}

\section{Observational Material} \label{sec2}

\subsection{Sample}

\citet{Zhuang&Ho2020ApJ} selected 453 AGNs with broad \Ha\ emission from the parent catalog of 14,584 low-redshift type~1 AGNs of \citet{Liu+2019ApJS}. They limited the sample to a narrow slice in redshift ($0.3-0.35$) to mitigate the effects of differential fiber coverage and Malmquist bias. To implement the method of SFR estimation based on the \OII\ $\lambda 3727$ and \OIII\ $\lambda 5007$ emission lines \citep{Ho2005, Zhuang&Ho2019ApJ}, we additionally required the sources to have high excitation (``Seyfert'' classification according to optical line-intensity ratio diagnostics; e.g., see \citealt{Ho2008}), as well as narrow \Ha\ and \Hb\ lines of sufficiently high signal-to-noise ratio to yield a reliable Balmer decrement, which was used for both extinction correction and estimation of molecular gas content \citep{Yesuf2019}.  Building upon a refined \OII\ SFR calibration for star-forming galaxies, \cite{Zhuang&Ho2019ApJ} proposed, on the basis of radiation pressure-dominated photoionization models, a new SFR estimator for AGNs that explicitly accounts for the contribution of the narrow-line region to \OII, using \OIII\ as the benchmark for the AGN. The sample consists of powerful AGNs with bolometric luminosity $L_{\rm bol} = 10^{44.3}-10^{47.4}$ erg~s$^{-1}$, BH masses $M_{\rm BH} = 10^{6.7}-10^{9.6}\, M_{\odot}$, and Eddington ratios $\lambda_{\rm E} = 10^{-2.4}-10^{1.6}$.  The median bolometric luminosity of the sample ($L_{\rm bol} = 10^{45.5}$ erg~s$^{-1}$) coincides with the historical $B$-band absolute magnitude defintion of a quasar ($M_B < -23$ mag; \citealt{Schmidt&Green1983ApJ}), upon adjusting to our adopted cosmology\footnote{We convert $M_B$ to $L_{\rm bol}$ using the composite quasar SED and bolometric correction at 5100 \AA\ from \citet{Richards+2006ApJS}.}.

\subsection{Pan-STARRS1 $3\pi$ Steradian Survey}

The $3\pi$ Steradian Survey of PS1 \citep{Chambers+2016arXiv} imaged three-quarters of the sky north of declination $-30\degr$ with five filters ($grizy$) and $\sim$\,12 epochs per filter. The stacked images from this nearly four-year survey have a median seeing of 1\farcs31, 1\farcs19, 1\farcs11, 1\farcs07, and 1\farcs02, and a mean $5\,\sigma$ point-source sensitivity of 23.3, 23.2, 23.1, 22.3, and 21.4~mag (AB) for $grizy$, respectively. The larger sky coverage, introduction of the near-infrared $y$ filter, finer pixel scale (0\farcs25 versus 0\farcs396), and higher sensitivity ($\sim$\,0.2, 0.5, 0.9, 1.6~mag deeper in the $griz$ bands) compared to the main SDSS photometric survey render the stacked images of the PS1 $3\pi$ survey extremely valuable for studying the photometric properties of large samples of galaxies, with the possibility of even contemplating galaxies of somewhat higher redshifts. We use the stacked images, combined in each of the five bands, from the second data release \citep[DR2;][]{Flewelling+2020ApJS} of the PS1 $3\pi$ survey, which includes an improved calibration of the third full reduction of all the data \citep{Waters+2020ApJS}. Each image cutout, centered on the SDSS position of the AGN, has a minimum size of $5\arcmin \times 5\arcmin$, which gradually increases in steps of 1\arcmin\ according to the number of stars that are available for PSF construction and testing (Section~\ref{sec2.3}). We also retrieve the stacked-object catalog from PS1 DR2 to obtain PSF and \cite{Kron1980} magnitudes ($m_{\rm PSF}$ and $m_{\rm Kron}$) for all detected objects within the field-of-view of the cutout images.

\subsection{PSF Construction and Verification} \label{sec2.3}

To perform accurate AGN-host galaxy decomposition, obtaining an accurate PSF model is of utmost priority. We select isolated point-like sources by requiring $m_{\rm PSF}<21$~mag and $m_{\rm PSF} - m_{\rm Kron}<0.05$~mag in the $i$ band \citep{Farrow+2014MNRAS} and removing objects with companions located within 33 pixels ($\sim$\,8 times the full width at half maximum of the PSF). At a radius of 16 pixels, light from the wings of the PSF is $\lesssim 5\times 10^{-4}$ smaller than that of the central pixel, and thus the contamination from a companion is negligible. Choosing a larger radius would decrease the number of point sources available for PSF construction. To construct the PSF model, for each star in each band, we cut out a stamp with size $33 \times 33$ pixels after removing the background.

We consider two methods to construct the PSF model, one by building a pixel-based model and the other by stacking star images. For the pixel-based model method, we use the Python package \texttt{photutils} \citep{photutils} and the software \texttt{PSFEx} \citep{Bertin2011ASPC}. Following the prescription of \citet{Anderson&King2000PASP}, \texttt{photutils} builds an ``effective'' PSF that distills a representative PSF model from a set of point samplings provided by arrays of point samplings of different stars. \texttt{PSFEx} constructs the PSF based on the output of \texttt{SExtractor} \citep{1996A&AS..117..393B}, using different bases generated with an algorithm similar to \texttt{photutils}.  These include a pixel basis, which is more general and is adopted here, the Gauss-Laguerre basis, which is recommended for PSFs that are nearly Gaussian, and a user-provided vector basis. No instrument has a perfectly stable PSF that does not vary in time or position on the focal plane. \texttt{PSFEx} considers the spatial variation of the PSF by implementing a polynomial dependence of the PSF vector basis on position. For the image-stacking method, we use the Python package \texttt{reproject}\footnote{https://reproject.readthedocs.io/en/stable/index.html} and the code \texttt{SWarp} \citep{Bertin+2002ASPC}. Three algorithms are provided to reproject images/data cubes. We use the flux-conserving scheme \texttt{spherical polygon intersection}, which follows the core algorithm of \texttt{Montage} \citep{Berriman+2008ASPC}, by treating pixels as four-sided spherical polygons and computing the exact overlap of pixels on the sky. Among the several options provided by \texttt{SWarp} to perform reprojection, we adopt the \texttt{lanczos3} routine, which is recommended as it does a good job in preserving the signal and creates relatively modest artifacts around image discontinuities. 

We construct five PSF models for each object in each band, with two models from \texttt{PSFEx}: \texttt{PSFEx constant} gives a position-constant PSF, while \texttt{PSFEx variable} considers the spatial variation of the PSF model across the field-of-view. Considering the size of our image cutout and the stacking nature of the image, we adopt a second-order polynomial dependence of the PSF on pixel coordinates, which has six vectors ($constant$, $x$, $x^2$, $xy$, $y$, and $y^2$, where $x$ and $y$ are orthogonal pixel coordinates) and requires at least six input stars. 

To obtain the best PSF model for each AGN image, we select ``test stars'' to evaluate the performance of the five PSF models. Test stars are point-like objects with $m_{\rm PSF}$ within $\sim$\,1~mag of the target AGN. We use all point-like objects (besides the science target) in the same image cutout to construct the PSF model for the test star. Figure~\ref{fig1} shows an example of a test star and five corresponding PSF models constructed for it. We then use \texttt{GALFIT} to fit the test star with its five PSF models to obtain the goodness-of-fit $\chi^{2}_{\nu}$ and best-fit magnitudes (Figure~\ref{fig2}). For the 2265 images of the 453 objects, we use a total of 16,551 test stars (2654, 2886, 3420, 3525, and 4066 for $grizy$, respectively) and 42,093 isolated point-like sources (7949, 8803, 8240, 8739, and 8362 for $grizy$, respectively) to construct PSFs. Figure~\ref{fig3} gives the distribution of $\chi^{2}_{\nu}$ and compares the magnitudes obtained from our PSF fitting with those from the PS1 DR2 catalog, for all test stars and five PSF models.  The PSF models constructed from the different methods generally are all capable of describing the light profile of the test stars. The median value of $\chi^{2}_{\nu}$ is close to 1. However, the PSF models using \texttt{photutils}, \texttt{SWarp}, and \texttt{reproject} perform significantly better than the two \texttt{PSFEx} models in recovering the magnitude of point-like sources, showing no systematic bias and a small scatter of $\lesssim0.04$~mag.  The overall good performance of the PSF models without consideration for their spatial variation suggests that this effect is not severe in the stacked images of PS1, at least for a relatively small field-of-view of $\sim$\,30 arcmin$^2$. To assign the optimal PSF for each target, for every image of each band we select the best PSF model based on the distribution of $\chi^{2}_{\nu}$ and magnitude comparisons performed using the test stars, with preference given, all else being equal, to PSF models generated using the image-stacking method.  Among all 2265 images, \texttt{SWarp} and \texttt{reproject} perform best (1236 and 791 cases, respectively, or a total of $\sim$\,90\% of the sample), \texttt{photutils} stands out in 209 cases, while \texttt{PSFEx constant} and \texttt{PSFEx variable} excel compared to others in merely 24 and 5 cases, respectively.

\begin{figure*}[t]
\centering
\includegraphics[width=1.0\textwidth]{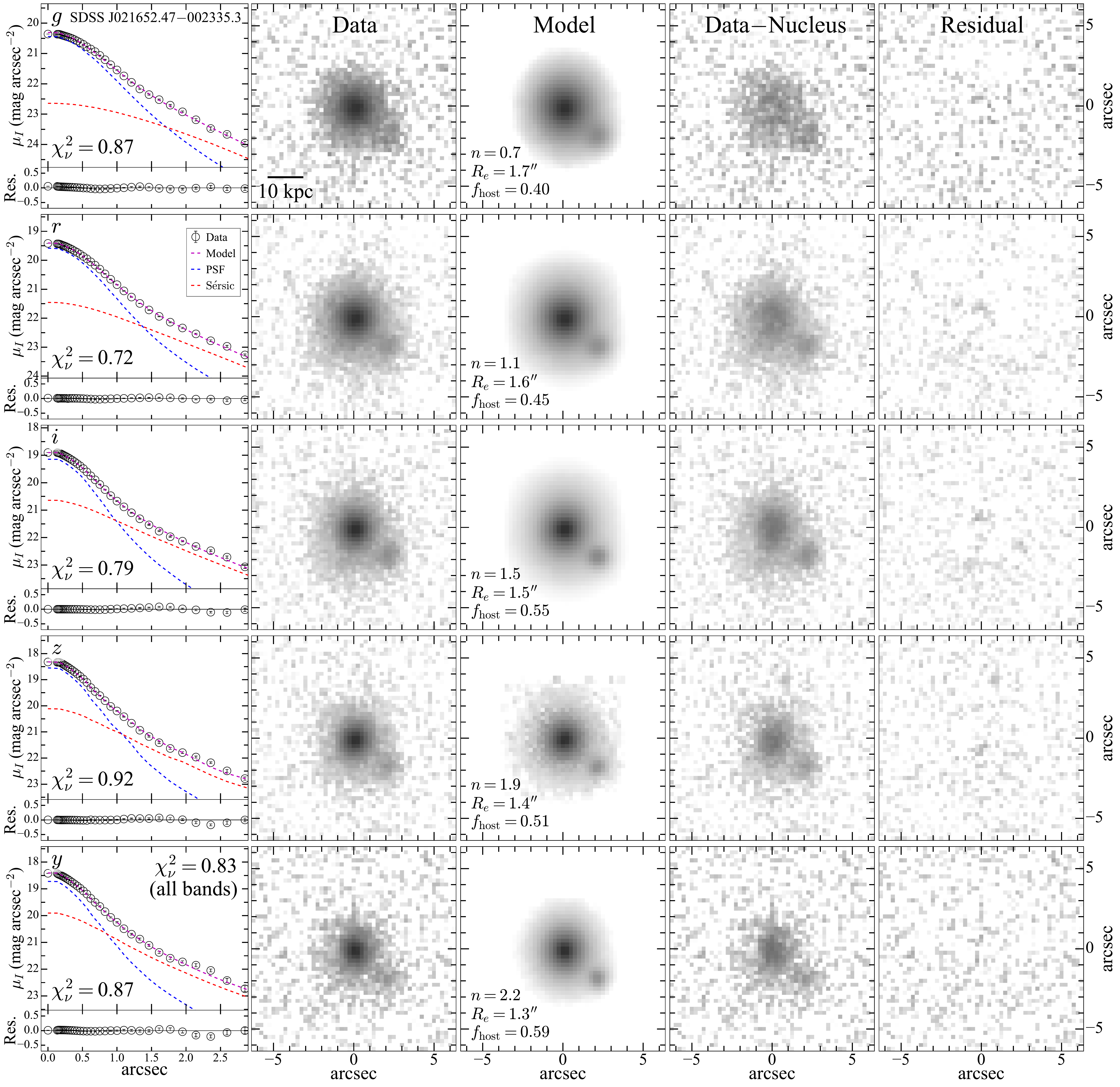}
\caption{An example of the simultaneous multiwavelength decomposition of SDSS~J021652.47$-$002335.3 (redshift $0.304$). Rows from top to bottom are the results for filters $grizy$. The upper panel of the left column shows the radial profile of the surface brightness (open circles with error bars), PSF component (blue), \sersic\ (red) component, and total model (purple; PSF $+$ \sersic).  The $\chi_{\nu}^2$ from \texttt{GALFITM} for each band is shown in the lower-left corner, while that for all five bands is shown in the upper-right corner of the first panel in the bottom row.  The lower sub-panel gives the residuals between the data and the model (data$-$model). The images show, from left to right, the original data, best-fit total model, data with the nucleus component modeled by a PSF subtracted, and residuals of the total model. The best-fit \sersic\ index,  half-light radius ($R_e$), and host fraction $f_{\rm host}  \equiv $ {\sersic\ / (PSF $+$ \sersic)} are shown in the third column.  Two other examples of objects with the lowest and highest $\chi_{\nu}^2$ in the sample (see Section~\ref{sec4.1}) are shown in Appendix~\ref{appendix1}. Decomposition results of the entire sample can be found at \dataset[https://doi.org/10.12149/101130]{https://doi.org/10.12149/101130}.}
\label{fig4}
\end{figure*}

\begin{figure}[t]
\centering
\includegraphics[width=0.48\textwidth]{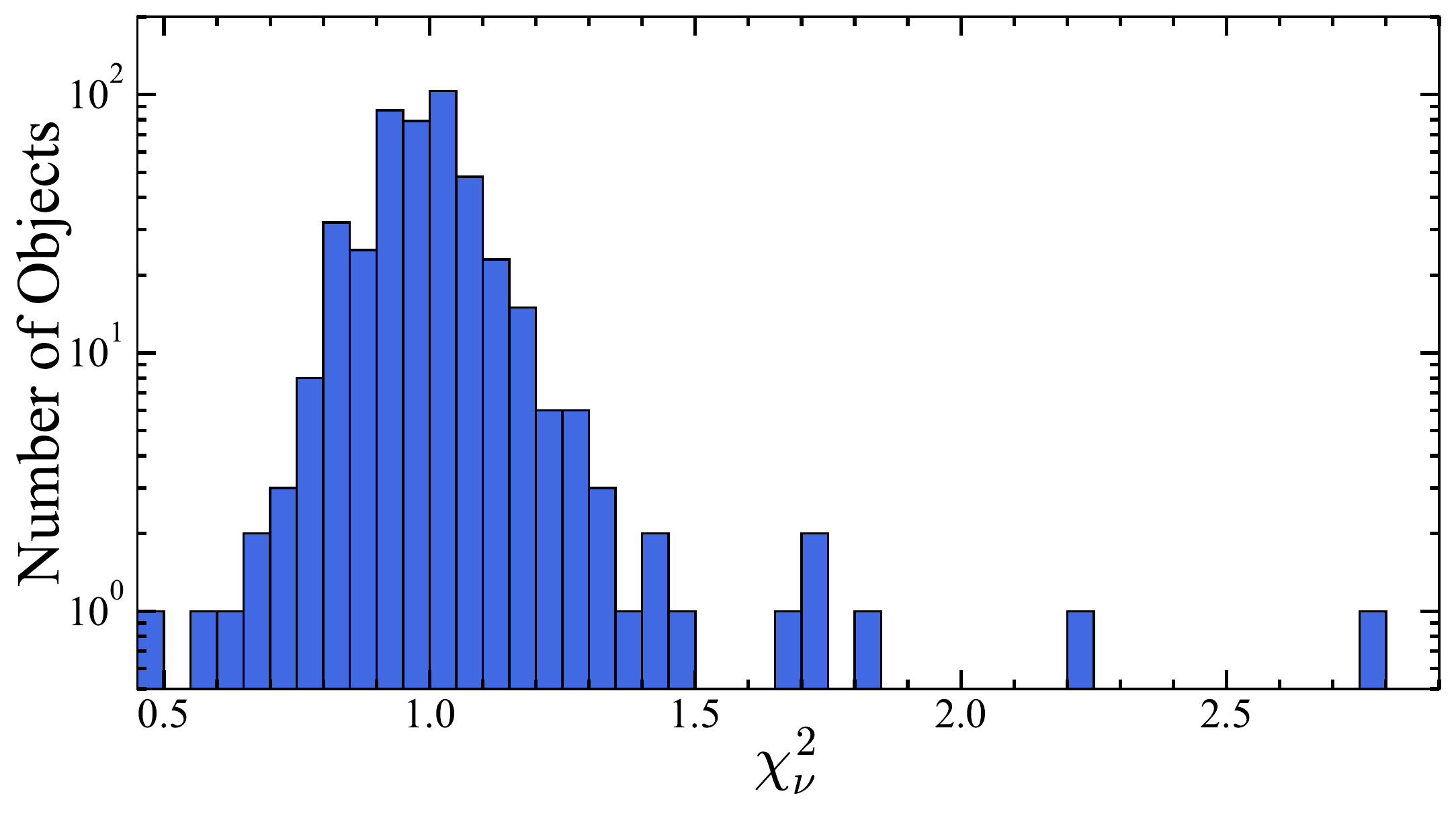}
\caption{Distribution of $\chi_{\nu}^2$ of the simultaneous fits to all five bands.}
\label{fig5}
\end{figure}

\section{Image Decomposition} \label{sec3}

\subsection{Two-dimensional Simultaneous Multiwavelength Fits} \label{sec3.1}

The morphology and structure of a galaxy may depend on wavelength as a result of internal variations in stellar population and dust attenuation. \citet{Kelvin+2012MNRAS} performed multi-band modeling of the images of 167,600 galaxies, fitting each galaxy independently with images covering nine filters. They find that the \sersic\ index and half-light radius of both early-type and late-type galaxies vary smoothly and systematically with wavelength: $n$ increases and $R_e$ decreases from the ultraviolet to the near-infrared. This reflects the fact that shorter wavelengths are more sensitive to dust attenuation and young stellar populations, while longer wavelengths better trace older stars. To account for bandpass variation of galaxy structure, we adopt \texttt{GALFITM} \citep{Haussler+2013MNRAS, Vika+2013MNRAS}, which can handle two-dimensional multiwavelength images under the familiar framework of \texttt{GALFIT} \citep{Peng+2002AJ, Peng+2010AJ} by introducing a wavelength-dependent function to each galaxy model parameter. \texttt{GALFITM} uses a series of Chebyshev polynomials to model the wavelength dependence.  The maximum order of each polynomial series can be specified by the user. All the free parameters of the model are fitted to the multi-filter data simultaneously by minimizing a single likelihood function, which significantly improves the reliability of the structural information extracted from bands with low signal-to-noise ratio \citep{Vika+2013MNRAS}. 

We describe each active galaxy with a two-component model comprising an unresolved nucleus mimicked by the PSF and a single \sersic\ function to represent the host galaxy. The fitting region is set to $51 \times 51$ pixels, which covers $\sim$\,$60 \times 60$ kpc. As discussed in Section~\ref{sec3.2.1}, a single \sersic\ component suffices to model the stellar emission of redshift $\sim$\,0.3 galaxies under the conditions of the PS1 survey. We follow \citet{Haussler+2013MNRAS} for the choice of the wavelength dependence of parameters. The ellipticity and position angle are constant with wavelength, but the magnitudes of the AGN and galaxy are free to vary with wavelength, described by a polynomial with a maximum order of 4. H{\"a}u{\ss}ler  et al.'s analysis of $\sim$\,$10^4$ nearby galaxies indicates that under most circumstances a linear function is enough to describe the wavelength dependence of \sersic\ index $n$ and half-light radius $R_e$. In a few cases, a higher second-order function is required to model the mild curvature mostly seen around the $H$ band ($\sim$\,1.6 \micron). Considering the rest-frame wavelength coverage of PS1 at redshift $\sim$\,0.3 (3600--7400 \AA), we only allow $n$ and $R_e$ to vary linearly with wavelength. We further tie the positions of the AGN and galaxy together, as generally done in previous works \citep[e.g.,][]{Kim+2008ApJS}, under the assumption that the supermassive BH is located in the center of the galaxy. Some objects have nearby companions within the fitting region. We simultaneously fit objects that are close to or overlap with our target AGN, using a PSF model for small, unresolved objects and a \sersic\ model for extended, resolved sources, all the while masking other sources. 

We adopt object-specific initial guesses for the position, magnitudes for the PSF model ($m_{\rm PSF}$) and \sersic\ model (\msersic), and general parameter constraints to improve and speed up the fitting process. We take the initial guess for $m_{\rm PSF}$ from the PS1 catalog, which derives from fitting a PSF model alone to the object. As for the host, an initial guess comes from the difference between $m_{\rm Kron}$ and $m_{\rm PSF}$, $m_{\rm s\acute{e}rsic}=-2.5\times \log(10^{-m_{\rm Kron}/2.5} - 10^{-m_{\rm PSF}/2.5})$, but we constrain it to be brighter than 26~mag, $\sim$\,4~mag weaker than the faintest detected object. The \sersic\ index is allowed to vary within the range $n = 0.3-8$ and the half-light radius $R_e = 0.5-20$ pixels, which corresponds to $\sim$\,$0.6-22.3$ kpc at redshift $0.3$. In practice, these boundaries are seldom reached during the fitting process. 

An example of our simultaneous, multi-object, multi-band decomposition is given in Figure~\ref{fig4}. The smooth residuals and low $\chi_{\nu}^2$ indicate that the fitting is quite successful. From the surface brightness profiles, we can clearly discern the extended emission from the host galaxy and distinguish it from the unresolved nucleus. Consistent with our expectations, we find that the host-to-total flux ratio ($f_{\rm host}=0.40-0.59$) and \sersic\ index ($n=0.7-2.2$) increase significantly and systematically from the $g$ band to the $y$ band. Note that at redshift $\sim$\,0.3 broad \Ha\ falls into the $z$ bandpass, causing $f_{\rm host}$ to decrease from 0.55 in the $i$ band to 0.51 in the $z$ band. 

For most of the objects, we achieve good fits overall with $\chi_{\nu}^2$ peaked at $\sim$\,1 (Figure~\ref{fig5}). Six outliers ($\chi_{\nu}^2>1.5$) have significant residuals and are excluded from further analysis.  Two additional objects are removed because they suffer from contamination by a nearby, bright star.

\begin{figure*}[t]
\centering
\includegraphics[width=0.8\textwidth]{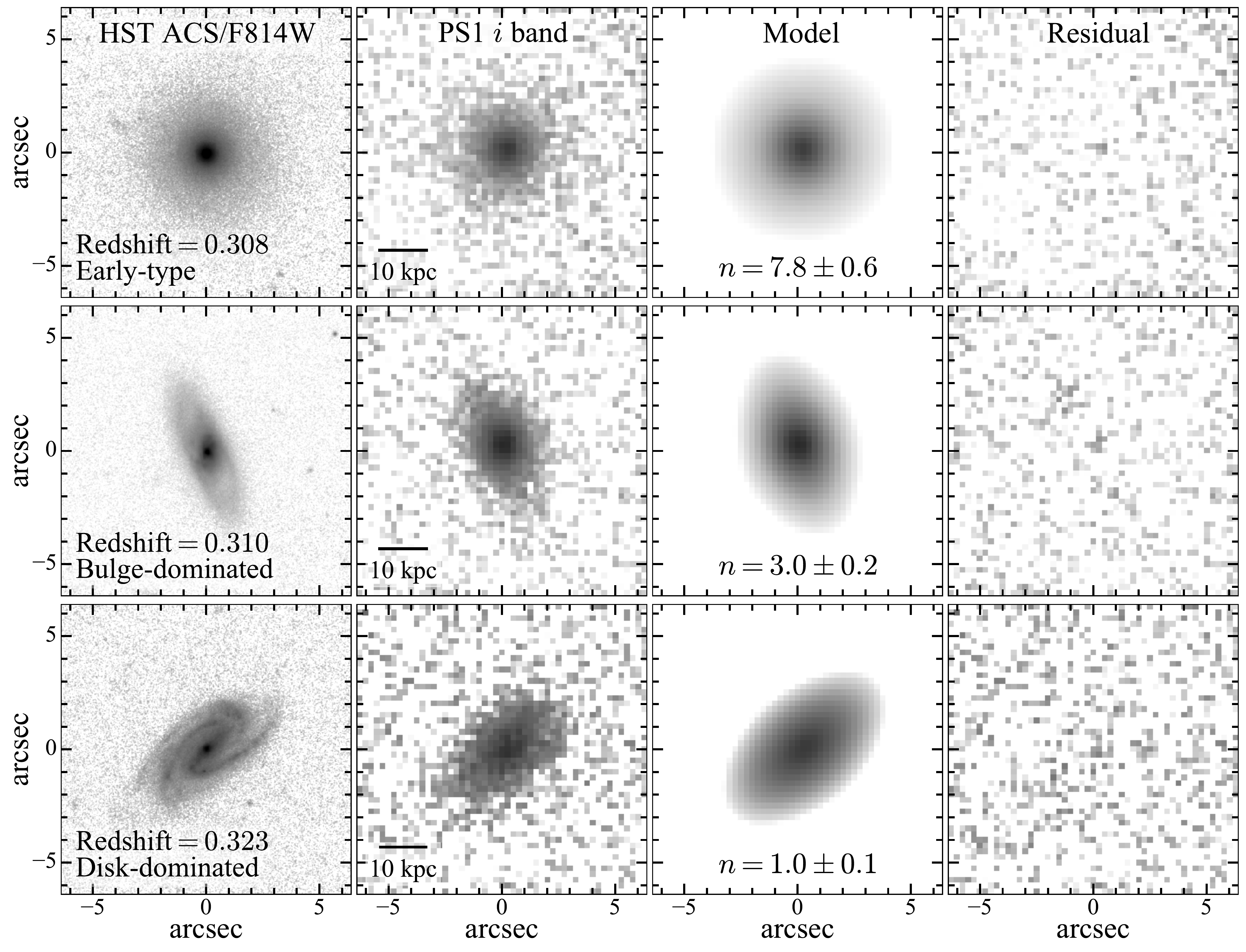}
\caption{Examples of images and fits using \texttt{GALFITM} for (top) early-type, (middle) bulge-dominated, and (bottom) disk-dominated galaxies. The columns, from left to right, show the HST ACS/F814W image, the PS1 $i$-band image, the best-fit single \sersic\ model for the $i$ band, and the residuals (data $-$ model). The best-fit \sersic\ index $n$ and its uncertainty are shown in the panel for the model.}
\label{fig6}
\end{figure*}

\begin{figure*}[t]
\centering
\includegraphics[width=0.9\textwidth]{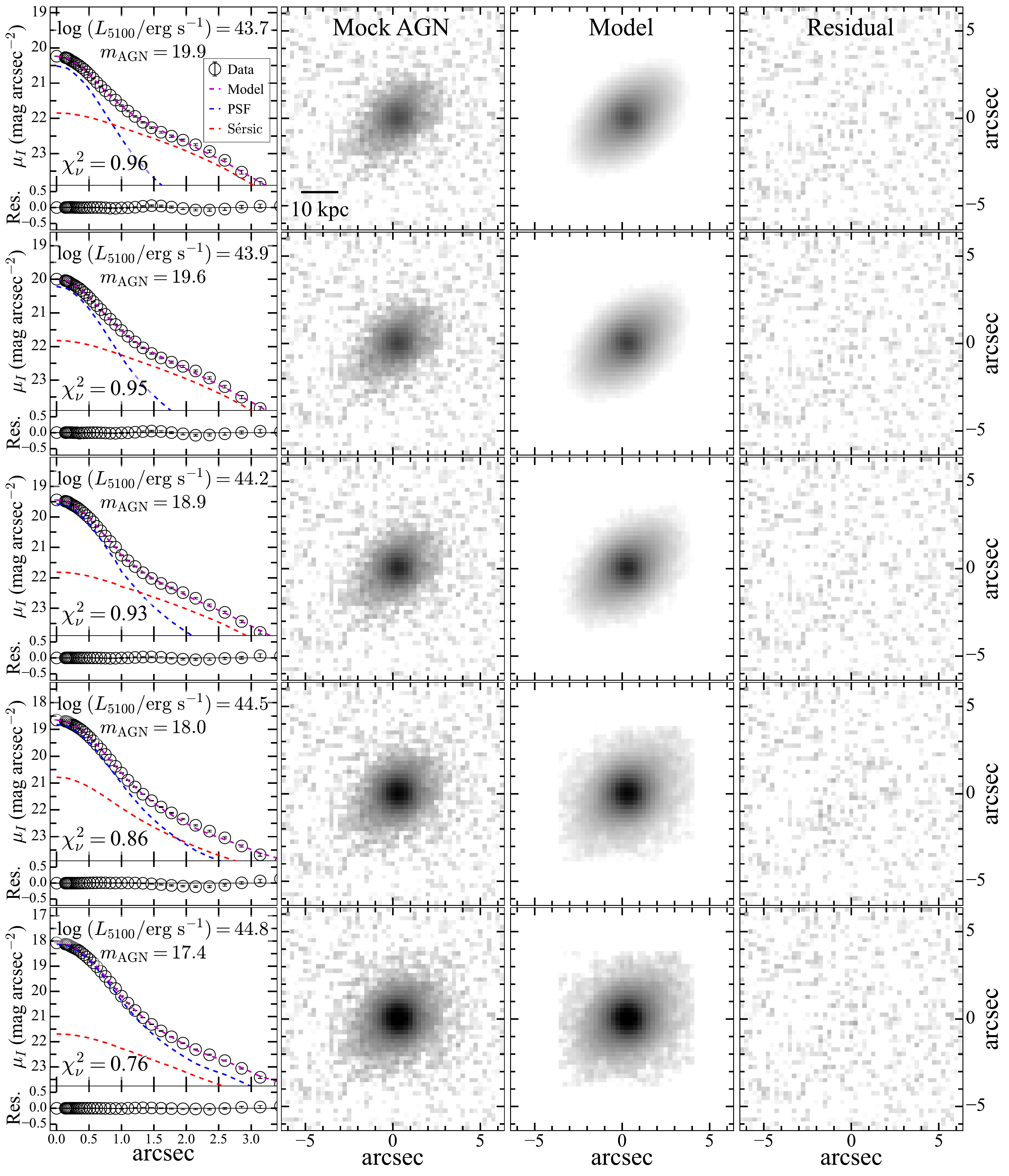}
\caption{Examples of fits to mock AGNs in the $i$ band generated from the third (disk-dominated) object in Figure~\ref{fig6} (\msersic\ $=19.1$~mag) with AGN strength increasing from $m_{\rm AGN} = 19.9$~mag (top) to 17.4~mag (bottom). Columns are similar to Figure~\ref{fig4}, showing the surface brightness profile, data of the mock AGN, the best-fit model, and the residuals (mock AGN $-$ model).}
\label{fig7}
\end{figure*}

\begin{figure*}[t]
\centering
\includegraphics[width=\textwidth]{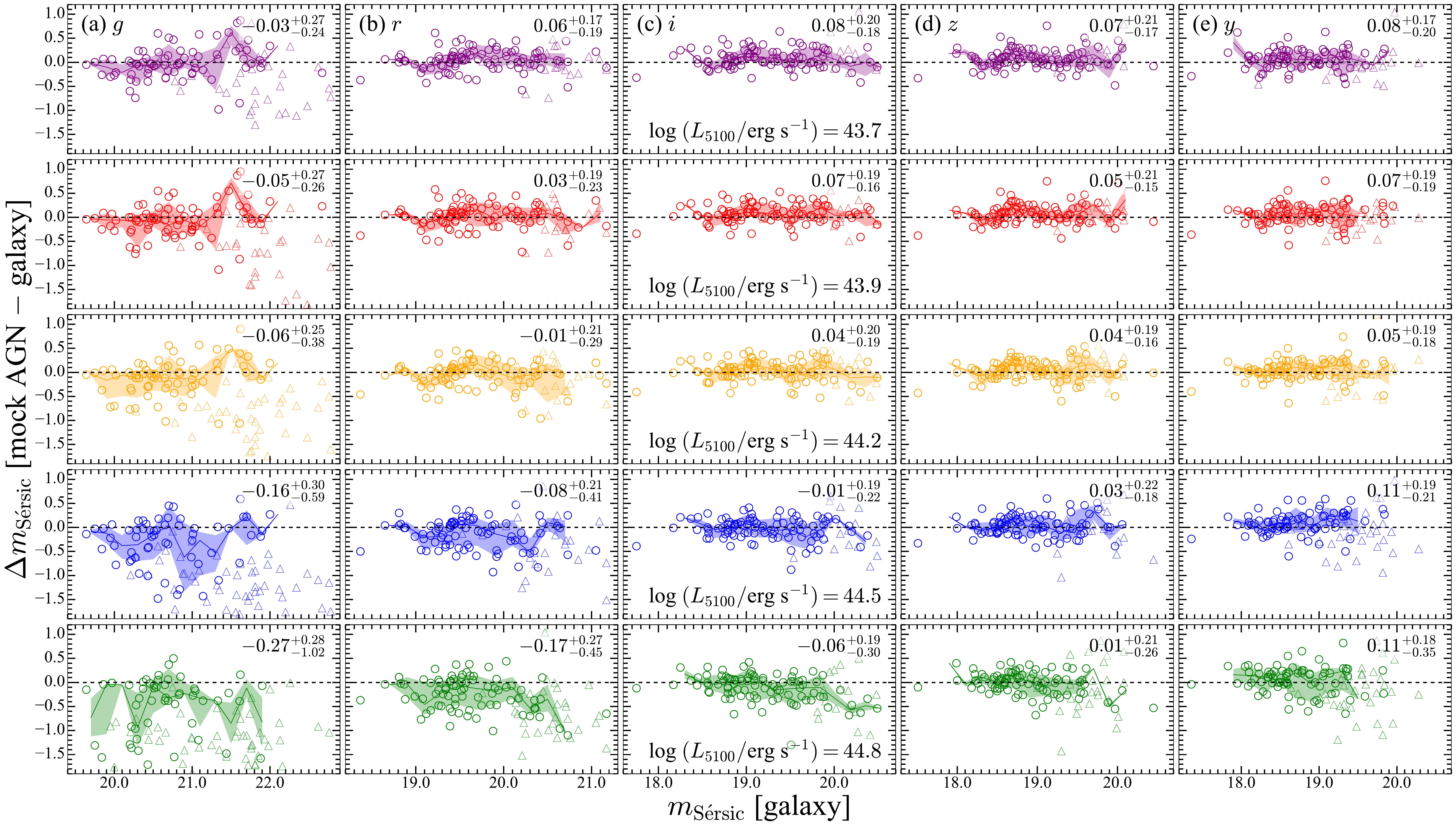}
\caption{Comparison of the difference in \msersic\ ($\Delta$\msersic) between the best-fit magnitude measured before and after adding the mock AGN, for galaxies selected from the COSMOS field as observed in the $grizy$ bands (panels a--e). Various colors represent AGNs with different strengths ($L_{5100}$), ranging from the 5th (purple), 16th (red), 50th (orange), 84th (blue), and 95th (green) percentiles of our redshift = 0.3 quasar sample. Circles represent objects with value at least 3 times its error, and faint triangles are for those with value less than 3 times its error. The upper-right corner of each panel gives the median and upper and lower $1\,\sigma$ (84th and 16th percentiles $-$ median) of $\Delta$\msersic\ of the circles.}
\label{fig8}
\end{figure*}

\begin{figure}[t]
\centering
\includegraphics[width=0.5\textwidth]{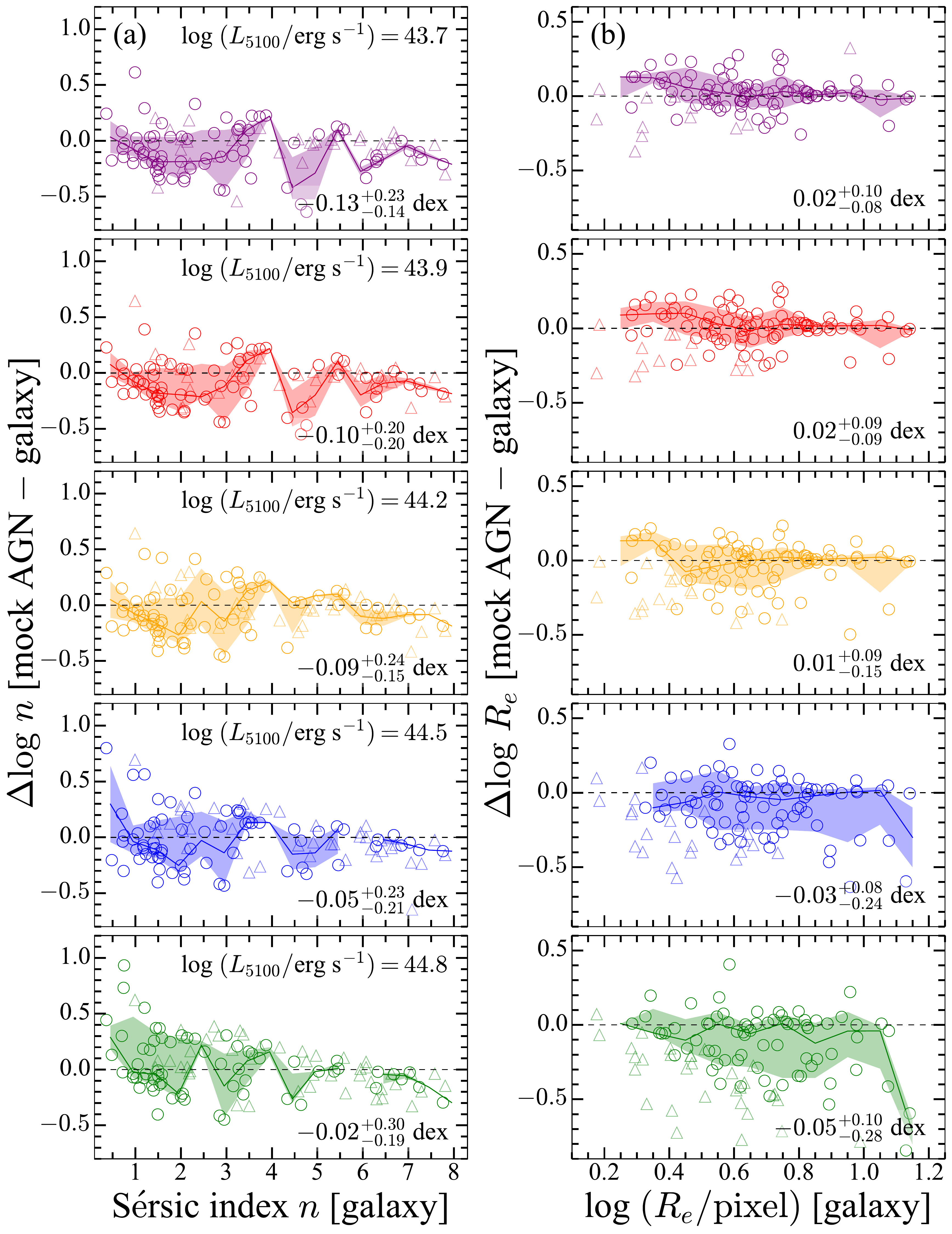}
\caption{Same as in Figure~\ref{fig8}, but for (a) \sersic\ index $n$ and (b) half-light radius $R_e$. We focus only on the results for the $i$ band.  The statistics in panel (a) are calculated after taking the logarithm of the value (i.e., $\log\,n$).}
\label{fig9}
\end{figure}

\subsection{Verification of Results}\label{sec3.2}

The rich morphological complexities of galaxies are difficult to describe by simplified parametric models.  Parameters returned by codes such as \texttt{GALFIT} and \texttt{GALFITM} may suffer from various degrees of systematic bias and degeneracy, which will result in underestimation of the true uncertainties, even if the formal statistical errors are small and the fitting residuals look acceptable \citep[e.g.,][]{Haussler+2007ApJS, Kim+2008ApJS, Vika+2013MNRAS, Gao&Ho2017, Zhao+2021ApJ}. We design realistic input-output experiments to verify our results using two sets of mock AGNs: one from galaxies actually observed at redshift $\sim$\,0.3 and another from model galaxies generated from best-fit parameters.

\subsubsection{Mock AGNs from Real Galaxies} \label{sec3.2.1}

Even though a single \sersic\ function can describe the overall emission profiles of AGN host galaxies at redshift $\sim$\,0.3 as observed under the conditions of PS1 (Figures~\ref{fig4} and \ref{fig5}), significant uncertainties can arise from model mismatch, as the assumed model vastly oversimplifies the true underlying galaxy, which can contain substructures such as rings, lenses, bars, and spiral arms. To investigate whether and to what extent our results are affected by these potential issues, we generate mock AGNs by adding an artificial point source to the center of real redshift $\sim$\,0.3 galaxies, and then compare the parameters measured before and after the addition of the point source. 

For this purpose, we select objects with spectroscopic redshifts from the zCOSMOS survey \citep{Lilly+2009ApJS} in the COSMOS field \citep{Scoville+2007ApJS}, which is covered in the PS1 survey footprint. These objects have abundant information, such as high-resolution images from the HST Advanced Camera for Surveys (ACS) \citep{Koekemoer+2007ApJS}, multiwavelength data from the X-rays to the radio, morphological classifications \citep{Scarlata+2007ApJS}, and stellar mass measurements \citep{Laigle+2016ApJS}. We choose 111 redshift $0.30-0.35$ galaxies with morphological classification and $M_* = 10^{10.4} - 10^{11.4}\, M_{\odot}$, which spans the 5th--95th percentile of the stellar mass range of our quasars, taking care to avoid objects close to bright stars. To ensure that these are genuine representations of inactive galaxies, we exclude objects with detected X-ray emission.

Since the image-stacking method provides overall better performance than the pixel-basis method (Section~\ref{sec2.3}), we only use \texttt{reproject} to construct the PSF models.  We then use \texttt{GALFITM} to perform multiwavelength simultaneous fitting to derive the parameters of the redshift $\sim$\,0.3 galaxies, following the same procedure as that employed for the AGNs. Figure~\ref{fig6} gives examples of high-resolution HST ACS/F814W images, PS1 $i$-band images, best-fit models, and residual images for galaxies of three different morphologies (early-type, bulge-dominated, and disk-dominated). The residuals are quite clean.  This exercise demonstrates that a single \sersic\ component adequately describes the galaxy light distribution at redshift $\sim$\,0.3, for images of galaxies that should closely resemble those of the underlying population of AGN hosts under consideration, as captured with the image quality and observational characteristics of PS1. 

To generate mock active galaxies, we add a point source to the center of each galaxy with AGN strength following the same distribution as the actual AGN sample. From the catalog of \citet{Liu+2019ApJS}, for our sample the 5th, 16th, 50th, 84th, and 95th percentile of the monochromatic AGN continuum luminosity at 5100~\AA\ is $\log \, (L_{5100}/{\rm erg~s^{-1}}) = 43.7, 43.9, 44.2, 44.5$, and 44.8, respectively. With the composite quasar SED of \citet{Richards+2006ApJS}, at a median redshift of 0.32 these values of $L_{5100}$ translate to $m_{\rm AGN} = 19.9, 19.6, 18.9, 18.0$, and 17.4~mag for the $i$ band.  In the end, we have in total 555 mock AGNs spanning the same parameter space as our true sample. Examples of fits of mock AGNs with different $m_{\rm AGN}$ are shown in Figure~\ref{fig7}. For the brightest case ($m_{\rm AGN}=17.4$~mag), the nucleus can dominate the observed total emission even out to a radius of $\sim$\,$3\arcsec$. Fortunately, even under these conditions the AGN point source can still be distinguished from the extended emission of the underlying host galaxy.

To test whether we can reliably recover the host galaxy parameters in the presence of an AGN, we compare \msersic\ before and after adding a simulated active nucleus to the galaxy image (Figure~\ref{fig8}). In general, \msersic\ can be recovered with quite good consistency in $rizy$, with a small median magnitude difference ($\lesssim 0.1$~mag) and $\sim$\,0.2~mag scatter within the 16th and 84th percentiles of the range of AGN strengths explored (three middle rows). The higher contrast between the AGN and its host galaxy in the $g$ band introduces a larger systematic difference (0.09~mag) and scatter ($\sim$\,0.3~mag; averaged over the 16th, 50th, and 84th percentiles of AGN strength). At the same time, the fraction of objects with measurement values less than 3 times its error (unreliable measurements) is much higher compared to the average of the other four bands ($39\%$ versus $21\%$).  The best performance is found in $i$, for which the fraction of unreliable measurements is only $13.5\%$. Even in the extreme case when the AGN strength is at the 95th percentile, we still achieve reasonable consistency in the $rizy$ bands, with the $g$ band, once again, faring the worst. 

Figure~\ref{fig9} examines the effect of the AGN on our ability to measure the \sersic\ index and half-light radius, focusing on the results in the $i$ band, which gives the best performance.  We find that the presence of an AGN generally leads to a systematic underestimate of $n$, on average by $\sim$\,0.08~dex. The effect of PSF convolution evidently reassigns a portion of the central galaxy flux to the point source, which lowers the value of $n$. The parameter least affected is $R_e$, with close to zero systematic offset and only $\sim$\,0.12~dex scatter between the 16th and 84th percentiles of AGN strength. 

We also assess the performance of separate single-band decomposition, as commonly practiced in previous works \citep[e.g.,][]{Matsuoka+2014ApJ, Ishino+2020PASJ, Li_Junyao+2021ApJ}.  With the same set of mock AGNs generated here, we fix the galaxy structure to that determined from the $i$ band and compare the behavior of separate single-band decomposition with our preferred method of simultaneous multiwavelength decomposition. The performance is notably inferior, as summarized in Appendix~\ref{appendix2}.

Our mock tests attempt to simulate, to the extent possible, conditions that resemble our actual observed sample.  We demonstrate that we can obtain reliable estimates of host galaxy magnitude and size.  Significant scatter remains, however, and the systematic uncertainties on \sersic\ index are particularly serious, as are difficulties with the $g$ band where AGN contamination becomes most acute, or in the $y$ band where the sensitivity sharply drops.  We apply systematic corrections to $n$ and \msersic\ (the latter only in the $g$ and $y$ bands) based on the mean offsets from the true values measured at the 16th, 50th, and 84th percentiles of AGN strength, and we include the corresponding mean scatter into the error budget as a quadrature sum. A summary of the uncertainties is given in Table~\ref{tableA1} of Appendix~\ref{appendix2}.  As expected, stronger AGNs in images of lower signal-to-noise (shorter exposure time) may completely dominate the observed fluxes due to high sky background level and lead to unreliable results. To help mitigate against this issue, we use SED fitting to help rule out objects with unrealistic SEDs (see Section~\ref{sec4.1}).

\subsubsection{Mock AGNs from Idealized Galaxies}\label{sec3.2.2}

\begin{figure}[t]
\centering
\includegraphics[width=0.4\textwidth]{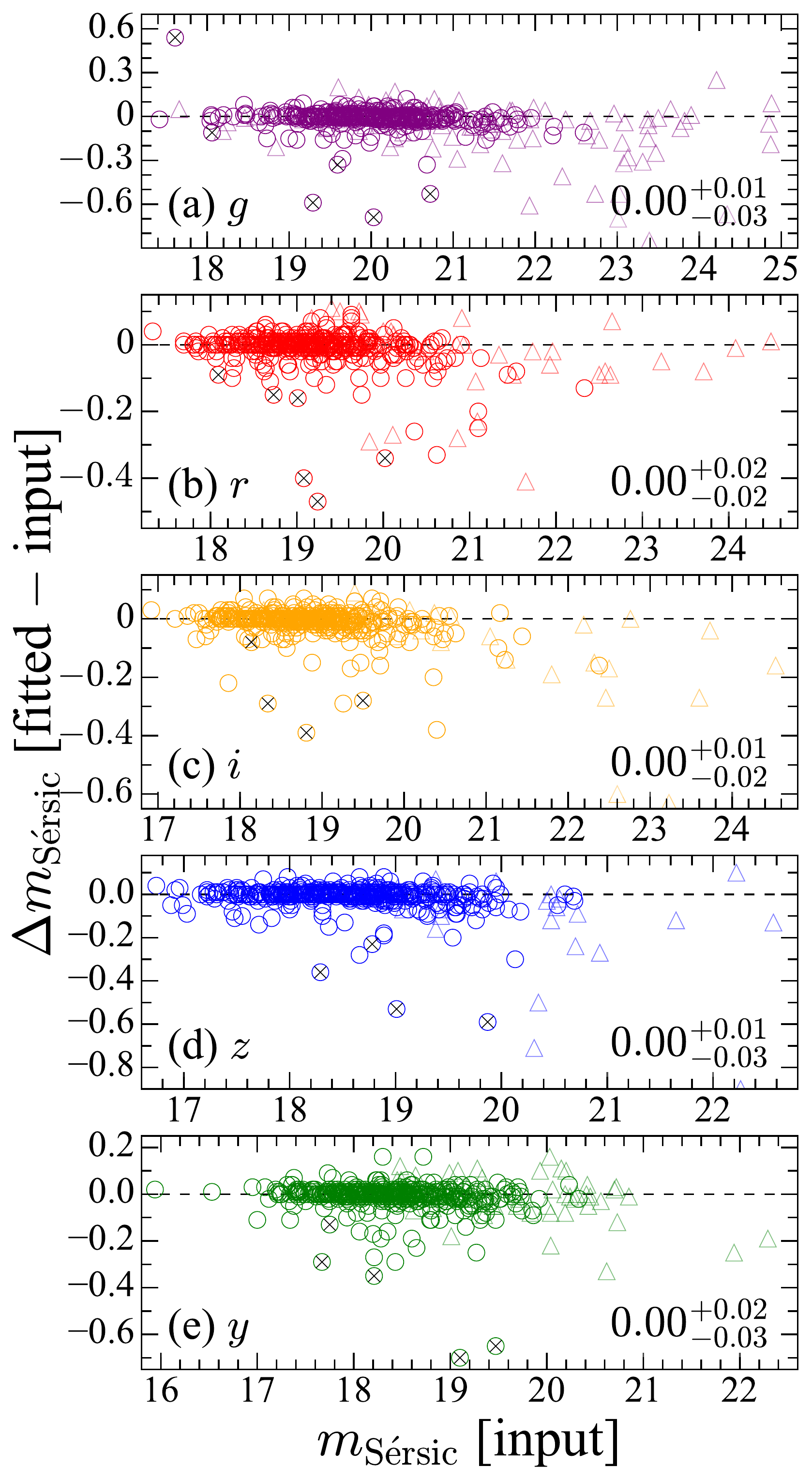}
\caption{Comparison of best-fit \msersic\ of the mock AGNs with those used as input parameters to generate mock AGNs in the five bands (panels a--e). Mock AGNs are generated using best-fit parameters from simultaneous fitting to observed five-band data for the real AGNs. Symbols are the same as in Figure~\ref{fig8}. Objects with $|\Delta$\msersic$|>3\,\delta$ \msersic\ [fitted] are marked with crosses, where $\delta$\msersic\ is the error of \msersic. The median and upper and lower $1\,\sigma$ (84th and 16th percentiles $-$ median) of the differences (fitted $-$ input) for the detections (circles excluding crosses) are shown in the lower-right corner of each panel. }
\label{fig10}
\end{figure}

\begin{figure}[t]
\centering
\includegraphics[width=0.45\textwidth]{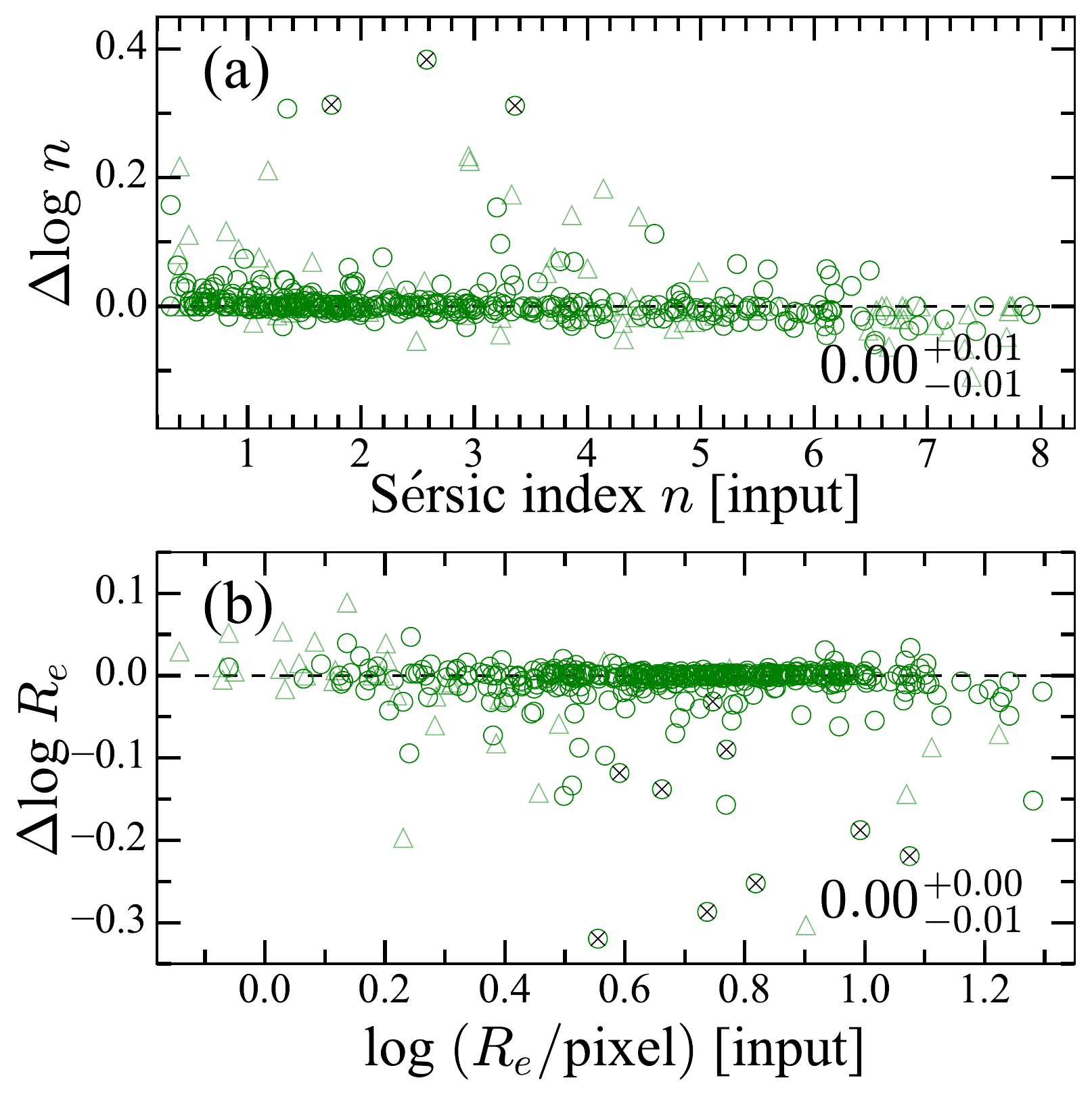}
\caption{Similar as Figure~\ref{fig10}, but for (a) \sersic\ index $n$ and (b) half-light radius $R_e$. We focus only on the results for the $i$ band.}
\label{fig11}
\end{figure}

The above simulations neglect the complication that the PS1 images do not have uniform depth. The exposure times of our sample range from a minimum of 390~s to a maximum of 3720~s, with a median of 817, 896, 1427, 720, and 800~s in the $grizy$ bands, respectively. For reference, the median exposure times for the matched galaxy sample in the COSMOS field discussed in Section~\ref{sec3.2.1} are longer in $gr$ and shorter in $izy$. This difference makes it difficult to assess the impact of signal-to-noise ratio on individual objects.  We address this issue using idealized simulations in which we use the best-fit parameters derived for each object to construct mock images that exactly mimic the object-specific parameters of the actual observation. For each object in each band, we generate 100 realizations of mock observations that account for the Poisson noise associated with the source, background Gaussian noise, and the properties of the specific image (gain, exposure time, and background variation).  Then, we use \texttt{GALFITM} to repeat the decomposition and adopt the median value and standard deviation of the 100 results as the measurement and its error. Excellent consistency is found with the input values (Figures~\ref{fig10} and \ref{fig11}), with no detectable systematic bias and very small scatter. A handful of objects (highlighted with crosses) deviate by more than 3 times its uncertainty. We exclude the objects with inconsistent fluxes when analyzing the SEDs in Section~\ref{sec4.1}, and those with highly discrepant $n$ or $R_e$ are deemed to have unreliable morphology in Section~\ref{sec4.2}. The uncertainties derived from this set of simulations are added in quadrature to the final error budget of each parameter.

\section{Results}\label{sec4}

\subsection{Derivation of Stellar Masses}\label{sec4.1}

We use the SED fitting code \texttt{CIGALE} \citep{2019A&A...622A.103Boquien+} to derive the stellar masses of the host galaxies. We first correct the photometry for Galactic foreground extinction using the Python package \texttt{dustmaps} \citep{2018JOSS....3..695Green} and the dust maps from \citet{1998ApJ...500..525Schlegel+}, converting $E(B-V)$ to extinction for the PS1 bands assuming $R_V=3.1$ \citep{2011ApJ...737..103Schlafly&Finkbeiner}. We only consider objects with host galaxy component detected in at least four bands with a signal-to-noise ratio of 3.  This reduces the sample to 366 objects (332 detected in all five bands, 34 detected in four bands), or $81\%$ of the original sample. Most of the non-detections are in the $g$ band, mostly attributable to non-uniform exposure time, intrinsic weakness of the galaxy, or high AGN-to-host contrast in the blue. 

\begin{figure}[t]
\centering
\includegraphics[width=0.48\textwidth]{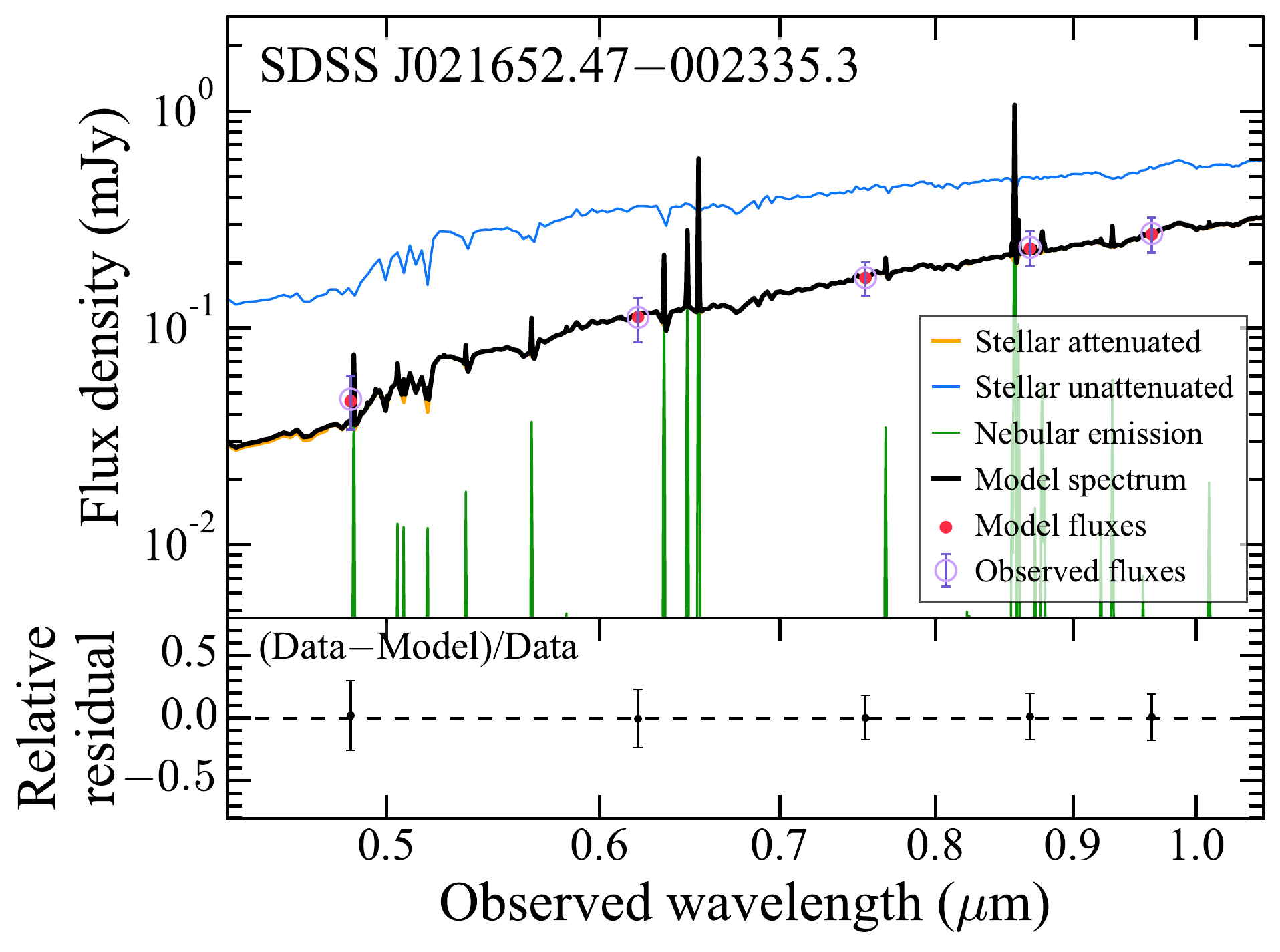}
\caption{SED fitting to the five PS1 bands of SDSS~J021652.47$-$002335.3 using \texttt{CIGALE}. The observed fluxes (purple open circles) have been subtracted of AGN contribution. The upper panel shows the best-fit total model spectrum (black solid curve), its components (stellar attenuated: yellow solid curve; nebular emission: green solid curve), the model fluxes (red filled circles), and the stellar emission before attenuation (blue curve). Relative residuals (data$-$model)/data are shown in the lower panel.}
\label{fig12}
\end{figure}

We adopt a ``delayed'' star formation history model, which has a functional form ${\rm SFR}(t) \propto t/\tau^2 \exp (-t/\tau)$, with $\tau$ the e-folding time of the stellar population, and an exponential burst to allow for a recent episode of star formation \citep[e.g.,][]{Kauffmann+2003MNRAS, Wild2007MNRAS, Kim&Ho2019ApJ}. The stellar component is represented using single stellar population models from \citet[BC03]{BC03}, with a \citet{Chabrier2003PASP} initial mass function and solar metallicity (0.02). Stellar masses are converted to our adopted \citet{Kroupa2001MNRAS} initial mass function by multiplying by a factor of 1.08 \citep{Madau&Dickinson2014ARA&A}. We assume a fixed ionization parameter ($\log\, U=-2$) for the nebular emission. Both the stellar continuum and the nebular emission are attenuated by dust using the \texttt{dustatt modified starburst} module, assuming the extinction law of \citet{Calzetti+2000ApJ} and a fixed ratio of 0.44 for the $E(B-V)$ of the stellar continuum to that of the line emission. The modules and input parameters used in this paper are given in Table~\ref{table1}. Figure~\ref{fig12} shows an example SED fit to the five-band PS1 photometry.

\begin{figure*}[t]
\centering
\includegraphics[width=0.45\textwidth]{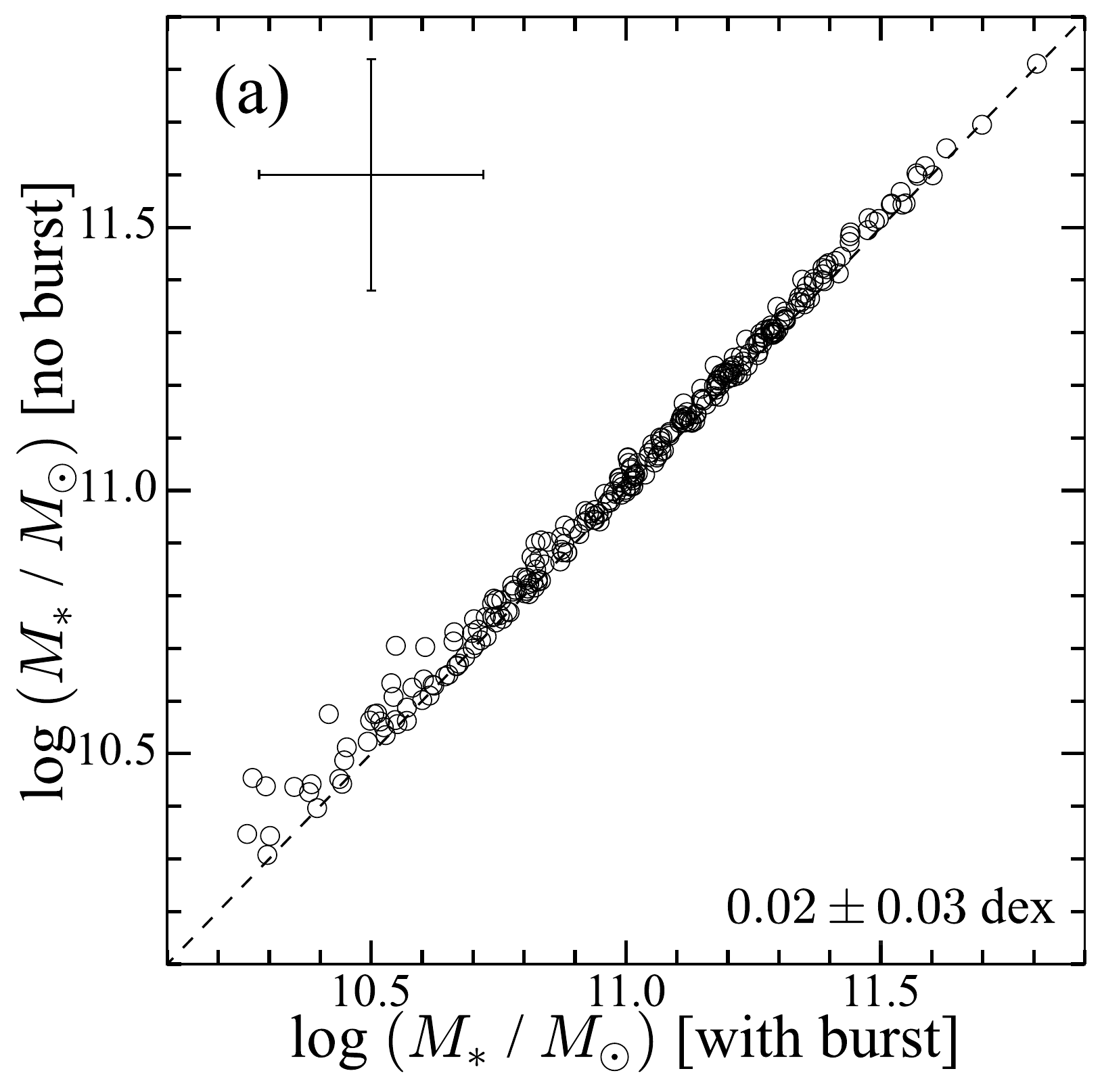}
\includegraphics[width=0.45\textwidth]{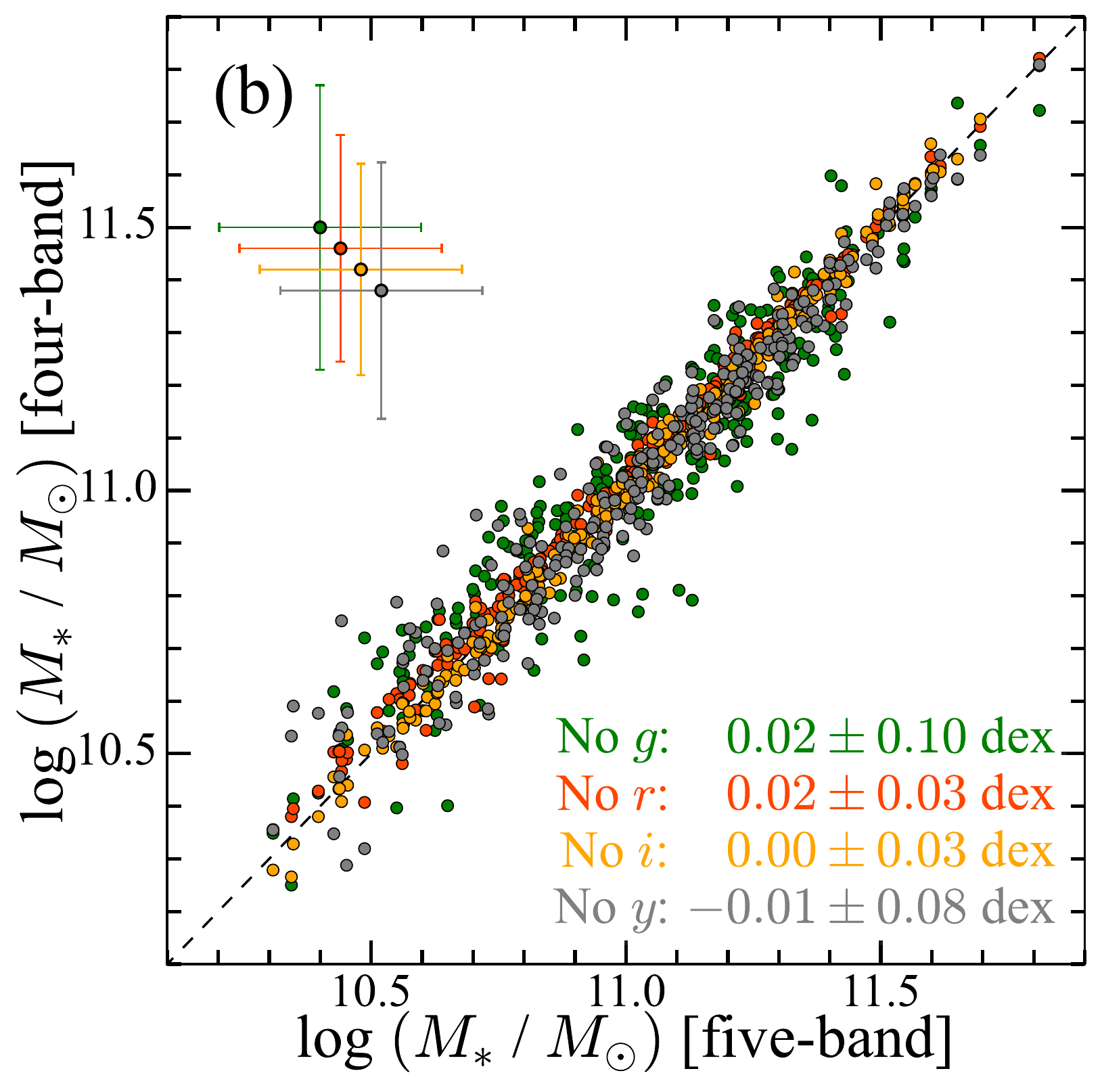}
\caption{Comparison of stellar masses for 286 objects with $\chi_{\nu}^2\leq1$ derived from \texttt{CIGALE} (a) with and without including a burst component in the star formation history model and (b) using complete (five-band) and incomplete (four-band) photometry after manually removing one of the bands. The green, red, orange, and gray dots represent objects with the $g$, $r$, $i$, or $y$ band removed, respectively.  Median differences (y-axis $-$ x-axis) and standard deviations for each comparison are shown in the lower-right corner; typical uncertainties are shown in the upper-left corner. The dashed line gives the 1:1 relation.}
\label{fig13}
\end{figure*}

We obtain the stellar mass from a Bayesian estimate provided by \texttt{CIGALE}, which is based on the probability density function of the likelihood associated with every template. Since our error is likely overestimated due to the statistical correction applied in Section~\ref{sec3.2} from analysis of mock AGNs, we only limit our attention to objects with high-quality SED fits ($\chi^2_{\nu} \leq 1$).  Incorporating an additional burst component has a negligible impact on the majority of the derived stellar masses ($0.02\pm0.03$~dex; Figure~\ref{fig13}a), and for the sake of reducing the number of free parameters, we adopt the star formation history model without a burst component.  A minority of the sample (34 objects) have incomplete SEDs, missing data in one of the bands.  We evaluate the significance of the gap in SED coverage by artificially removing a single band from the objects having complete SEDs.  The $g$ and $y$ bands are affected the most (scatter 0.10 and 0.08~dex, respectively), but in general the effects are minimal, with median systematic differences of $|\Delta{\log\, M_*}| \leq 0.02$~dex for all four bands (Figure~\ref{fig13}b).  We add the corresponding uncertainty into the error budget for the stellar masses of the objects lacking photometry in the $g$ or $y$ band.

\begin{figure}[t]
\centering
\includegraphics[width=0.5\textwidth]{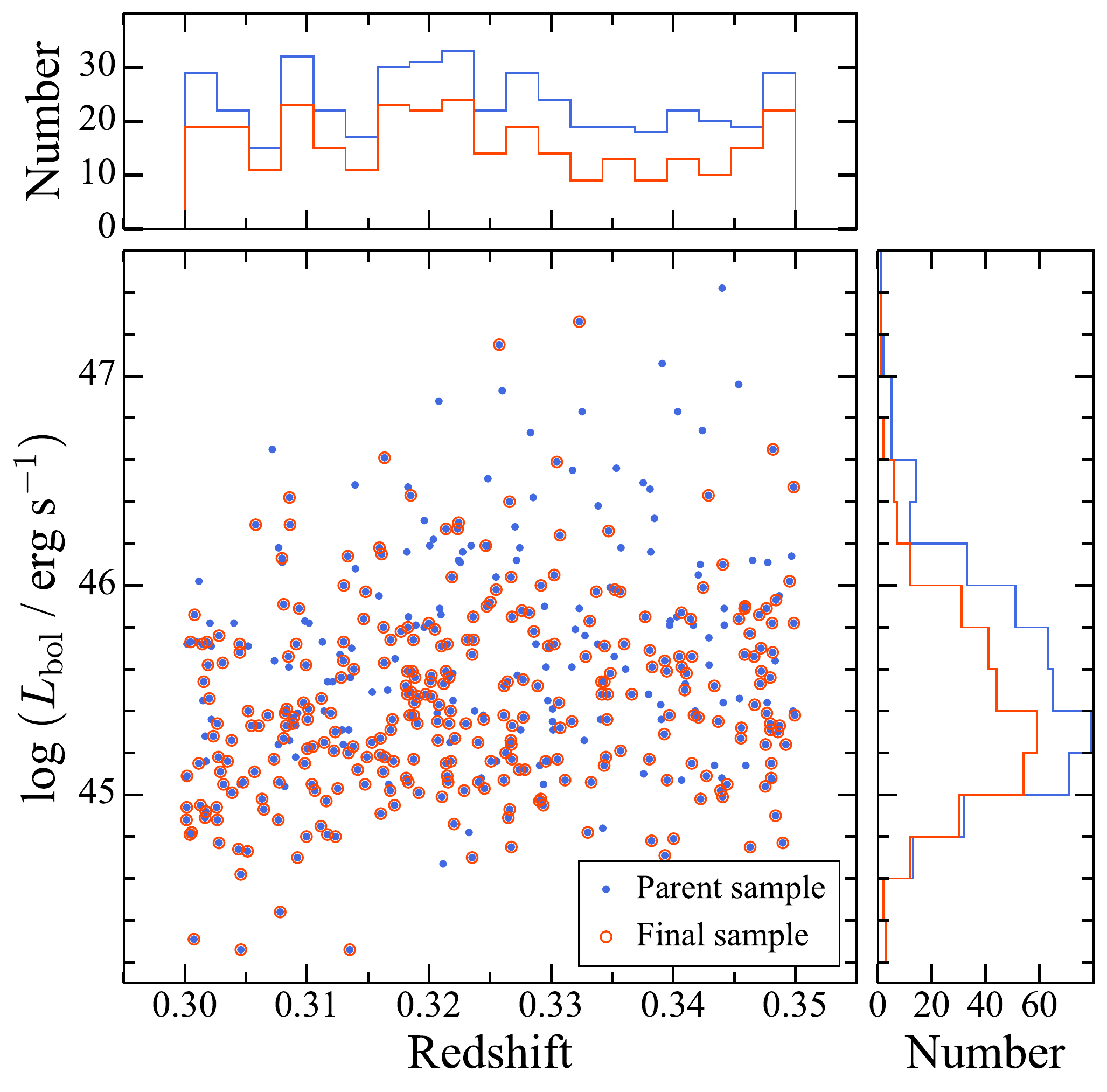}
\caption{Comparison of redshifts and AGN bolometric luminosities ($L_{\rm bol}$) for the parent sample (453 objects) and final sample (305 objects) that has \texttt{CIGALE} fits with $\chi^2_{\nu}\leq1$.}
\label{fig14}
\end{figure}

Finally, we successfully obtained reasonably accurate stellar masses for 305 of the 366 ($\sim$\,84\%) objects with good fits ($\chi^2_{\nu} \leq 1$). The median uncertainty of the stellar mass is $0.2$ dex. This final sample, which has a similar distribution of redshift and AGN bolometric luminosity as the parent sample (Figure~\ref{fig14}), is by far the largest sample of quasars with derived stellar masses at redshift $\sim$\,0.3. We present the basic properties of the final sample in Table~\ref{table2}. Table~\ref{table3} gives the fluxes of the host galaxies in the five PS1 bands, along with derived physical properties.

\begin{deluxetable*}{llc}
\caption{Parameters of the SED Fits \label{table1}}
\tablehead{
\colhead{Model} & \colhead{Parameter} & \colhead{Value}
}
\startdata
Star formation history:& e-folding time of the main stellar population in Gyr ($\tau_{\rm main}$)&0.001, 0.1, 0.25, 0.5, 0.75, 1.0, 1.5, 2.0, \\ 
delayed& & 2.5, 3.0, 4.0, 5.0, 6.0, 8.0, 10, 12, 15, 20\\
 &Age of the main stellar population in Gyr (${\rm age_{main}}$) &9.0\\
 &e-folding time of the late starburst population in Myr ($\tau_{\rm burst}$) & 5, 10, 25, 50, 100, 250, 500, 750, 1000\\
 &Age of the late burst in Myr (${\rm age_{burst}}$) &10, 25, 50, 100, 200, 250, 500, 750, 1000\\
 &Mass fraction of the late burst population (${f_{\rm burst}}$)* & $-10^{-5}$, 0, 0.005, 0.01, 0.015, 0.02, \\
 & & 0.04, 0.07, 0.10, 0.15, 0.20, 0.30\\
 \hline
Single stellar population: & Initial mass function & Chabrier \\
BC03&Metallicity & 0.02 (solar)\\
 \hline
 Nebular emission & Ionization parameter (log $U$) & $-2.0$\\
 \hline
 Dust attenuation: modified & The color excess of the nebular lines in mag [$E(B-V)$] & 0.0, 0.001, 0.005, 0.01, 0.03, 0.05, \\
starburst attenuation law  & & 0.1, 0.2, 0.3, 0.4, 0.5, 0.6, 0.8\\
& Ratio of $E(B-V)$ between stellar continuum and emission line& 0.44\\
\enddata
\tablecomments{*A very small negative value is used to increase the parameter sampling close to 0.}
\end{deluxetable*}

{\catcode`\&=11
\gdef\KimandHo2019ApJ{\citet{Kim&Ho2019ApJ}}}
{\catcode`\&=11
\gdef\ZhuangandHo2020ApJ{\citet{Zhuang&Ho2020ApJ}}}

\begin{deluxetable*}{cllRRcLRRL}
\caption{Basic Properties of the Final Sample \label{table2}}
\tablehead{
\colhead{Index} & \colhead{Name} & \colhead{Redshift} & \colhead{R. A.} & \colhead{Dec.} & \colhead{log $M_{\rm BH}$} & \colhead{log $L_{\rm bol}$} & \colhead{log $\dot{M}_{\rm BH}$} & \colhead{log $\lambda_{\rm E}$} & \colhead{log SFR}\\
\nocolhead{ } & \nocolhead{ } & \nocolhead{ } & \colhead{($^{\circ}$)} & \colhead{($^{\circ}$)} & \colhead{($M_{\odot}$)} & \colhead{(erg s$^{-1}$)} & \colhead{($M_{\odot}$ yr$^{-1}$)} & \nocolhead{ } & \colhead{($M_{\odot}$ yr$^{-1}$)}\\
\colhead{(1)} & \colhead{(2)} & \colhead{(3)} & \colhead{(4)} & \colhead{(5)} & \colhead{(6)} & \colhead{(7)} & \colhead{(8)} & \colhead{(9)}& \colhead{(10)}
}
\startdata
1 & SDSS J002822.54$-$103903.5  &  0.301712  &  7.0939008  &     -10.650981  &  7.85  &  44.92 \pm 0.15  &   -0.9 \pm 0.15  &  -1.03  &   0.73 \pm 0.18\\
2 & SDSS J003321.77$+$141626.5  &  0.300153  &   8.340729  &      14.274034  &   7.50  &  44.94 \pm 0.25  &  -0.89 \pm 0.25  &  -0.66  &   1.13 \pm 0.36\\
3 & SDSS J004214.71$+$153148.2  &  0.316244  &   10.56129  &      15.530053  &  7.51  &  45.11 \pm 0.24  &  -0.72 \pm 0.24  &  -0.50  &   1.23 \pm 0.33\\
4 & SDSS J004319.75$+$005115.3  &  0.308341  &  10.832272  &     0.85425062  &  9.55  &  45.41 \pm 0.16  &  -0.41 \pm 0.16  &  -2.24  &   1.03 \pm 0.21\\
5 & SDSS J004458.67$+$004319.9  &  0.349953  &  11.244478  &     0.72219575  &  8.13  &  45.38 \pm 0.16  &  -0.44 \pm 0.16  &  -0.85  &   0.62 \pm  0.30\\
6 & SDSS J012256.19$-$000252.6  &  0.340495  &  20.734137  &    -0.04794548  &  7.96  &  45.66 \pm 0.18  &  -0.16 \pm 0.18  &  -0.40  &   1.54 \pm 0.24\\
7 & SDSS J013352.65$+$011345.4  &  0.308083  &  23.469391  &      1.2292659  &  8.31  &  45.27 \pm 0.16  &  -0.56 \pm 0.16  &  -1.14  &   0.14 \pm 0.55\\
8 & SDSS J015957.64$+$003310.5  &  0.311674  &  29.990148  &     0.55291729  &   8.00  &  44.81 \pm 0.14  &  -1.01 \pm 0.14  &  -1.29  &   1.01 \pm 0.12\\
9 & SDSS J021652.47$-$002335.3  &  0.304493  &   34.21864  &    -0.39314734  &  7.25  &  45.72 \pm  0.20  &   -0.1 \pm  0.20  &   0.37  &   1.57 \pm 0.28\\
10 & SDSS J022138.74$+$005048.3  &  0.306465  &  35.411399  &      0.8467546  &  7.66  &  44.93 \pm 0.18  &   -0.9 \pm 0.18  &  -0.83  &   0.87 \pm 0.23\\
\enddata
\tablecomments{Col. (1): Object index. Col. (2): Object name. Col. (3): Right ascension (J2000). Col. (4): Declination (J2000). Col. (5): Redshift. Col. (6): BH mass estimated using broad \Ha\ from \citet{Liu+2019ApJS}. We adopt a conservative uncertainty of 0.5~dex following \KimandHo2019ApJ. Col. (7): AGN bolometric luminosity. Col. (8): BH accretion rate. Col. (9): Eddington ratio, whose uncertainty is dominated by the uncertainty of $M_{\rm BH}$. Col. (10): Star formation rate. Cols. (7)--(10) are from \ZhuangandHo2020ApJ. (Table~\ref{table2} is published in its entirety in the machine-readable format.)}
\end{deluxetable*}

\begin{deluxetable*}{cCCCCCCCRc}
\tablecaption{Fluxes and Derived Properties of the Final Sample \label{table3}}
\tabletypesize{\small}
\tablehead{
\colhead{Index} & \colhead{$f_g$} & \colhead{$f_r$} & \colhead{$f_i$} & \colhead{$f_z$} & \colhead{$f_y$} & \colhead{log $M_*$} & \colhead{$n$} & \colhead{$R_e$} & \colhead{Morphology}\\
\nocolhead{ } & \colhead{(mJy)} & \colhead{(mJy)} & \colhead{(mJy)} & \colhead{(mJy)} & \colhead{(mJy)} & \colhead{($M_{\odot})$} & \colhead{ } & \colhead{(kpc)} & \nocolhead{ }\\
\colhead{(1)} & \colhead{(2)} & \colhead{(3)} & \colhead{(4)} & \colhead{(5)} & \colhead{(6)} & \colhead{(7)} & \colhead{(8)} & \colhead{(9)} & \colhead{(10)}
}
\startdata
1  &  0.049 \pm  0.014  &  0.112 \pm  0.026  &  0.166 \pm  0.029  &  0.207 \pm  0.037  &  0.231 \pm  0.042  &  11.06 \pm 0.20  &  1.71 \pm  0.70  &   5.99 \pm  1.67  &      late\\
2  &  0.036 \pm  0.011  &  0.087 \pm  0.020  &  0.160 \pm  0.028  &  0.201 \pm  0.037  &  0.184 \pm  0.036  &  11.09 \pm 0.18  &  2.19 \pm  0.90  &   4.72 \pm  1.31  &     early\\
 3  &  0.034 \pm  0.014  &  0.077 \pm  0.024  &  0.092 \pm  0.020  &  0.161 \pm  0.038  &  0.089 \pm  0.033  &  10.67 \pm 0.26  &  \nodata &   7.29 \pm  2.30  & uncertain\\
4  &  0.058 \pm  0.023  &  \nodata &  0.221 \pm  0.046  &  0.334 \pm  0.072  &  0.308 \pm  0.084  &  11.33 \pm 0.24  &  7.36 \pm  3.38  &   9.77 \pm  3.15  &     early\\
 5  &  0.058 \pm  0.017  &  0.132 \pm  0.031  &  0.164 \pm  0.030  &  0.211 \pm  0.039  &  0.189 \pm  0.039  &  10.97 \pm 0.20  &  3.52 \pm  1.47  &   5.62 \pm  1.57  &     early\\
6  &  0.043 \pm  0.016  &  0.086 \pm  0.024  &  0.128 \pm  0.026  &  0.126 \pm  0.030  &  0.113 \pm  0.039  &  10.77 \pm 0.24  &  5.73 \pm  2.49  &   2.83 \pm  0.85  &     early\\
7  &  0.075 \pm  0.023  &  0.119 \pm  0.029  &  0.203 \pm  0.039  &  0.398 \pm  0.076  &  0.437 \pm  0.088  &  11.46 \pm 0.24  &  6.26 \pm  2.66  &  10.11 \pm  2.96  &     early\\
8  &  0.034 \pm  0.012  &  0.109 \pm  0.028  &  0.128 \pm  0.027  &  0.163 \pm  0.045  &  0.151 \pm  0.060  &  10.96 \pm 0.25  &  4.97 \pm  2.19  &   1.75 \pm  0.58  &     early\\
9  &  0.047 \pm  0.013  &  0.112 \pm  0.026  &  0.171 \pm  0.030  &  0.236 \pm  0.043  &  0.273 \pm  0.050  &  11.20 \pm 0.18  &  1.85 \pm  0.77  &   6.64 \pm  1.86  &      late\\
10  &  0.037 \pm  0.011  &  0.101 \pm  0.024  &  0.160 \pm  0.029  &  0.219 \pm  0.040  &  0.310 \pm  0.057  &  11.31 \pm 0.18  &  3.07 \pm  1.27  &   6.52 \pm  1.82  &     early\\
\enddata
\tablecomments{Col. (1): Object index. Cols. (2)--(6): Flux density of AGN host galaxy in the $grizy$ bands after correction for Galactic extinction and systematic effects (Section~\ref{sec3.2}). Col. (7): Stellar mass from SED fitting. Cols. (8)--(9): \sersic\ index $n$ and half-light radius in the $i$ band after correction for systematic effects (Section~\ref{sec3.2}). Col. (10): Morphological classification based on $n$ and $R_e$. (Table~\ref{table3} is published in its entirety in the machine-readable format.)}
\end{deluxetable*}

\subsection{Estimation of Host Galaxy Morphologies} \label{sec4.2}

\begin{figure}[t]
\centering
\includegraphics[width=0.5\textwidth]{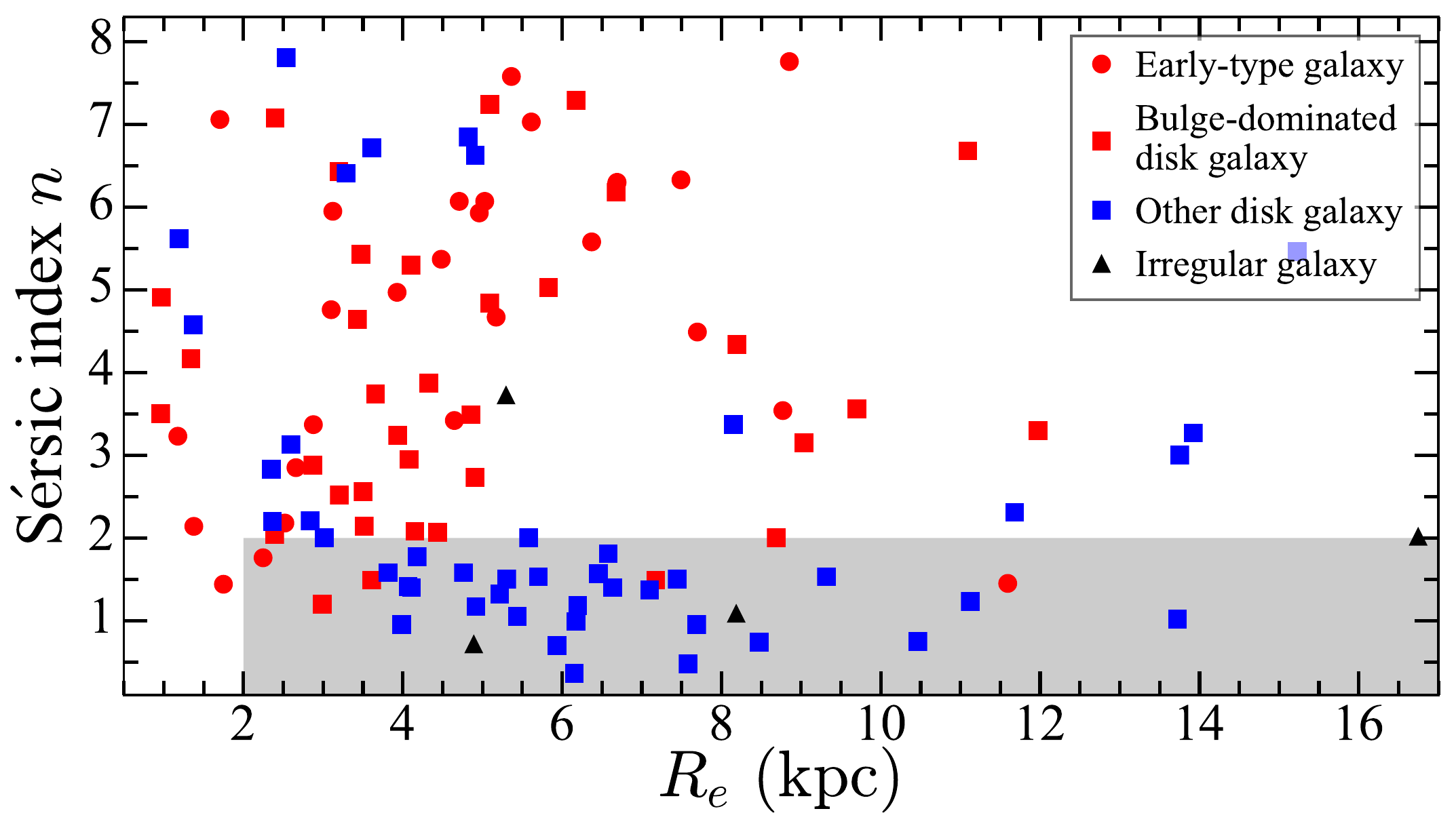}
\caption{Distribution of \sersic\ index ($n$) versus effective radius ($R_e$) measured in the PS1 $i$ band for redshift $\sim$\,0.3 galaxies selected from the COSMOS field (Section~\ref{sec3.2.1}). Morphological type information is from the Zurich Estimator of Structural Type (ZEST) catalog \citep{Scarlata+2007ApJS}, which is based on high-resolution HST images. Circles, squares, and triangles represent early-type, disk, and irregular galaxies, respectively, as given in the ZEST catalog. Red and blue symbols are early-type and late-type galaxies according to our definition; the shaded area denotes the region dominated by late-type galaxies.}
\label{fig15}
\end{figure}

Well-resolved images are required for rigorous morphological classification of galaxies, be they nearby \citep[e.g.,][]{Gao+2019ApJS} or distant \citep[e.g.,][]{van_der_Wel+2012ApJS, Davari2017}.  At redshifts $\sim$\,0.3, the arcsecond-scale resolution of the PS1 images subtends a physical scale of $\sim$\,2.3 kpc in radius, which is inadequate for reliable bulge-to-disk decomposition. The presence of a bright nucleus doubly exacerbates the situation.  As a consequence, we have confined our image decomposition to the simplest single-component model to fit the host galaxy. Fortunately, the galaxies in the COSMOS field have detailed morphological classifications based on high-resolution HST ACS/F814W images from the Zurich Estimator of Structural Type (ZEST) catalog \citep{Scarlata+2007ApJS}. ZEST classified galaxies into three structural types (early-type, disk, and irregular galaxies) using a three-dimensional space defined by the three main eigenvectors from a principal component analysis of five nonparametric quantities. With a more detailed subtype (``bulgeness'') for disk galaxies refined using the help of the source's \sersic\ index, which is loosely correlated with the bulge-to-disk ratio, the catalog distinguishes the disk galaxies into more detailed subtypes (bulge-dominated disk, intermediate-bulge disk, and pure disk).  To evaluate the feasibility of estimating morphological types from our data, we take advantage of the available classifications from ZEST, in conjunction with the $i$-band \sersic\ indices and half-light radii measured from simultaneous multiwavelength fits of the PS1 images of a subset of redshift $\sim 0.3$ galaxies in the COSMOS field. For this work, we only define two broad bins of morphological types: (1) early-type, which includes galaxies classified by ZEST as ellipticals, spheroidals, and bulge-dominated disk galaxies, and (2) late-type, which combines all other sources classified by ZEST as disk galaxies, including intermediate-bulge disk galaxies and pure disk galaxies.  Figure~\ref{fig15} shows that late-type galaxies preferentially have smaller $n$ and larger $R_e$. A combination of a horizontal line with $n\leq2$ and a vertical line with $R_e\geq2$ kpc approximately separates the two galaxy types, with a false positive (misclassification) rate of $\sim$\,24\% for early-type and $22\%$ for late-type galaxies. A few outliers with high $n$ ($\ga4$) and small $R_e$ ($\la5$ kpc) are classified as late-type galaxies according to ZEST. Visual inspection of their HST images, however, reveals that these galaxies indeed have a prominent bulge component besides a disk, and they should be better described as early-type. Therefore, the actual false positive rates may be somewhat more favorable than the above statistics indicate.  We note that 4/111 objects are classified as irregular by ZEST. Of these, two exhibit features that may be associated with interactions. The combination of $n$ and $R_e$ still gives a reasonable estimate of their morphologies (three late-type and 1 early-type systems).

Bearing in mind the obvious limitations of our crude method and the large uncertainty of the individual values of $n$, we apply our empirical classification scheme to estimate the host galaxy morphology of our quasar sample (Table~\ref{table3}). Among 305 quasars, 9 ($3\%$) have highly uncertain $n$ or/and $R_e$ based on the mock tests using idealized galaxies in Section~\ref{sec3.2.2}, and we label their morphologies as ``unreliable.'' For the remaining 296 quasars ($97\%$ of the total sample), 175 ($57\%$) reside in early-type galaxies and 121 ($40\%$) are hosted by late-type galaxies, consistent with the significant late-type fraction found in previous studies of low-redshift, optically selected type~1 quasars \citep[e.g.,][]{Falomo+2014MNRAS, Kim+2017ApJS, Yue+2018ApJ, Li_Junyao+2021ApJ, Zhao+2021ApJ}, optically selected type~2 quasars \citep{Zhao+2019ApJ}, and X-ray-select AGNs \citep[e.g.,][]{Gabor+2009ApJ, Kim+2021ApJS}.

\subsection{Mass-size Relation}\label{sec4.3}

\begin{figure}[t]
\centering
\includegraphics[width=0.48\textwidth]{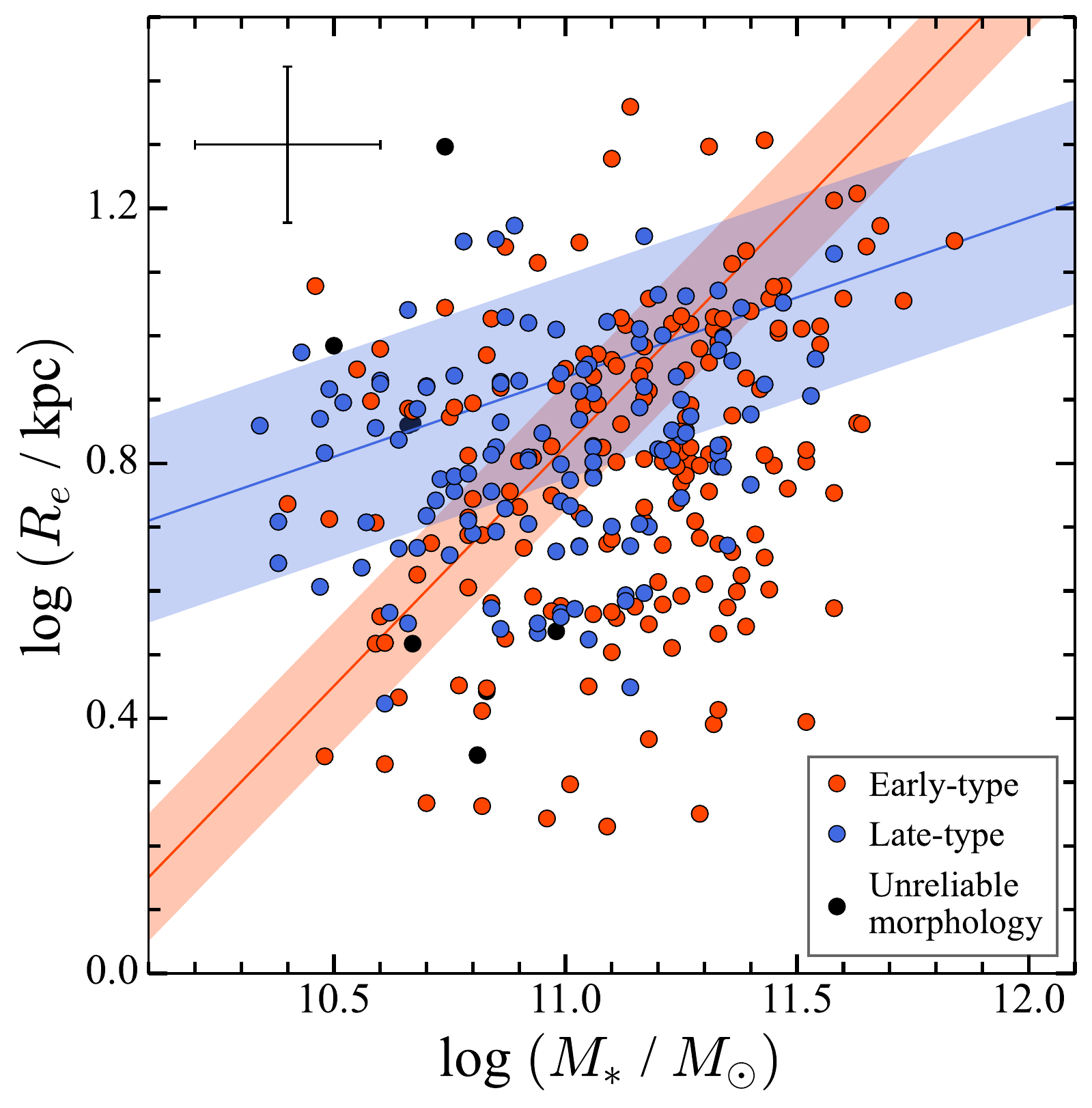}
\caption{Relation between stellar mass ($M_*$) and effective radius ($R_e$) of AGN host galaxies of early-type (red), late-type (blue), and unreliable morphology (black).  Blue and red lines represent the relations of early-type and late-type galaxies at redshift $0.25$ from \citet{van_der_Wel2014ApJ}, with shaded area indicating the intrinsic scatter of $R_e$ from their fits. Typical uncertainties are shown in the upper-left corner.}
\label{fig16}
\end{figure}

The stellar mass of a galaxy and how that mass is distributed spatially, commonly described by the stellar mass-size relation, encode vital clues about the galaxy's evolution and assembly history. The mass-size relation differs between galaxies of early and late type, and it evolves strongly with redshift \citep[see, e.g., review in ][]{Somerville&Dave2015ARA&A}.  Figure~\ref{fig16} illustrates that the early-type and late-type host galaxies of red quasars follow the same $M_*-R_e$ relations defined by the inactive galaxy population at a similar redshift \citep[0.25;][]{van_der_Wel2014ApJ}\footnote{Because the $i$ band corresponds to rest-frame $\sim$\,5700 \AA\ at redshift $0.3$, we do not apply any correction to the galaxy sizes given by \cite{van_der_Wel2014ApJ} which have already been converted to rest-frame 5000 \AA.}. As with the regular galaxy population, at the same $M_*$ early-type galaxies are systematically more compact (smaller $R_e$) than late-type galaxies.  Closer inspection reveals something more: $\sim$\,1/3 (107/305) of the quasars lie below the $M_*-R_e$ relation of either galaxy type by at least $1\,\sigma$ (the blue and red shaded areas), suggesting that a substantial fraction of quasars tend to reside in more concentrated galaxies.  Note that our classification of galaxy type is based on morphology, instead of star formation activity, as commonly practiced in the literature when studying the mass-size relation (e.g., \citealt{van_der_Wel2014ApJ} used the $UVJ$ diagram to distinguish early-type from late-type galaxies).  As Section~\ref{sec5.1} will show, our quasars indeed have high SFRs and are predominately located on and above the galaxy star-forming MS. In this respect, we may regard our sample as comprising compact star-forming galaxies, similar to AGNs found at similar and higher redshifts \citep[e.g.,][]{Barro+2014ApJ, Kocevski+2017ApJ, Silverman+2019ApJ, Li_Junyao+2021ApJ, Stacey+2021MNRAS}.

We do not find a significant secondary dependence of the scatter of the $M_*-R_e$ relation with $L_{\rm bol}$, $M_{\rm BH}$, or $\lambda_{\rm E}$. This suggests that quasars do not have a direct, instantaneous impact on host galaxy size, as might be expected for scenarios of AGN feedback that involve rapid and extensive gas expulsion \citep{Fan+2008ApJ, Fan+2010ApJ}. On the contrary, nearby quasars have an abundant cold gas reservoir \citep[e.g.,][]{Husemann+2017MNRAS, Shangguan2018, Jarvis+2020MNRAS, Shangguan+2020aApJ, Koss+2021ApJS}, which is compact \citep{Molina2021} and participates in vigorous \citep{Xie+2021ApJ}, centrally concentrate star formation \citep{Zhuang&Ho2020ApJ}.

\section{Implications}\label{sec5}

\subsection{Quasar Hosts and the Star-forming Main Sequence}\label{sec5.1}

\begin{figure}[t]
\centering
\includegraphics[width=0.48\textwidth]{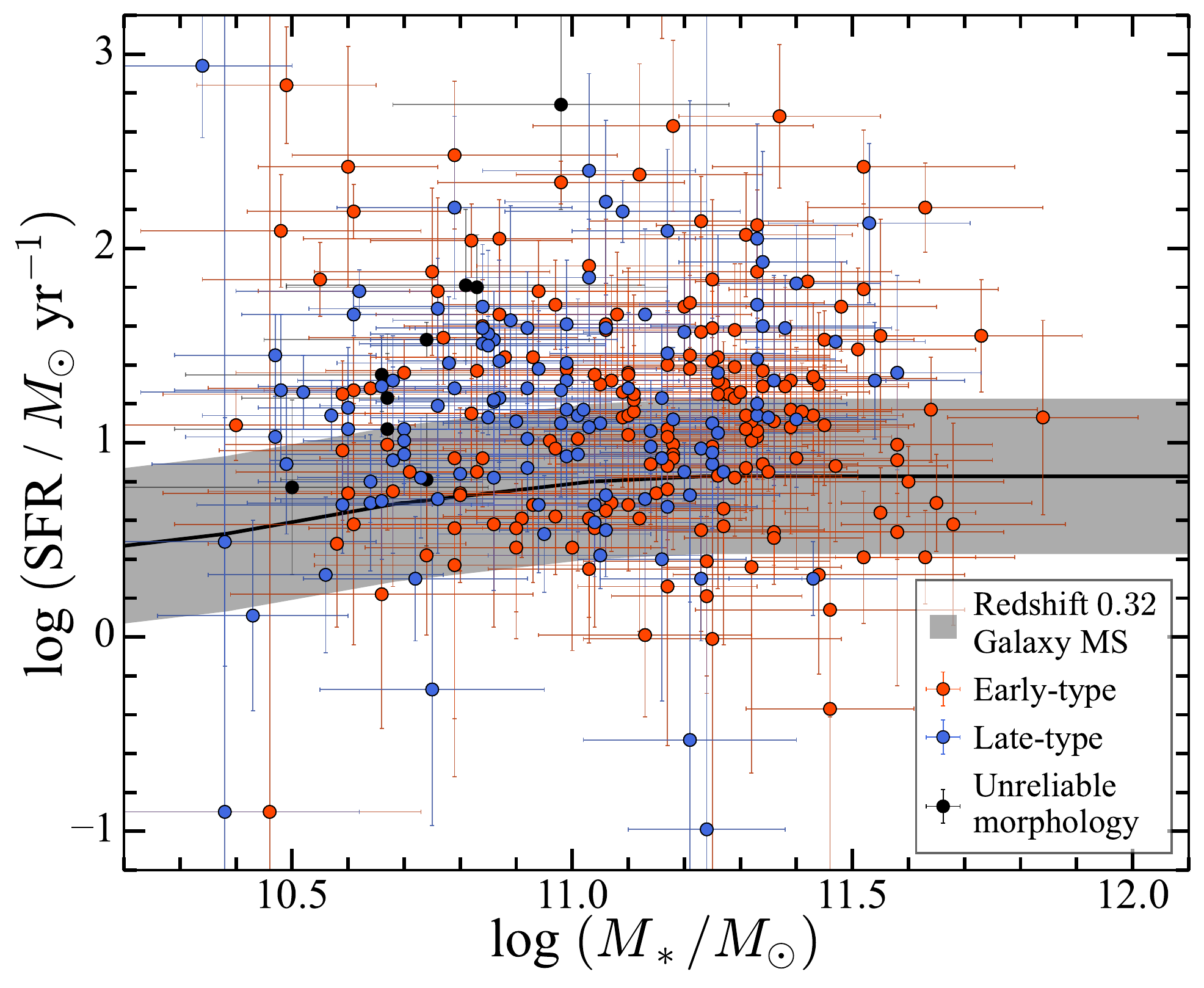}
\caption{Comparison of star formation rate (SFR) versus stellar mass ($M_*$) of our redshift $\sim$\,0.3 quasar sample with the galaxy MS at redshift $0.32$ from \citet{2019MNRAS.483.3213Popesso+, 2019MNRAS.490.5285Popesso+}, shown as a black curve whose grey-shaded stripe ($\pm0.4$~dex) indicates the width of the MS.  AGN host galaxies are color-coded by their morphology: early-type (red), late-type (blue), and unreliable morphology (black).
\label{fig17}}
\end{figure}

\begin{figure*}[t]
\centering
\includegraphics[width=0.85\textwidth]{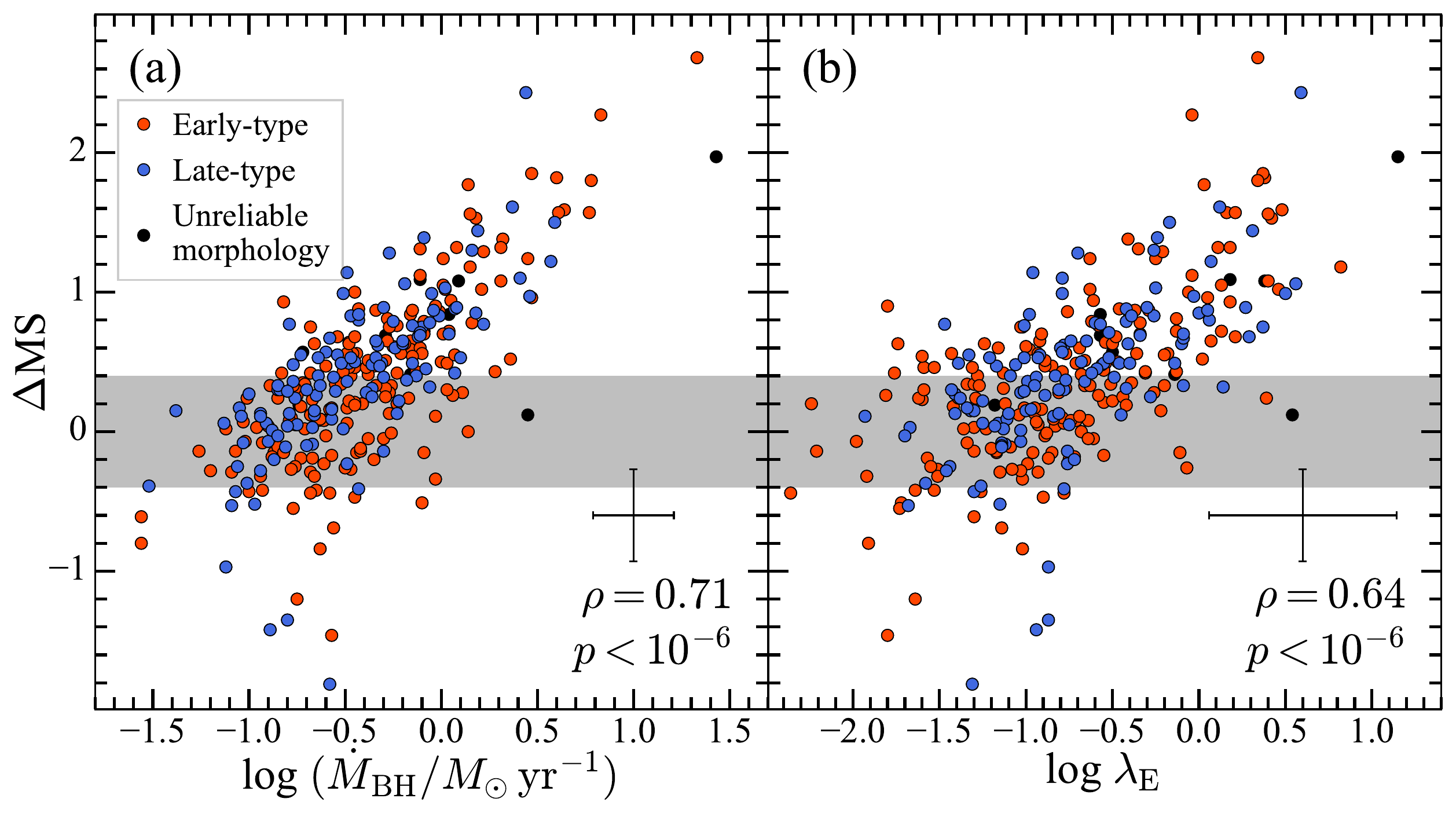}
\caption{Distance to the galaxy star-forming main sequence ($\Delta{\rm MS} \equiv {\rm log\ SFR - log\ SFR_{MS}}$) versus (a) BH accretion rate and (b) Eddington ratio with colors indicating their morphology: early-type (red), late-type (blue), and unreliable morphology (black). The lower-right corner of each panel shows the typical uncertainties and the Spearman correlation strength ($\rho$) and $p$-value for the whole sample.  The grey-shaded stripe ($\pm0.4$~dex) indicates the width of the MS at redshift $0.32$ from \citet{2019MNRAS.483.3213Popesso+, 2019MNRAS.490.5285Popesso+}.  }
\label{fig18}
\end{figure*}

The location of the host galaxies of quasars on the star-forming MS provides important clues regarding the possible connection between BH accretion and galaxy evolution.  Existing studies, as discussed in Section~\ref{sec1}, present a rather bewildering assortment of conflicting results.  With the stellar masses in hand, the present sample, on account not only of its size but also the robust SFR measurements from \citet{Zhuang&Ho2020ApJ}, offers an opportunity to examine the nature of the star-forming MS for active galaxies (Figure~\ref{fig17}).  As a benchmark for comparison with the general galaxy population, we adopt the MS and its redshift evolution from \citet{2019MNRAS.483.3213Popesso+, 2019MNRAS.490.5285Popesso+}. Conservatively setting the scatter to $\pm 0.4$~dex (width of 0.8~dex), the fraction of quasars that lie on or above the above is 43\% (132/305) and 50\% (151/305), respectively; and only 7\% (22/305) sit below the MS.  This result is robust against the choice of the MS used for comparison. For example, if we adopt the steeper MS relation of \citet{Speagle+2014ApJS}, which increases the SFR across the entire $M_*$ range covered here, $\sim$\,85\% of the quasars are still located on or above the MS.  Defining the distance to the galaxy star-forming MS as $\Delta{\rm MS} \equiv {\rm log\ SFR - log\ SFR_{MS}}$, $\Delta{\rm MS}$ positively correlates with both $\dot{M}_{\rm BH}$ and $\lambda_{\rm E}$ (Figure~\ref{fig18}), indicating a close connection between BH accretion in the innermost regions of the galaxy and star formation activity on much larger scales.

\begin{figure*}[t]
\centering
\includegraphics[width=0.98\textwidth]{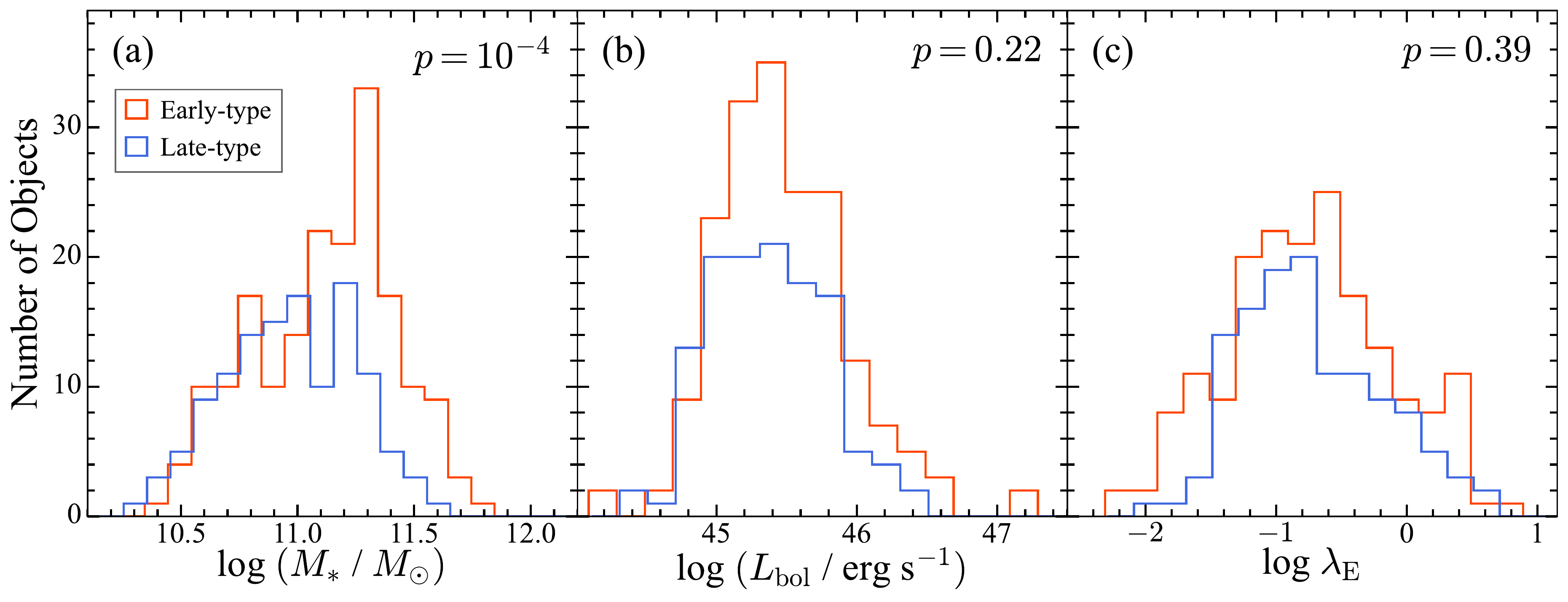}
\caption{Distribution of (a) stellar mass, (b) AGN bolometric luminosity, and (c) Eddington ratio for early-type (red) and late-type (blue) galaxies.  The $p$-value from the two-sample Kolmogorov-Smirnov test is shown in the upper-right corner of each panel.}
\label{fig19}
\end{figure*}

There have been numerous attempts to quantify the connection between BH accretion and star formation activity.  The situation is complex and depends strongly on sample selection (Section~\ref{sec1}).  Recent studies have emphasized the correlation between host galaxy SFR and the luminosity (or mass accretion rate) and Eddington ratio of the AGN, a trend reported in samples with diverse levels of AGN activity \citep[e.g.,][]{Chen+2013ApJ, Dai+2018MNRAS, Woo+2020ApJ}, including that from which the current study is drawn \citep{Zhuang&Ho2020ApJ}.  The results in Figure~\ref{fig17} bear a close resemblance to the situation for low-redshift ($\lesssim 0.5$) Palomar-Green quasars, many of which form stars prodigiously enough to qualify as starburst systems based on their star formation efficiencies (gas depletion timescales) and offset above the MS \citep{Shangguan+2020aApJ, Shangguan+2020bApJ, Xie+2021ApJ}.  Vigorous star formation goes hand-in-hand with BH accretion.  But why?  The connection may be largely indirect, a mere reflection of the fact that a common, contemporaneous gas supply is prerequisite to fuel both processes \citep[e.g.,][]{Jarvis+2020MNRAS, Shangguan+2020aApJ, Yesuf2020}. In support of this interpretation, \citet{Zhuang+2021ApJ} demonstrate that the strong correlation observed between $\dot{M}_{\rm BH}$ and SFR is driven mainly by the mutual dependence of these two parameters on molecular gas mass. Still, some investigators have argued that the two phenomena may be causally related.  Star formation may be triggered directly by positive AGN feedback, as evidenced by the detection of in situ star formation in AGN-driven outflows and the tendency for the SFR in outflows to increase with higher mass outflow rate \citep[e.g.,][]{Maiolino+2017Natur, Gallagher+2019MNRAS}.  More luminous AGNs exhibit more frequent and stronger outflows (e.g., \citealt{Fiore+2017A&A, Rakshit&Woo2018ApJ, Fluetsch+2019MNRAS}; but see \citealt{Shangguan+2020bApJ, Molina2022}). 

We remain agnostic on this unsettled debate.  In our estimation, it is premature to judge which AGN property ($\dot{M}_{\rm BH}$ or $\lambda_{\rm E}$), if any, is fundamentally correlated with enhanced star formation activity. As \citet{Zhuang&Ho2020ApJ} show, current sample selection effects preclude us from obtaining truly independent estimates of $\dot{M}_{\rm BH}$ and $\lambda_{\rm E}$: a relatively narrow range of $M_{\rm BH}$ can produce an artificial correlation between $\dot{M}_{\rm BH}$ and $\lambda_{\rm E}$, and thus their correlation with $\Delta{\rm MS}$. AGN samples with larger dynamical ranges in $M_{\rm BH}$, $\dot{M}_{\rm BH}$, and $\lambda_{\rm E}$ are required to fully investigate these issues.

\subsection{Host Galaxy Morphology}

Within our sample, the relative fraction of early-type and late-type hosts depends only on stellar mass, not on AGN properties such as $L_{\rm bol}$ and $\lambda_{\rm E}$.  While the two types of hosts differ significantly ($p \approx 10^{-4}$) according to the two-sample Kolmogorov-Smirnov test, they are very similar ($p > 0.2$) in terms of $L_{\rm bol}$ and $\lambda_{\rm E}$ (Figure~\ref{fig19}). Dividing the sample according to the median stellar mass of $M_* = 10^{11}\, M_{\odot}$, we find that the fraction of quasars hosted in late-type galaxies drops from $50\%$ in less massive systems to $\sim$\,30\% in the more massive bin. Approximately 70\% of the AGNs live in early-type galaxies, qualitatively resembling the demographics of the host galaxies of less powerful AGNs \citep{Ho+1997, Kauffmann+2003MNRAS}. The preference for quasars to reside in massive, bulge-dominated systems, of course, simply reflects the close link between BHs and galactic bulges. 

However, disk-dominated hosts are still quite prevalent. They make up $\sim$\,40\% of the entire sample, and as much as $\sim$\,50\% of the objects with $M_*\la10^{11}\, M_{\odot}$.  These results, together with the lack of correlation between galaxy morphology and the level of AGN activity, contributes to the long-standing debate on the triggering mechanism of AGNs.  While gas-rich major mergers offer a natural framework to unify many aspects of AGNs and their host galaxies \citep[e.g.,][]{Hopkins+2008ApJS, Alexander&Hickox2012NewAR}, it has become increasingly recognized that more commonplace, less violent processes, such as minor mergers and stochastic gas accretion from bar and disk instability \citep{Kormendy&Kennicutt2004ARA&A, Steinborn+2018MNRAS}, suffice to fuel most of the gamut of AGN activity (e.g., \citealt{Ho2009, Cisternas2011, Villforth+2017MNRAS, Zhao2022}).  Indeed, a large fraction of nearby luminous AGNs are hosted in barred galaxies \citep[e.g.,][]{Kim+2017ApJS, Zhao+2019ApJ, Zhao+2021ApJ}.  To the extent that major mergers destroy disks \citep{Toomre&Toomre1972ApJ, Barnes1996}, and even if under some circumstances disks can survive \citep{Hopkins2009} or reform \citep{Scannapieco2009}, the sizable fraction of disk-dominated galaxies in our sample constitutes additional evidence of the secondary importance of major mergers in regulating AGN activity, even for the relatively high luminosities contained in our sample (median $L_{\rm bol} = 10^{45.4}\,{\rm erg\,s^{-1}}$).  Recent cosmological simulations find that the fraction of AGNs triggered by mergers is not enhanced in massive ($M_*>10^{11}\, M_{\odot}$) galaxies, and that powerful AGNs are not more likely to be found in mergers compared to inactive galaxies \citep[e.g.,][]{Steinborn+2018MNRAS, McAlpine+2020MNRAS, Sharma2022}. Mergers contribute $\lesssim 40\%$ to the growth of both the central BH and the bulge in galaxies of $M_*\approx10^{11} \, M_{\odot}$, highlighting the importance of gas fueling by secular processes to the coevolution of supermassive BHs and their host galaxies \citep[e.g.,][]{Parry+2009MNRAS, Martin+2018MNRAS, McAlpine+2020MNRAS}.

The issue of the role of major mergers gains additional relevance owing to the large population of sources that lie significantly above the star-forming galaxy MS (Figure~\ref{fig17}).  Such clear-cut starburst systems, especially in view of the large masses involved ($M_* \approx 10^{11.0\pm0.5}\,M_\odot$), ordinarily would be associated with unambiguous, gas-rich major mergers (e.g., \citealt{Petty2014, Shangguan2019}). Our results do not conform to this expectation, judging by the broad representation of late-type galaxies among the population of $\Delta{\rm MS}$ sources.

\section{Summary}\label{sec6}

We perform two-dimensional, simultaneous, multiwavelength image decomposition of Pan-STARRS1 $grizy$ images of 453 redshift $\sim$\,0.3 quasars, the largest sample studied to date, by explicitly considering the wavelength-dependence of galaxy structure.  We obtain robust five-band photometry, structural parameters (half-light radii, \sersic\ indices), and morphological classifications.  Analysis of mock AGNs generated from both idealized and real galaxy images allows us to assess the uncertainties and systematic biases of the image decomposition. We combine SFRs from \cite{Zhuang&Ho2020ApJ} with stellar masses derived from SED fitting of the multi-band photometry to investigate the star formation activity of quasar host galaxies in the context of the galaxy star-forming main sequence.  

Our main results are as follows:

\begin{itemize}

\item{The vast majority of the objects ($\sim 90\%$) are located on or above the galaxy star-forming main sequence. The distance to the main sequence positively correlates with BH accretion rate and Eddington ratio, with more rapidly accreting BHs exhibiting larger departures above the main sequence. This suggests a possible connection between BH accretion and star formation activity on galactic scales, but we caution against over-interpretation in view of their mutual dependence on gas supply.}

\item{Using properly matched galaxies selected from the HST COSMOS field that have independently derived morphological types as a calibration sample, we classify our quasar hosts into late-type (39\%) and early-type (58\%) galaxies.  The high fraction of late-type hosts, in conjunction with the lack of a clear dependence between host morphology and AGN properties, highlights the importance of internal secular processes in regulating AGN activity.}

\item{While the majority of quasar host galaxies roughly follow the stellar mass-size relation of inactive galaxies, both for late-type and early-type galaxies, $\sim$\,1/3 of the quasar hosts fall below the $M_* - R_e$ relation. Together with the evidence for intense star formation, the host galaxies of nearby quasars structurally resemble compact star-forming galaxies at higher redshift.}

\end{itemize}

This paper demonstrates the feasibility and tremendous potential of simultaneous multiwavelength image decomposition of nearby and even moderately distant galaxies, including those with strong nuclear activity.  Apart from deep existing imaging surveys such as SDSS Stripe 82, Pan-STARRS1, and the Hyper Suprime-Cam Subaru Strategic Program, we can look forward to the upcoming Legacy Survey of Space and Time \citep[LSST;][]{Ivezic+2019ApJ} of the Vera C. Rubin Observatory.

\begin{acknowledgments}
We thank the anonymous referee and Yingjie Peng for helpful suggestions. This work was supported by the National Science Foundation of China (11721303, 11991052, 12192220, and 12192222) and the China Manned Space Project (CMS-CSST-2021-A04 and CMS-CSST-2021-A06).
\end{acknowledgments}

\vspace{5mm}

\software{\texttt{Astropy} \citep{2013A&A...558A..33A,2018AJ....156..123A},
          \texttt{CIGALE} \citep{2019A&A...622A.103Boquien+},
          \texttt{GALFIT} \citep{Peng+2002AJ, Peng+2010AJ},
          \texttt{GALFITM} \citep{Haussler+2013MNRAS, Vika+2013MNRAS},
          \texttt{Matplotlib} \citep{Hunter2007},
          \texttt{Numpy} \citep{Harris2020},
          \texttt{photutils} \citep{photutils},
          \texttt{PSFEx} \citep{Bertin2011ASPC},
          \texttt{reproject} (https://reproject.readthedocs.io/en/stable/index.html),
          \texttt{SExtractor} \citep{1996A&AS..117..393B},
          \texttt{SWarp} \citep{Bertin+2002ASPC}}

\appendix

\begin{figure*}[t]
\figurenum{A1}
\centering
\includegraphics[width=\textwidth]{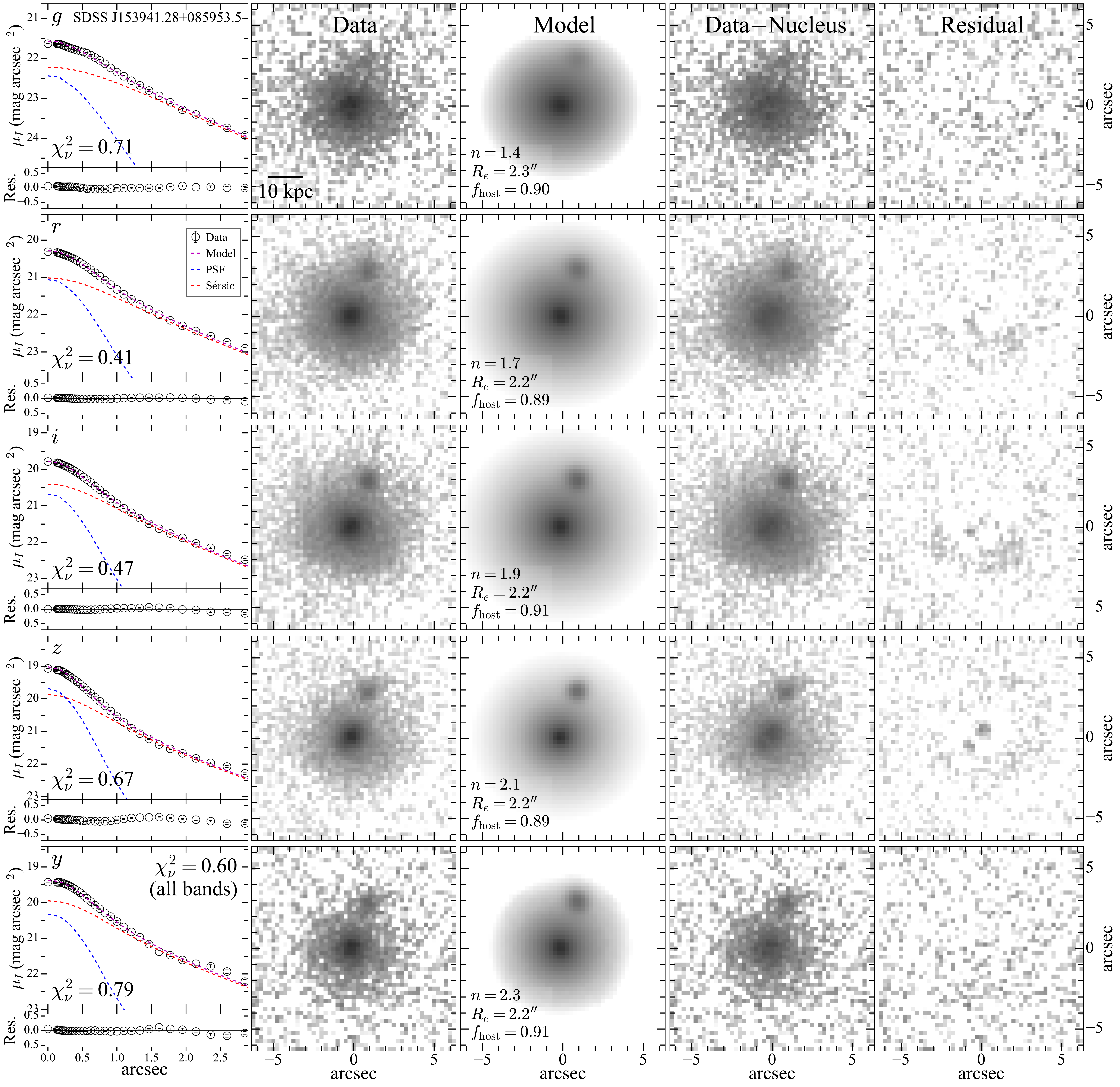}
\caption{{Same as in Figure~\ref{fig4}, but for SDSS J153941.28+085953.5 (redshift 0.324), which has the lowest $\chi^2_{\nu}$ in the final sample.}}
\label{figA1}
\end{figure*}

\begin{figure}[t]
\figurenum{A2}
\centering
\includegraphics[width=\textwidth]{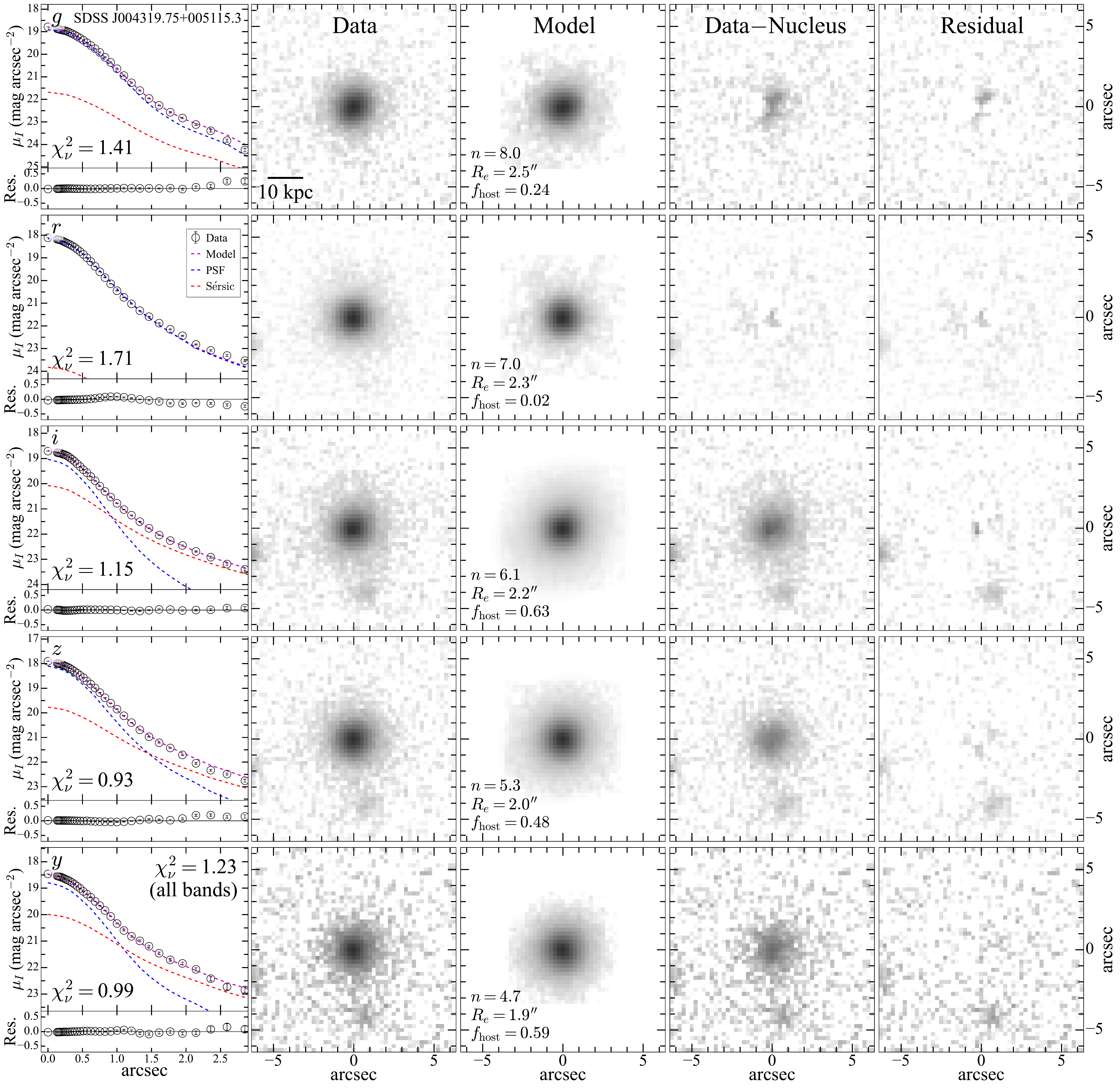}
\caption{{Same as in Figure~\ref{fig4}, but for SDSS J004319.75+005115.3 (redshift 0.308), which has the highest $\chi^2_{\nu}$ in the final sample.}}
\label{figA2}
\end{figure}

\section{Image Decomposition Results}\label{appendix1}

We show examples of decomposition results of two AGNs with the lowest (0.60) and highest (1.23) $\chi^2_{\nu}$ in the final sample of 305 objects. Results for the entire sample can be found at \dataset[https://doi.org/10.12149/101130]{https://doi.org/10.12149/101130}.

\begin{figure*}[t]
\figurenum{A3}
\centering
\includegraphics[width=\textwidth]{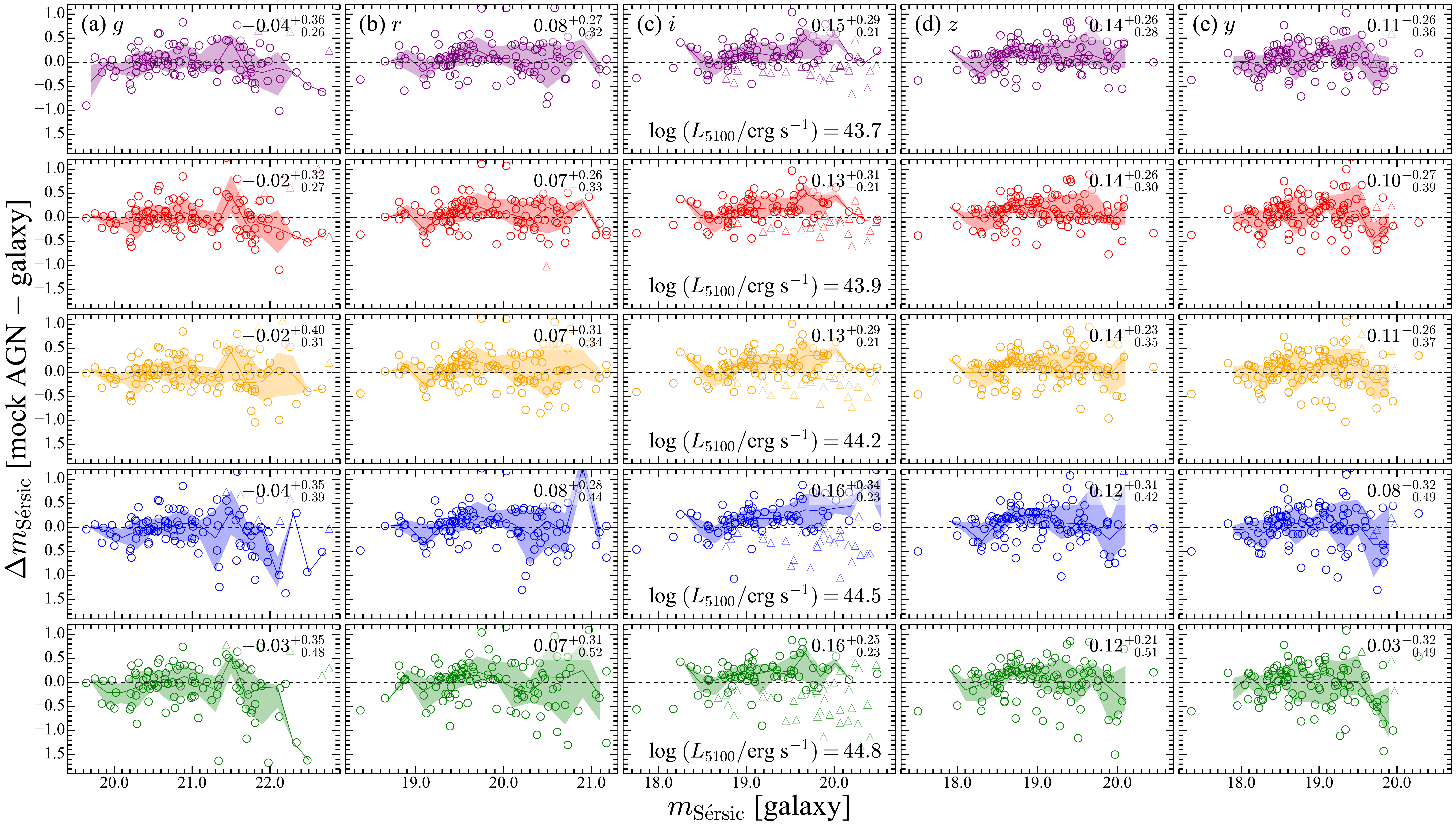}
\caption{Same as in Figure~\ref{fig8}, but for \msersic\ obtained using separate, \texttt{GALFIT} single-band fits, fixing the structure of the galaxy to that derived in the $i$ band.}
\label{figA1}
\end{figure*}

\begin{figure}[t]
\figurenum{A4}
\centering
\includegraphics[width=0.5\textwidth]{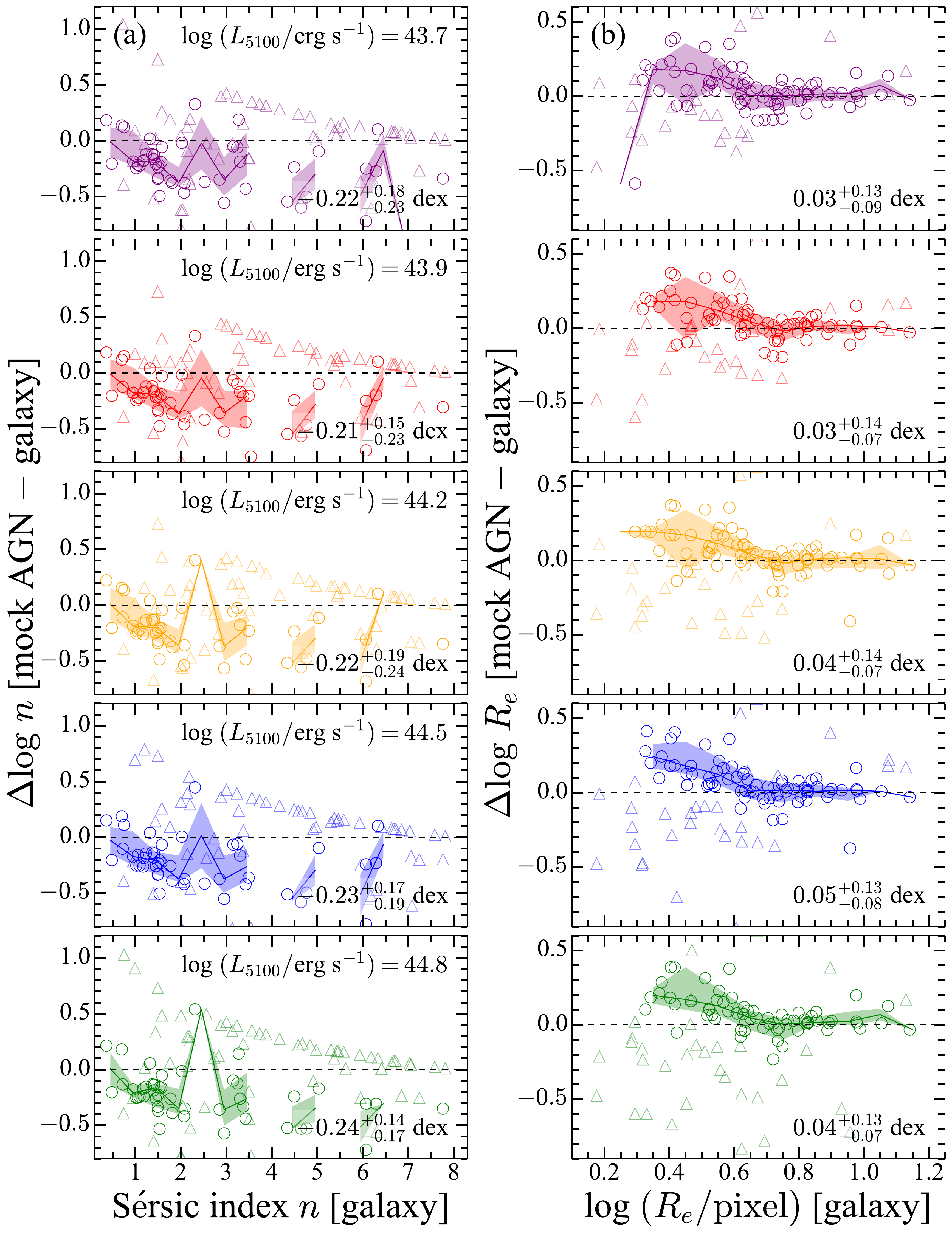}
\caption{Same as in Figure~\ref{fig9}, but for the \sersic\ index $n$ and half-light radius ($R_e$) obtained from \texttt{GALFIT} decomposition of the $i$-band image of the mock AGNs.}
\label{figA2}
\end{figure}

\begin{deluxetable*}{lccccccc}[h!]
\tablenum{A1}
\caption{Statistical Comparison of Two Methods of Decomposition \label{tableA1}}
\tablehead{
\colhead{Statistics} & \colhead{\msersic\ [$g$]} & \colhead{\msersic\ [$r$]} & \colhead{\msersic\ [$i$]} & \colhead{\msersic\ [$z$]} & \colhead{\msersic\ [$y$]} & \colhead{$\log n$} & \colhead{$\log R_e$}\\
\nocolhead{Statistics*} & \colhead{(mag)} & \colhead{(mag)} & \colhead{(mag)} & \colhead{(mag)} & \colhead{(mag)} & \nocolhead{$\log n$} & \colhead{(pixel)}
}
\startdata
{\it Simultaneous Multiwavelength Decomposition} &&&&&&\\
\hline
Median &0.09& $-0.02$ &0.03&0.04&0.08& $-0.08$ & 0.00\\ 
Scatter&0.30&0.25&0.19&0.19&0.19& 0.21&0.12\\
Detection fraction & $61\%$ & $78\%$ & $86\%$ & $82\%$ & $70\%$ & $65\%$ & $77\%$ \\
\hline
{\it Separate Single-band Decomposition} &&&&&&\\
\hline
Median &$-0.03$&0.07&0.14&0.13&0.10& $-0.22$& 0.04 \\ 
Scatter &0.34&0.33&0.27&0.31&0.35& 0.20 & 0.11\\
Detection fraction & \nodata & \nodata & $72\%$ & \nodata & \nodata & $43\%$ & $67\%$ \\
\enddata
\tablecomments{Statistics are averaged over the 16th, 50th, and 84th percentiles of AGN strength.}
\end{deluxetable*}

\section{Tests for Separate Single-band Decomposition}\label{appendix2}

AGN host galaxies images are usually analyzed one band at a time. When more than one band is being considered, it is common practice to fix certain parameters of the model based on the results obtained from another band (e.g., \citealt{Zhao+2021ApJ}).  In this Appendix, we compare the results of this traditional approach with those of our new strategy of performing simultaneous, multiwavelength decomposition.  As in Section~\ref{sec3.2.1}, we perform mock tests using a set of galaxies selected from the COSMOS field.  We first fit the $i$-band image independently with a nucleus plus \sersic\ model.  Then, fixing the best-fit position, structural ($R_e$, $n$), and geometric (position angle, axis ratio) parameters, we fit the same two-component model separately to the images of each of the other four ($grzy$) bands, allowing only the flux normalization to vary.  The results for \msersic\ for the five bands are given in Figure~\ref{figA1}, and Figure~\ref{figA2} assesses $n$ and $R_e$, highlighting only the results for the $i$ band.

The statistical results, averaged over the 16th, 50th, and 84th percentiles of AGN strength, are given Table~\ref{tableA1}, which summarizes the median difference, scatter, and detection fraction (fraction of objects with value at least 3 times its error) for both simultaneous multiwavelength decomposition (Section~\ref{sec3.2.1}) and for separate single-band decomposition (this Appendix).  While single-band decomposition still yields useful results, its overall performance, as judged by either the absolute median difference or scatter, is worse than that of simultaneous multiwavelength decomposition. The most notable advantage of simultaneous multiwavelength decomposition lies in its significantly higher success rate of returning reliable detections.

\vfill\eject

\end{CJK*}

\begin{thebibliography}{}
\expandafter\ifx\csname natexlab\endcsname\relax\def\natexlab#1{#1}\fi
\providecommand{\url}[1]{\href{#1}{#1}}
\providecommand{\dodoi}[1]{doi:~\href{http://doi.org/#1}{\nolinkurl{#1}}}
\providecommand{\doeprint}[1]{\href{http://ascl.net/#1}{\nolinkurl{http://ascl.net/#1}}}
\providecommand{\doarXiv}[1]{\href{https://arxiv.org/abs/#1}{\nolinkurl{https://arxiv.org/abs/#1}}}

\bibitem[{{Aihara} {et~al.}(2018){Aihara}, {Arimoto}, {Armstrong}, {Arnouts},
  {Bahcall}, {Bickerton}, {Bosch}, {Bundy}, {Capak}, {Chan}, {Chiba}, {Coupon},
  {Egami}, {Enoki}, {Finet}, {Fujimori}, {Fujimoto}, {Furusawa}, {Furusawa},
  {Goto}, {Goulding}, {Greco}, {Greene}, {Gunn}, {Hamana}, {Harikane},
  {Hashimoto}, {Hattori}, {Hayashi}, {Hayashi}, {He{\l}miniak}, {Higuchi},
  {Hikage}, {Ho}, {Hsieh}, {Huang}, {Huang}, {Ikeda}, {Imanishi}, {Inoue},
  {Iwasawa}, {Iwata}, {Jaelani}, {Jian}, {Kamata}, {Karoji}, {Kashikawa},
  {Katayama}, {Kawanomoto}, {Kayo}, {Koda}, {Koike}, {Kojima}, {Komiyama},
  {Konno}, {Koshida}, {Koyama}, {Kusakabe}, {Leauthaud}, {Lee}, {Lin}, {Lin},
  {Lupton}, {Mandelbaum}, {Matsuoka}, {Medezinski}, {Mineo}, {Miyama},
  {Miyatake}, {Miyazaki}, {Momose}, {More}, {More}, {Moritani}, {Moriya},
  {Morokuma}, {Mukae}, {Murata}, {Murayama}, {Nagao}, {Nakata}, {Niida},
  {Niikura}, {Nishizawa}, {Obuchi}, {Oguri}, {Oishi}, {Okabe}, {Okamoto},
  {Okura}, {Ono}, {Onodera}, {Onoue}, {Osato}, {Ouchi}, {Price}, {Pyo}, {Sako},
  {Sawicki}, {Shibuya}, {Shimasaku}, {Shimono}, {Shirasaki}, {Silverman},
  {Simet}, {Speagle}, {Spergel}, {Strauss}, {Sugahara}, {Sugiyama}, {Suto},
  {Suyu}, {Suzuki}, {Tait}, {Takada}, {Takata}, {Tamura}, {Tanaka}, {Tanaka},
  {Tanaka}, {Tanaka}, {Terai}, {Terashima}, {Toba}, {Tominaga}, {Toshikawa},
  {Turner}, {Uchida}, {Uchiyama}, {Umetsu}, {Uraguchi}, {Urata}, {Usuda},
  {Utsumi}, {Wang}, {Wang}, {Wong}, {Yabe}, {Yamada}, {Yamanoi}, {Yasuda},
  {Yeh}, {Yonehara}, \& {Yuma}}]{Aihara+2018PASJ}
{Aihara}, H., {Arimoto}, N., {Armstrong}, R., {et~al.} 2018, \pasj, 70, S4

\bibitem[{{Alexander} \& {Hickox}(2012)}]{Alexander&Hickox2012NewAR}
{Alexander}, D.~M., \& {Hickox}, R.~C. 2012, \nar, 56, 93

\bibitem[{{Anderson} \& {King}(2000)}]{Anderson&King2000PASP}
{Anderson}, J., \& {King}, I.~R. 2000, \pasp, 112, 1360

\bibitem[{{Annis} {et~al.}(2014){Annis}, {Soares-Santos}, {Strauss}, {Becker},
  {Dodelson}, {Fan}, {Gunn}, {Hao}, {Ivezi{\'c}}, {Jester}, {Jiang},
  {Johnston}, {Kubo}, {Lampeitl}, {Lin}, {Lupton}, {Miknaitis}, {Seo}, {Simet},
  \& {Yanny}}]{Annis+2014ApJ}
{Annis}, J., {Soares-Santos}, M., {Strauss}, M.~A., {et~al.} 2014, \apj, 794,
  120

\bibitem[{{Astropy Collaboration} {et~al.}(2018){Astropy Collaboration},
  {Price-Whelan}, {Sip{\H{o}}cz}, {G{\"u}nther}, {Lim}, {Crawford}, {Conseil},
  {Shupe}, {Craig}, {Dencheva}, {Ginsburg}, {VanderPlas}, {Bradley},
  {P{\'e}rez-Su{\'a}rez}, {de Val-Borro}, {Aldcroft}, {Cruz}, {Robitaille},
  {Tollerud}, {Ardelean}, {Babej}, {Bach}, {Bachetti}, {Bakanov}, {Bamford},
  {Barentsen}, {Barmby}, {Baumbach}, {Berry}, {Biscani}, {Boquien}, {Bostroem},
  {Bouma}, {Brammer}, {Bray}, {Breytenbach}, {Buddelmeijer}, {Burke},
  {Calderone}, {Cano Rodr{\'\i}guez}, {Cara}, {Cardoso}, {Cheedella}, {Copin},
  {Corrales}, {Crichton}, {D'Avella}, {Deil}, {Depagne}, {Dietrich}, {Donath},
  {Droettboom}, {Earl}, {Erben}, {Fabbro}, {Ferreira}, {Finethy}, {Fox},
  {Garrison}, {Gibbons}, {Goldstein}, {Gommers}, {Greco}, {Greenfield},
  {Groener}, {Grollier}, {Hagen}, {Hirst}, {Homeier}, {Horton}, {Hosseinzadeh},
  {Hu}, {Hunkeler}, {Ivezi{\'c}}, {Jain}, {Jenness}, {Kanarek}, {Kendrew},
  {Kern}, {Kerzendorf}, {Khvalko}, {King}, {Kirkby}, {Kulkarni}, {Kumar},
  {Lee}, {Lenz}, {Littlefair}, {Ma}, {Macleod}, {Mastropietro}, {McCully},
  {Montagnac}, {Morris}, {Mueller}, {Mumford}, {Muna}, {Murphy}, {Nelson},
  {Nguyen}, {Ninan}, {N{\"o}the}, {Ogaz}, {Oh}, {Parejko}, {Parley}, {Pascual},
  {Patil}, {Patil}, {Plunkett}, {Prochaska}, {Rastogi}, {Reddy Janga},
  {Sabater}, {Sakurikar}, {Seifert}, {Sherbert}, {Sherwood-Taylor}, {Shih},
  {Sick}, {Silbiger}, {Singanamalla}, {Singer}, {Sladen}, {Sooley},
  {Sornarajah}, {Streicher}, {Teuben}, {Thomas}, {Tremblay}, {Turner},
  {Terr{\'o}n}, {van Kerkwijk}, {de la Vega}, {Watkins}, {Weaver}, {Whitmore},
  {Woillez}, {Zabalza}, \& {Astropy Contributors}}]{2018AJ....156..123A}
{Astropy Collaboration}, {Price-Whelan}, A.~M., {Sip{\H{o}}cz}, B.~M., {et~al.}
  2018, \aj, 156, 123

\bibitem[{{Astropy Collaboration} {et~al.}(2013){Astropy Collaboration}, {Robitaille}, {Tollerud}, {Greenfield}, {Droettboom}, {Bray}, {Aldcroft}, {Davis}, {Ginsburg}, {Price-Whelan}, {Kerzendorf}, {Conley}, {Crighton}, {Barbary}, {Muna}, {Ferguson}, {Grollier}, {Parikh}, {Nair}, {Unther}, {Deil}, {Woillez}, {Conseil}, {Kramer}, {Turner}, {Singer}, {Fox}, {Weaver}, {Zabalza}, {Edwards}, {Azalee Bostroem}, {Burke}, {Casey}, {Crawford}, {Dencheva}, {Ely}, {Jenness}, {Labrie}, {Lim}, {Pierfederici}, {Pontzen}, {Ptak}, {Refsdal}, {Servillat}, \& {Streicher}}]{2013A&A...558A..33A} {Astropy Collaboration}, {Robitaille}, T.~P., {Tollerud}, E.~J., {et~al.} 2013, \aap, 558, A33

\bibitem[Barnes \& Hernquist (1996)]{Barnes1996} Barnes, J. E., \& Hernquist, L. E. 1996, \apj, 471, 115

\bibitem[{{Barro} {et~al.}(2014){Barro}, {Faber}, {P{\'e}rez-Gonz{\'a}lez},
  {Pacifici}, {Trump}, {Koo}, {Wuyts}, {Guo}, {Bell}, {Dekel}, {Porter},
  {Primack}, {Ferguson}, {Ashby}, {Caputi}, {Ceverino}, {Croton}, {Fazio},
  {Giavalisco}, {Hsu}, {Kocevski}, {Koekemoer}, {Kurczynski}, {Kollipara},
  {Lee}, {McIntosh}, {McGrath}, {Moody}, {Somerville}, {Papovich}, {Salvato},
  {Santini}, {Tal}, {van der Wel}, {Williams}, {Willner}, \&
  {Zolotov}}]{Barro+2014ApJ}
{Barro}, G., {Faber}, S.~M., {P{\'e}rez-Gonz{\'a}lez}, P.~G., {et~al.} 2014,
  \apj, 791, 52

\bibitem[{{Bell} {et~al.}(2003){Bell}, {McIntosh}, {Katz}, \&
  {Weinberg}}]{Bell+2003ApJS}
{Bell}, E.~F., {McIntosh}, D.~H., {Katz}, N., \& {Weinberg}, M.~D. 2003, \apjs, 149, 289

\bibitem[{{Bennert} {et~al.}(2021){Bennert}, {Treu}, {Ding}, {Stomberg}, {Birrer}, {Snyder}, {Malkan}, {Stephens}, \& {Auger}}]{Bennert+2021ApJ} {Bennert}, V.~N., {Treu}, T., {Ding}, X., {et~al.} 2021, \apj, 921, 36

\bibitem[{{Bentz} \& {Manne-Nicholas}(2018)}]{Bentz+Manne-Nicholas2018} 
{Bentz}, M. C., \& {Manne-Nicholas}, E. 2018, \apj, 864, 146

\bibitem[{{Berriman} {et~al.}(2008){Berriman}, {Good}, {Laity}, \&
  {Kong}}]{Berriman+2008ASPC}
{Berriman}, G.~B., {Good}, J.~C., {Laity}, A.~C., \& {Kong}, M. 2008, in ASP Conf. Ser. 394, Astronomical Data Analysis Software and Systems XVII, ed. R.~W. {Argyle}, P.~S.  {Bunclark}, \& J.~R. {Lewis} (San Francisco: ASP), 83

\bibitem[{{Bertin}(2011)}]{Bertin2011ASPC}
{Bertin}, E. 2011, in ASP Conf. Ser.  442, Astronomical Data Analysis Software and Systems XX, ed. I.~N.  {Evans}, A.~{Accomazzi}, D.~J. {Mink}, \& A.~H. {Rots} (San Francisco: ASP), 435

\bibitem[{{Bertin} \& {Arnouts}(1996)}]{1996A&AS..117..393B}
{Bertin}, E., \& {Arnouts}, S. 1996, \aaps, 117, 393

\bibitem[{{Bertin} {et~al.}(2002){Bertin}, {Mellier}, {Radovich}, {Missonnier},
  {Didelon}, \& {Morin}}]{Bertin+2002ASPC}
{Bertin}, E., {Mellier}, Y., {Radovich}, M., {et~al.} 2002, in ASP Conf. Ser. 281, Astronomical Data Analysis Software and Systems XI, ed. D.~A. {Bohlender}, D.~{Durand}, \& T.~H. {Handley} (San Francisco: ASP), 228

\bibitem[{{Birrer} \& {Amara}(2018)}]{Birrer&Amara2018PDU}
{Birrer}, S., \& {Amara}, A. 2018, Physics of the Dark Universe, 22, 189

\bibitem[{{Boquien} {et~al.}(2019){Boquien}, {Burgarella}, {Roehlly}, {Buat},
  {Ciesla}, {Corre}, {Inoue}, \& {Salas}}]{2019A&A...622A.103Boquien+}
{Boquien}, M., {Burgarella}, D., {Roehlly}, Y., {et~al.} 2019, \aap, 622, A103


\bibitem[{Bradley {et~al.}(2020)Bradley, Sip{\H o}cz, Robitaille, Tollerud,
  Vin{\'{\i}}cius, Deil, Barbary, Wilson, Busko, G{\"u}nther, Cara, Conseil,
  Bostroem, Droettboom, Bray, Bratholm, Lim, Barentsen, Craig, Pascual, Perren,
  Greco, Donath, de~Val-Borro, Kerzendorf, Bach, Weaver, D'Eugenio, Souchereau,
  \& Ferreira}]{photutils}
Bradley, L., Sip{\H o}cz, B., Robitaille, T., {et~al.} 2020, astropy/photutils:
  1.0.1, 1.0.1,  Zenodo

\bibitem[{{Brinchmann} {et~al.}(2004){Brinchmann}, {Charlot}, {White},
  {Tremonti}, {Kauffmann}, {Heckman}, \& {Brinkmann}}]{Brinchmann+2004MNRAS}
{Brinchmann}, J., {Charlot}, S., {White}, S.~D.~M., {et~al.} 2004, \mnras, 351,
  1151

\bibitem[Br{\"u}ggen \& Kaiser(2002)]{Bruggen&Kaiser2002Natur} Br{\"u}ggen, M., \& Kaiser, C.~R.\ 2002, \nat, 418, 301

\bibitem[{{Bruzual} \& {Charlot}(2003)}]{BC03}
{Bruzual}, G., \& {Charlot}, S. 2003, \mnras, 344, 1000

\bibitem[Calistro Rivera et al.(2016)]{Calistro_Rivera+2016ApJ} Calistro Rivera, G., Lusso, E., Hennawi, J.~F., et al.\ 2016, \apj, 833, 98

\bibitem[{{Calzetti} {et~al.}(2000){Calzetti}, {Armus}, {Bohlin}, {Kinney},
  {Koornneef}, \& {Storchi-Bergmann}}]{Calzetti+2000ApJ}
{Calzetti}, D., {Armus}, L., {Bohlin}, R.~C., {et~al.} 2000, \apj, 533, 682

\bibitem[{{Chabrier}(2003)}]{Chabrier2003PASP}
{Chabrier}, G. 2003, \pasp, 115, 763

\bibitem[{{Chambers} {et~al.}(2016){Chambers}, {Magnier}, {Metcalfe},
  {Flewelling}, {Huber}, {Waters}, {Denneau}, {Draper}, {Farrow}, {Finkbeiner},
  {Holmberg}, {Koppenhoefer}, {Price}, {Rest}, {Saglia}, {Schlafly}, {Smartt},
  {Sweeney}, {Wainscoat}, {Burgett}, {Chastel}, {Grav}, {Heasley}, {Hodapp},
  {Jedicke}, {Kaiser}, {Kudritzki}, {Luppino}, {Lupton}, {Monet}, {Morgan},
  {Onaka}, {Shiao}, {Stubbs}, {Tonry}, {White}, {Ba{\~n}ados}, {Bell},
  {Bender}, {Bernard}, {Boegner}, {Boffi}, {Botticella}, {Calamida},
  {Casertano}, {Chen}, {Chen}, {Cole}, {Deacon}, {Frenk}, {Fitzsimmons},
  {Gezari}, {Gibbs}, {Goessl}, {Goggia}, {Gourgue}, {Goldman}, {Grant},
  {Grebel}, {Hambly}, {Hasinger}, {Heavens}, {Heckman}, {Henderson}, {Henning},
  {Holman}, {Hopp}, {Ip}, {Isani}, {Jackson}, {Keyes}, {Koekemoer}, {Kotak},
  {Le}, {Liska}, {Long}, {Lucey}, {Liu}, {Martin}, {Masci}, {McLean}, {Mindel},
  {Misra}, {Morganson}, {Murphy}, {Obaika}, {Narayan}, {Nieto-Santisteban},
  {Norberg}, {Peacock}, {Pier}, {Postman}, {Primak}, {Rae}, {Rai}, {Riess},
  {Riffeser}, {Rix}, {R{\"o}ser}, {Russel}, {Rutz}, {Schilbach}, {Schultz},
  {Scolnic}, {Strolger}, {Szalay}, {Seitz}, {Small}, {Smith}, {Soderblom},
  {Taylor}, {Thomson}, {Taylor}, {Thakar}, {Thiel}, {Thilker}, {Unger},
  {Urata}, {Valenti}, {Wagner}, {Walder}, {Walter}, {Watters}, {Werner},
  {Wood-Vasey}, \& {Wyse}}]{Chambers+2016arXiv}
{Chambers}, K.~C., {Magnier}, E.~A., {Metcalfe}, N., {et~al.} 2016, arXiv:1612.05560

\bibitem[{{Chen} {et~al.}(2013){Chen}, {Hickox}, {Alberts}, {Brodwin}, {Jones},
  {Murray}, {Alexander}, {Assef}, {Brown}, {Dey}, {Forman}, {Gorjian},
  {Goulding}, {Le Floc'h}, {Jannuzi}, {Mullaney}, \& {Pope}}]{Chen+2013ApJ}
{Chen}, C.-T.~J., {Hickox}, R.~C., {Alberts}, S., {et~al.} 2013, \apj, 773, 3

\bibitem[{{Ciesla} {et~al.}(2015){Ciesla}, {Charmandaris}, {Georgakakis},
  {Bernhard}, {Mitchell}, {Buat}, {Elbaz}, {LeFloc'h}, {Lacey}, {Magdis}, \&
  {Xilouris}}]{Ciesla+2015A&A}
{Ciesla}, L., {Charmandaris}, V., {Georgakakis}, A., {et~al.} 2015, \aap, 576,
  A10

\bibitem[Cisternas et al. (2011)]{Cisternas2011} Cisternas, M., Jahnke, K., Inskip, K. J., et al. 2011, \apj, 726, 57

\bibitem[{{Conroy} (2013)}]{Conroy2013} {Conroy}, C. 2013, ARA\&A, 51, 393

\bibitem[Cresci et al.(2015)]{Cresci+2015ApJ} Cresci, G., Mainieri, V., Brusa, M., et al.\ 2015, \apj, 799, 82

\bibitem[{{Dai} {et~al.}(2018){Dai}, {Wilkes}, {Bergeron}, {Kuraszkiewicz}, {Omont}, {Atanas}, \& {Teplitz}}]{Dai+2018MNRAS} {Dai}, Y.~S., {Wilkes}, B.~J., {Bergeron}, J., {et~al.} 2018, \mnras, 478, 4238

\bibitem[Davari et al. (2017)]{Davari2017} Davari, R., Ho, L. C., Mobasher, B., \& Canalizo, G. 2017, \apj, 836, 75

\bibitem[Di Matteo et al.(2005)]{Di_Matteo+2005Natur} Di Matteo, T., Springel, V., \& Hernquist, L.\ 2005, \nat, 433, 604

\bibitem[{{Ding} {et~al.}(2020){Ding}, {Silverman}, {Treu}, {Schulze},
  {Schramm}, {Birrer}, {Park}, {Jahnke}, {Bennert}, {Kartaltepe}, {Koekemoer},
  {Malkan}, \& {Sanders}}]{Ding+2020ApJ}
{Ding}, X., {Silverman}, J., {Treu}, T., {et~al.} 2020, \apj, 888, 37

\bibitem[Dunlop et al. (2003)]{Dunlop2003} Dunlop, J. S., McLure, R. J., Kukula, M. J., et al. 2003, \mnras, 340, 1095

\bibitem[{{Elbaz} {et~al.}(2011){Elbaz}, {Dickinson}, {Hwang},
  {D{\'\i}az-Santos}, {Magdis}, {Magnelli}, {Le Borgne}, {Galliano},
  {Pannella}, {Chanial}, {Armus}, {Charmandaris}, {Daddi}, {Aussel}, {Popesso},
  {Kartaltepe}, {Altieri}, {Valtchanov}, {Coia}, {Dannerbauer}, {Dasyra},
  {Leiton}, {Mazzarella}, {Alexander}, {Buat}, {Burgarella}, {Chary}, {Gilli},
  {Ivison}, {Juneau}, {Le Floc'h}, {Lutz}, {Morrison}, {Mullaney}, {Murphy},
  {Pope}, {Scott}, {Brodwin}, {Calzetti}, {Cesarsky}, {Charlot}, {Dole},
  {Eisenhardt}, {Ferguson}, {F{\"o}rster Schreiber}, {Frayer}, {Giavalisco},
  {Huynh}, {Koekemoer}, {Papovich}, {Reddy}, {Surace}, {Teplitz}, {Yun}, \&
  {Wilson}}]{Elbaz+2011A&A}
{Elbaz}, D., {Dickinson}, M., {Hwang}, H.~S., {et~al.} 2011, \aap, 533, A119

\bibitem[{{Ellison} {et~al.}(2016){Ellison}, {Teimoorinia}, {Rosario}, \&
  {Mendel}}]{Ellison+2016MNRAS}
{Ellison}, S.~L., {Teimoorinia}, H., {Rosario}, D.~J., \& {Mendel}, J.~T. 2016,
  \mnras, 458, L34

\bibitem[{{Erwin}(2015)}]{Erwin2015ApJ} {Erwin}, P. 2015, \apj, 799, 226

\bibitem[{{Falomo} {et~al.}(2014){Falomo}, {Bettoni}, {Karhunen}, {Kotilainen},
  \& {Uslenghi}}]{Falomo+2014MNRAS}
{Falomo}, R., {Bettoni}, D., {Karhunen}, K., {Kotilainen}, J.~K., \&
  {Uslenghi}, M. 2014, \mnras, 440, 476

\bibitem[{{Fan} {et~al.}(2010){Fan}, {Lapi}, {Bressan}, {Bernardi}, {De Zotti},
  \& {Danese}}]{Fan+2010ApJ}
{Fan}, L., {Lapi}, A., {Bressan}, A., {et~al.} 2010, \apj, 718, 1460

\bibitem[{{Fan} {et~al.}(2008){Fan}, {Lapi}, {De Zotti}, \&
  {Danese}}]{Fan+2008ApJ}
{Fan}, L., {Lapi}, A., {De Zotti}, G., \& {Danese}, L. 2008, \apjl, 689, L101

\bibitem[{{Farrow} {et~al.}(2014){Farrow}, {Cole}, {Metcalfe}, {Draper},
  {Norberg}, {Foucaud}, {Burgett}, {Chambers}, {Kaiser}, {Kudritzki},
  {Magnier}, {Price}, {Tonry}, \& {Waters}}]{Farrow+2014MNRAS}
{Farrow}, D.~J., {Cole}, S., {Metcalfe}, N., {et~al.} 2014, \mnras, 437, 748

\bibitem[{{Ferrarese} \& {Merritt}(2000)}]{Ferrarese2000}
{Ferrarese}, L., \& {Merritt}, D. 2000, \apj, 539, L9

\bibitem[Feruglio et al.(2010)]{Feruglio+2010A&A} Feruglio, C., Maiolino, R., Piconcelli, E., et al.\ 2010, \aap, 518, L155

\bibitem[{{Fiore} {et~al.}(2017){Fiore}, {Feruglio}, {Shankar}, {Bischetti},
  {Bongiorno}, {Brusa}, {Carniani}, {Cicone}, {Duras}, {Lamastra}, {Mainieri},
  {Marconi}, {Menci}, {Maiolino}, {Piconcelli}, {Vietri}, \&
  {Zappacosta}}]{Fiore+2017A&A}
{Fiore}, F., {Feruglio}, C., {Shankar}, F., {et~al.} 2017, \aap, 601, A143

\bibitem[{{Flewelling} {et~al.}(2020){Flewelling}, {Magnier}, {Chambers},
  {Heasley}, {Holmberg}, {Huber}, {Sweeney}, {Waters}, {Calamida}, {Casertano},
  {Chen}, {Farrow}, {Hasinger}, {Henderson}, {Long}, {Metcalfe}, {Narayan},
  {Nieto-Santisteban}, {Norberg}, {Rest}, {Saglia}, {Szalay}, {Thakar},
  {Tonry}, {Valenti}, {Werner}, {White}, {Denneau}, {Draper}, {Hodapp},
  {Jedicke}, {Kaiser}, {Kudritzki}, {Price}, {Wainscoat}, {Chastel}, {McLean},
  {Postman}, \& {Shiao}}]{Flewelling+2020ApJS}
{Flewelling}, H.~A., {Magnier}, E.~A., {Chambers}, K.~C., {et~al.} 2020, \apjs,
  251, 7

\bibitem[{{Fluetsch} {et~al.}(2019){Fluetsch}, {Maiolino}, {Carniani},
  {Marconi}, {Cicone}, {Bourne}, {Costa}, {Fabian}, {Ishibashi}, \&
  {Venturi}}]{Fluetsch+2019MNRAS}
{Fluetsch}, A., {Maiolino}, R., {Carniani}, S., {et~al.} 2019, \mnras, 483,
  4586

\bibitem[{{Gabor} {et~al.}(2009){Gabor}, {Impey}, {Jahnke}, {Simmons}, {Trump},
  {Koekemoer}, {Brusa}, {Cappelluti}, {Schinnerer}, {Smol{\v{c}}i{\'c}},
  {Salvato}, {Rhodes}, {Mobasher}, {Capak}, {Massey}, {Leauthaud}, \&
  {Scoville}}]{Gabor+2009ApJ}
{Gabor}, J.~M., {Impey}, C.~D., {Jahnke}, K., {et~al.} 2009, \apj, 691, 705

\bibitem[{{Gallagher} {et~al.}(2019){Gallagher}, {Maiolino}, {Belfiore},
  {Drory}, {Riffel}, \& {Riffel}}]{Gallagher+2019MNRAS}
{Gallagher}, R., {Maiolino}, R., {Belfiore}, F., {et~al.} 2019, \mnras, 485,
  3409

\bibitem[{{Gao} \& {Ho}(2017)}]{Gao&Ho2017} Gao, H., \& Ho, L. C. 2017, \apj, 845, 114

\bibitem[{{Gao} {et~al.}(2019){Gao}, {Ho}, {Barth}, \& {Li}}]{Gao+2019ApJS} {Gao}, H., {Ho}, L.~C., {Barth}, A.~J., \& {Li}, Z.-Y. 2019, \apjs, 244, 34

\bibitem[{{Gebhardt} {et~al.}(2000){Gebhardt}, {Bender}, {Bower}, {Dressler},
  {Faber}, {Filippenko}, {Green}, {Grillmair}, {Ho}, {Kormendy}, {Lauer},
  {Magorrian}, {Pinkney}, {Richstone}, \& {Tremaine}}]{Gebhardt+2000ApJ}
{Gebhardt}, K., {Bender}, R., {Bower}, G., {et~al.} 2000, \apj, 539, L13

\bibitem[{{Green}(2018)}]{2018JOSS....3..695Green}
{Green}, G.~M. 2018, The Journal of Open Source Software, 3, 695

\bibitem[{{Greene} {et al.}(2008)}]{Greene2008} 
Greene, J. E., Ho, L. C., \& Barth, A. J. 2008, \apj, 688, 159

\bibitem[{Harris {et~al.}(2020)Harris, Millman, van~der Walt, Gommers,
  Virtanen, Cournapeau, Wieser, Taylor, Berg, Smith, Kern, Picus, Hoyer, van
  Kerkwijk, Brett, Haldane, del R{\'{i}}o, Wiebe, Peterson,
  G{\'{e}}rard-Marchant, Sheppard, Reddy, Weckesser, Abbasi, Gohlke, \&
  Oliphant}]{Harris2020}
Harris, C.~R., Millman, K.~J., van~der Walt, S.~J., {et~al.} 2020, Nature, 585,
  357

\bibitem[{{H{\"a}u{\ss}ler} {et~al.}(2013){H{\"a}u{\ss}ler}, {Bamford}, {Vika}, {Rojas}, {Barden}, {Kelvin}, {Alpaslan}, {Robotham}, {Driver}, {Baldry}, {Brough}, {Hopkins}, {Liske}, {Nichol}, {Popescu}, \& {Tuffs}}]{Haussler+2013MNRAS} {H{\"a}u{\ss}ler}, B., {Bamford}, S.~P., {Vika}, M., {et~al.} 2013, \mnras, 430, 330

\bibitem[{{H{\"a}ussler} {et~al.}(2007){H{\"a}ussler}, {McIntosh}, {Barden},
  {Bell}, {Rix}, {Borch}, {Beckwith}, {Caldwell}, {Heymans}, {Jahnke}, {Jogee},
  {Koposov}, {Meisenheimer}, {S{\'a}nchez}, {Somerville}, {Wisotzki}, \&
  {Wolf}}]{Haussler+2007ApJS}
{H{\"a}u{\ss}ler}, B., {McIntosh}, D.~H., {Barden}, M., {et~al.} 2007, \apjs, 172,
  615

\bibitem[{{Heckman} \& {Best}(2014)}]{Heckman&Best2014ARA&A}
{Heckman}, T.~M., \& {Best}, P.~N. 2014, ARA\&A, 52, 589

\bibitem[{{Ho}(2004)}]{Ho2004coa..book}
{Ho}, L. C. 2004,  ed., Carnegie Observatories Astrophysics Series, Vol. 1: Coevolution of Black Holes and Galaxies (Cambridge: Cambridge Univ. Press)

\bibitem[{{Ho}(2005)}]{Ho2005}
Ho, L. C. 2005, \apj, 629, 680

\bibitem[{{Ho}(2008)}]{Ho2008} Ho, L. C. 2008, ARA\&A, 46, 475

\bibitem[Ho (2009)]{Ho2009} Ho, L. C. 2009, \apj, 699, 626

\bibitem[{Ho et al. (1997)}]{Ho+1997} Ho, L. C., Filippenko, A. V., \& Sargent, W. L. W. 1997, \apj, 487, 568

\bibitem[{Ho et al. (2003)}]{Ho2003} Ho, L. C., Filippenko, A. V., \& Sargent, W. L. W. 2003, \apj, 583, 159

\bibitem[H{\"o}nig \& Kishimoto(2017)]{Honig&Kishimoto2017ApJL} H{\"o}nig, S.~F. \& Kishimoto, M.\ 2017, \apjl, 838, L20

\bibitem[Hopkins et~al.(2009)]{Hopkins2009} Hopkins, P. F., Cox, T. J., Younger, J. D., \& Hernquist, L. 2009, \apj, 691, 1168

\bibitem[{{Hopkins} {et~al.}(2008){Hopkins}, {Hernquist}, {Cox}, \& {Kere{\v{s}}}}]{Hopkins+2008ApJS} {Hopkins}, P.~F., {Hernquist}, L., {Cox}, T.~J., \& {Kere{\v{s}}}, D. 2008, \apjs, 175, 356

\bibitem[{Hunter(2007)}]{Hunter2007}
Hunter, J.~D. 2007, Computing in Science \& Engineering, 9, 90

\bibitem[{{Husemann} {et~al.}(2017){Husemann}, {Davis}, {Jahnke},
  {Dannerbauer}, {Urrutia}, \& {Hodge}}]{Husemann+2017MNRAS}
{Husemann}, B., {Davis}, T.~A., {Jahnke}, K., {et~al.} 2017, \mnras, 470, 1570

\bibitem[{{Husemann} {et~al.}(2014){Husemann}, {Jahnke}, {S{\'a}nchez},
  {Wisotzki}, {Nugroho}, {Kupko}, \& {Schramm}}]{Husemann+2014MNRAS}
{Husemann}, B., {Jahnke}, K., {S{\'a}nchez}, S.~F., {et~al.} 2014, \mnras, 443,
  755

\bibitem[{{Ishino} {et~al.}(2020){Ishino}, {Matsuoka}, {Koyama}, {Saeda},
  {Strauss}, {Goulding}, {Imanishi}, {Kawaguchi}, {Minezaki}, {Nagao},
  {Noboriguchi}, {Schramm}, {Silverman}, {Taniguchi}, \&
  {Toba}}]{Ishino+2020PASJ}
{Ishino}, T., {Matsuoka}, Y., {Koyama}, S., {et~al.} 2020, \pasj, 72, 83

\bibitem[{{Ivezi{\'c}} {et~al.}(2019){Ivezi{\'c}}, {Kahn}, {Tyson}, {Abel},
  {Acosta}, {Allsman}, {Alonso}, {AlSayyad}, {Anderson}, {Andrew}, {Angel},
  {Angeli}, {Ansari}, {Antilogus}, {Araujo}, {Armstrong}, {Arndt}, {Astier},
  {Aubourg}, {Auza}, {Axelrod}, {Bard}, {Barr}, {Barrau}, {Bartlett}, {Bauer},
  {Bauman}, {Baumont}, {Bechtol}, {Bechtol}, {Becker}, {Becla}, {Beldica},
  {Bellavia}, {Bianco}, {Biswas}, {Blanc}, {Blazek}, {Blandford}, {Bloom},
  {Bogart}, {Bond}, {Booth}, {Borgland}, {Borne}, {Bosch}, {Boutigny},
  {Brackett}, {Bradshaw}, {Brandt}, {Brown}, {Bullock}, {Burchat}, {Burke},
  {Cagnoli}, {Calabrese}, {Callahan}, {Callen}, {Carlin}, {Carlson},
  {Chandrasekharan}, {Charles-Emerson}, {Chesley}, {Cheu}, {Chiang}, {Chiang},
  {Chirino}, {Chow}, {Ciardi}, {Claver}, {Cohen-Tanugi}, {Cockrum}, {Coles},
  {Connolly}, {Cook}, {Cooray}, {Covey}, {Cribbs}, {Cui}, {Cutri}, {Daly},
  {Daniel}, {Daruich}, {Daubard}, {Daues}, {Dawson}, {Delgado}, {Dellapenna},
  {de Peyster}, {de Val-Borro}, {Digel}, {Doherty}, {Dubois},
  {Dubois-Felsmann}, {Durech}, {Economou}, {Eifler}, {Eracleous}, {Emmons},
  {Fausti Neto}, {Ferguson}, {Figueroa}, {Fisher-Levine}, {Focke}, {Foss},
  {Frank}, {Freemon}, {Gangler}, {Gawiser}, {Geary}, {Gee}, {Geha}, {Gessner},
  {Gibson}, {Gilmore}, {Glanzman}, {Glick}, {Goldina}, {Goldstein}, {Goodenow},
  {Graham}, {Gressler}, {Gris}, {Guy}, {Guyonnet}, {Haller}, {Harris},
  {Hascall}, {Haupt}, {Hernandez}, {Herrmann}, {Hileman}, {Hoblitt}, {Hodgson},
  {Hogan}, {Howard}, {Huang}, {Huffer}, {Ingraham}, {Innes}, {Jacoby}, {Jain},
  {Jammes}, {Jee}, {Jenness}, {Jernigan}, {Jevremovi{\'c}}, {Johns}, {Johnson},
  {Johnson}, {Jones}, {Juramy-Gilles}, {Juri{\'c}}, {Kalirai}, {Kallivayalil},
  {Kalmbach}, {Kantor}, {Karst}, {Kasliwal}, {Kelly}, {Kessler}, {Kinnison},
  {Kirkby}, {Knox}, {Kotov}, {Krabbendam}, {Krughoff}, {Kub{\'a}nek},
  {Kuczewski}, {Kulkarni}, {Ku}, {Kurita}, {Lage}, {Lambert}, {Lange},
  {Langton}, {Le Guillou}, {Levine}, {Liang}, {Lim}, {Lintott}, {Long},
  {Lopez}, {Lotz}, {Lupton}, {Lust}, {MacArthur}, {Mahabal}, {Mandelbaum},
  {Markiewicz}, {Marsh}, {Marshall}, {Marshall}, {May}, {McKercher}, {McQueen},
  {Meyers}, {Migliore}, {Miller}, {Mills}, {Miraval}, {Moeyens}, {Moolekamp},
  {Monet}, {Moniez}, {Monkewitz}, {Montgomery}, {Morrison}, {Mueller},
  {Muller}, {Mu{\~n}oz Arancibia}, {Neill}, {Newbry}, {Nief}, {Nomerotski},
  {Nordby}, {O'Connor}, {Oliver}, {Olivier}, {Olsen}, {O'Mullane}, {Ortiz},
  {Osier}, {Owen}, {Pain}, {Palecek}, {Parejko}, {Parsons}, {Pease},
  {Peterson}, {Peterson}, {Petravick}, {Libby Petrick}, {Petry},
  {Pierfederici}, {Pietrowicz}, {Pike}, {Pinto}, {Plante}, {Plate}, {Plutchak},
  {Price}, {Prouza}, {Radeka}, {Rajagopal}, {Rasmussen}, {Regnault}, {Reil},
  {Reiss}, {Reuter}, {Ridgway}, {Riot}, {Ritz}, {Robinson}, {Roby}, {Roodman},
  {Rosing}, {Roucelle}, {Rumore}, {Russo}, {Saha}, {Sassolas}, {Schalk},
  {Schellart}, {Schindler}, {Schmidt}, {Schneider}, {Schneider}, {Schoening},
  {Schumacher}, {Schwamb}, {Sebag}, {Selvy}, {Sembroski}, {Seppala}, {Serio},
  {Serrano}, {Shaw}, {Shipsey}, {Sick}, {Silvestri}, {Slater}, {Smith},
  {Smith}, {Sobhani}, {Soldahl}, {Storrie-Lombardi}, {Stover}, {Strauss},
  {Street}, {Stubbs}, {Sullivan}, {Sweeney}, {Swinbank}, {Szalay}, {Takacs},
  {Tether}, {Thaler}, {Thayer}, {Thomas}, {Thornton}, {Thukral}, {Tice},
  {Trilling}, {Turri}, {Van Berg}, {Vanden Berk}, {Vetter}, {Virieux},
  {Vucina}, {Wahl}, {Walkowicz}, {Walsh}, {Walter}, {Wang}, {Wang}, {Warner},
  {Wiecha}, {Willman}, {Winters}, {Wittman}, {Wolff}, {Wood-Vasey}, {Wu},
  {Xin}, {Yoachim}, \& {Zhan}}]{Ivezic+2019ApJ}
{Ivezi{\'c}}, {\v{Z}}., {Kahn}, S.~M., {Tyson}, J.~A., {et~al.} 2019, \apj,
  873, 111

\bibitem[{{Jarvis} {et~al.}(2020){Jarvis}, {Harrison}, {Mainieri}, {Calistro Rivera}, {Jethwa}, {Zhang}, {Alexander}, {Circosta}, {Costa}, {De Breuck}, {Kakkad}, {Kharb}, {Lansbury}, \& {Thomson}}]{Jarvis+2020MNRAS} {Jarvis}, M.~E., {Harrison}, C.~M., {Mainieri}, V., {et~al.} 2020, \mnras, 498, 1560

\bibitem[{{Jiang} {et~al.}(2011)}]{Jiang2011} 
Jiang, Y.-F., Greene, J. E., Ho, L. C., Xiao, T., \& Barth, A. J. 2011, \apj, 742, 68

\bibitem[{{Kauffmann} {et~al.}(2003){Kauffmann}, {Heckman}, {Tremonti},
  {Brinchmann}, {Charlot}, {White}, {Ridgway}, {Brinkmann}, {Fukugita}, {Hall},
  {Ivezi{\'c}}, {Richards}, \& {Schneider}}]{Kauffmann+2003MNRAS}
{Kauffmann}, G., {Heckman}, T.~M., {Tremonti}, C., {et~al.} 2003, \mnras, 346,
  1055

\bibitem[{{Kelvin} {et~al.}(2012){Kelvin}, {Driver}, {Robotham}, {Hill},
  {Alpaslan}, {Baldry}, {Bamford}, {Bland-Hawthorn}, {Brough}, {Graham},
  {H{\"a}ussler}, {Hopkins}, {Liske}, {Loveday}, {Norberg}, {Phillipps},
  {Popescu}, {Prescott}, {Taylor}, \& {Tuffs}}]{Kelvin+2012MNRAS}
{Kelvin}, L.~S., {Driver}, S.~P., {Robotham}, A. S.~G., {et~al.} 2012, \mnras,
  421, 1007

\bibitem[{{Kim} {et~al.}(2021){Kim}, {Barth}, {Ho}, \& {Son}}]{Kim+2021ApJS}
{Kim}, M., {Barth}, A.~J., {Ho}, L.~C., \& {Son}, S. 2021, \apjs, 256, 40

\bibitem[{{Kim} \& {Ho}(2019)}]{Kim&Ho2019ApJ}
{Kim}, M., \& {Ho}, L.~C. 2019, \apj, 876, 35

\bibitem[Kim et al.(2006)]{Kim2006} 
Kim, M., Ho, L. C., \& Im, M. 2006, \apj, 642, 702

\bibitem[{{Kim} {et~al.}(2008){Kim}, {Ho}, {Peng}, {Barth}, \&
  {Im}}]{Kim+2008ApJS}
{Kim}, M., {Ho}, L.~C., {Peng}, C.~Y., {Barth}, A.~J., \& {Im}, M. 2008, \apjs,
  179, 283

\bibitem[{{Kim} {et~al.}(2017){Kim}, {Ho}, {Peng}, {Barth}, \&
  {Im}}]{Kim+2017ApJS}
{Kim}, M., {Ho}, L.~C., {Peng}, C.~Y., {Barth}, A.~J., \& {Im}, M. 2017, \apjs, 232, 21

\bibitem[{{Kocevski} {et~al.}(2017){Kocevski}, {Barro}, {Faber}, {Dekel},
  {Somerville}, {Young}, {Williams}, {McIntosh}, {Georgakakis}, {Hasinger},
  {Nandra}, {Civano}, {Alexander}, {Almaini}, {Conselice}, {Donley},
  {Ferguson}, {Giavalisco}, {Grogin}, {Hathi}, {Hawkins}, {Koekemoer}, {Koo},
  {McGrath}, {Mobasher}, {P{\'e}rez Gonz{\'a}lez}, {Pforr}, {Primack},
  {Santini}, {Stefanon}, {Trump}, {van der Wel}, {Wuyts}, \&
  {Yan}}]{Kocevski+2017ApJ}
{Kocevski}, D.~D., {Barro}, G., {Faber}, S.~M., {et~al.} 2017, \apj, 846, 112

\bibitem[{{Koekemoer} {et~al.}(2007){Koekemoer}, {Aussel}, {Calzetti}, {Capak},
  {Giavalisco}, {Kneib}, {Leauthaud}, {Le F{\`e}vre}, {McCracken}, {Massey},
  {Mobasher}, {Rhodes}, {Scoville}, \& {Shopbell}}]{Koekemoer+2007ApJS}
{Koekemoer}, A.~M., {Aussel}, H., {Calzetti}, D., {et~al.} 2007, \apjs, 172,
  196

\bibitem[{{Kormendy} \& {Ho}(2013)}]{Kormendy&Ho2013ARA&A}
{Kormendy}, J., \& {Ho}, L.~C. 2013, \araa, 51, 511

\bibitem[{{Kormendy} \& {Kennicutt}(2004)}]{Kormendy&Kennicutt2004ARA&A}
{Kormendy}, J., \& {Kennicutt} Jr., R.~C., 2004, \araa, 42, 603

\bibitem[{{Koss} {et~al.}(2021){Koss}, {Strittmatter}, {Lamperti}, {Shimizu},
  {Trakhtenbrot}, {Saintonge}, {Treister}, {Cicone}, {Mushotzky}, {Oh},
  {Ricci}, {Stern}, {Ananna}, {Bauer}, {Privon}, {B{\"a}r}, {De Breuck},
  {Harrison}, {Ichikawa}, {Powell}, {Rosario}, {Sanders}, {Schawinski}, {Shao},
  {Megan Urry}, \& {Veilleux}}]{Koss+2021ApJS}
{Koss}, M.~J., {Strittmatter}, B., {Lamperti}, I., {et~al.} 2021, \apjs, 252,
  29
 
\bibitem[Koutoulidis et al.(2022)]{Koutoulidis+2022A&A} Koutoulidis, L., Mountrichas, G., Georgantopoulos, I., et al.\ 2022, \aap, 658, A35

\bibitem[{{Kron}(1980)}]{Kron1980}
Kron, R. G. 1980, \apjs, 43, 305

\bibitem[{{Kroupa}(2001)}]{Kroupa2001MNRAS}
{Kroupa}, P. 2001, \mnras, 322, 231

\bibitem[{{Laigle} {et~al.}(2016){Laigle}, {McCracken}, {Ilbert}, {Hsieh},
  {Davidzon}, {Capak}, {Hasinger}, {Silverman}, {Pichon}, {Coupon}, {Aussel},
  {Le Borgne}, {Caputi}, {Cassata}, {Chang}, {Civano}, {Dunlop}, {Fynbo},
  {Kartaltepe}, {Koekemoer}, {Le F{\`e}vre}, {Le Floc'h}, {Leauthaud}, {Lilly},
  {Lin}, {Marchesi}, {Milvang-Jensen}, {Salvato}, {Sanders}, {Scoville},
  {Smolcic}, {Stockmann}, {Taniguchi}, {Tasca}, {Toft}, {Vaccari}, \&
  {Zabl}}]{Laigle+2016ApJS}
{Laigle}, C., {McCracken}, H.~J., {Ilbert}, O., {et~al.} 2016, \apjs, 224, 24

\bibitem[{{Leslie} {et~al.}(2016){Leslie}, {Kewley}, {Sanders}, \&
  {Lee}}]{Leslie+2016MNRAS}
{Leslie}, S.~K., {Kewley}, L.~J., {Sanders}, D.~B., \& {Lee}, N. 2016, \mnras,
  455, L82

\bibitem[{{Li} {et~al.}(2021{\natexlab{a}}){Li}, {Silverman}, {Ding},
  {Strauss}, {Goulding}, {Birrer}, {Yesuf}, {Xue}, {Kawinwanichakij},
  {Matsuoka}, {Toba}, {Nagao}, {Schramm}, \& {Inayoshi}}]{Li_Junyao+2021ApJ}
{Li}, J., {Silverman}, J.~D., {Ding}, X., {et~al.} 2021{\natexlab{a}}, \apj, 918, 22

\bibitem[{{Li} {et~al.}(2021{\natexlab{b}}){Li}, {Shen}, {Ho}, {Brandt}, {Dalla
  Bont{\`a}}, {Fonseca Alvarez}, {Grier}, {Hernandez Santisteban}, {Homayouni},
  {Horne}, {Peterson}, {Schneider}, \& {Trump}}]{Li_Jennifer+2021ApJ}
{Li}, J. I.-H., {Shen}, Y., {Ho}, L.~C., {et~al.} 2021{\natexlab{b}}, \apj,
  906, 103

\bibitem[{{Lilly} {et~al.}(2009){Lilly}, {Le Brun}, {Maier}, {Mainieri},
  {Mignoli}, {Scodeggio}, {Zamorani}, {Carollo}, {Contini}, {Kneib}, {Le
  F{\`e}vre}, {Renzini}, {Bardelli}, {Bolzonella}, {Bongiorno}, {Caputi},
  {Coppa}, {Cucciati}, {de la Torre}, {de Ravel}, {Franzetti}, {Garilli},
  {Iovino}, {Kampczyk}, {Kovac}, {Knobel}, {Lamareille}, {Le Borgne}, {Pello},
  {Peng}, {P{\'e}rez-Montero}, {Ricciardelli}, {Silverman}, {Tanaka}, {Tasca},
  {Tresse}, {Vergani}, {Zucca}, {Ilbert}, {Salvato}, {Oesch}, {Abbas},
  {Bottini}, {Capak}, {Cappi}, {Cassata}, {Cimatti}, {Elvis}, {Fumana},
  {Guzzo}, {Hasinger}, {Koekemoer}, {Leauthaud}, {Maccagni}, {Marinoni},
  {McCracken}, {Memeo}, {Meneux}, {Porciani}, {Pozzetti}, {Sanders},
  {Scaramella}, {Scarlata}, {Scoville}, {Shopbell}, \&
  {Taniguchi}}]{Lilly+2009ApJS}
{Lilly}, S.~J., {Le Brun}, V., {Maier}, C., {et~al.} 2009, \apjs, 184, 218

\bibitem[{{Liu} {et~al.}(2019){Liu}, {Liu}, {Dong}, {Zhou}, {Wang}, {Lu}, \&
  {Yuan}}]{Liu+2019ApJS}
{Liu}, H.-Y., {Liu}, W.-J., {Dong}, X.-B., {et~al.} 2019, \apjs, 243, 21

\bibitem[{{Madau} \& {Dickinson}(2014)}]{Madau&Dickinson2014ARA&A}
{Madau}, P., \& {Dickinson}, M. 2014, \araa, 52, 415

\bibitem[{{Magorrian} {et~al.}(1998){Magorrian}, {Tremaine}, {Richstone},
  {Bender}, {Bower}, {Dressler}, {Faber}, {Gebhardt}, {Green}, {Grillmair},
  {Kormendy}, \& {Lauer}}]{Magorrian+1998AJ}
{Magorrian}, J., {Tremaine}, S., {Richstone}, D., {et~al.} 1998, \aj, 115, 2285

\bibitem[{{Maiolino} {et~al.}(2017){Maiolino}, {Russell}, {Fabian}, {Carniani},
  {Gallagher}, {Cazzoli}, {Arribas}, {Belfiore}, {Bellocchi}, {Colina},
  {Cresci}, {Ishibashi}, {Marconi}, {Mannucci}, {Oliva}, \&
  {Sturm}}]{Maiolino+2017Natur}
{Maiolino}, R., {Russell}, H.~R., {Fabian}, A.~C., {et~al.} 2017, \nat, 544,
  202

\bibitem[Mandal et al.(2021)]{Mandal+2021MNRAS} Mandal, A., Mukherjee, D., Federrath, C., et al.\ 2021, \mnras, 508, 4738

\bibitem[Martin et al.(2018)]{Martin+2018MNRAS} Martin, G., Kaviraj, S., Volonteri, M., et al.\ 2018, \mnras, 476, 2801

\bibitem[{{Matsuoka} {et~al.}(2014){Matsuoka}, {Strauss}, {Price}, \&
  {DiDonato}}]{Matsuoka+2014ApJ}
{Matsuoka}, Y., {Strauss}, M.~A., {Price}, Ted~N., I., \& {DiDonato}, M.~S.
  2014, \apj, 780, 162

\bibitem[{{Matsuoka} {et~al.}(2015){Matsuoka}, {Strauss}, {Shen}, {Brandt},
  {Greene}, {Ho}, {Schneider}, {Sun}, \& {Trump}}]{Matsuoka+2015ApJ}
{Matsuoka}, Y., {Strauss}, M.~A., {Shen}, Y., {et~al.} 2015, \apj, 811, 91

\bibitem[McAlpine et al.(2020)]{McAlpine+2020MNRAS} McAlpine, S., Harrison, C.~M., Rosario, D.~J., et al.\ 2020, \mnras, 494, 5713

\bibitem[McLure et al. (1999)]{McLure1999} McLure, R. J., Dunlop, J. S., Kukula, M. J., et al. 1999, \mnras, 308, 377

\bibitem[McNamara \& Nulsen(2007)]{McNamara&Nulsen2007ARA&A} McNamara, B.~R. \& Nulsen, P.~E.~J.\ 2007, \araa, 45, 117

\bibitem[{{Merloni} {et~al.}(2010){Merloni}, {Bongiorno}, {Bolzonella}, {Brusa}, {Civano}, {Comastri}, {Elvis}, {Fiore}, {Gilli}, {Hao}, {Jahnke}, {Koekemoer}, {Lusso}, {Mainieri}, {Mignoli}, {Miyaji}, {Renzini}, {Salvato}, {Silverman}, {Trump}, {Vignali}, {Zamorani}, {Capak}, {Lilly}, {Sanders}, {Taniguchi}, {Bardelli}, {Carollo}, {Caputi}, {Contini}, {Coppa}, {Cucciati}, {de la Torre}, {de Ravel}, {Franzetti}, {Garilli}, {Hasinger}, {Impey}, {Iovino}, {Iwasawa}, {Kampczyk}, {Kneib}, {Knobel}, {Kova{\v{c}}}, {Lamareille}, {Le Borgne}, {Le Brun}, {Le F{\`e}vre}, {Maier}, {Pello}, {Peng}, {Perez Montero}, {Ricciardelli}, {Scodeggio}, {Tanaka}, {Tasca}, {Tresse}, {Vergani}, \& {Zucca}}]{Merloni+2010ApJ} {Merloni}, A., {Bongiorno}, A., {Bolzonella}, M., {et~al.} 2010, \apj, 708, 137

\bibitem[Molina et~al. (2022)]{Molina2022} Molina, J., Ho, L. C., Wang, R., et al. 2022, \apj, submitted

\bibitem[{{Molina} {et~al.}(2021){Molina}, {J.}, {Wang}, {R.}, {Shangguan}, {J.}, {Ho}, {C.}, {Bauer}, {E.}, {Treister}, {E.}, {Shao}, \& {Y}}]{Molina2021} {Molina}, J., {Wang}, R., {Shangguan}, J., et al. 2021, \apj, 908, 231

\bibitem[Parry et al.(2009)]{Parry+2009MNRAS} Parry, O.~H., Eke, V.~R., \& Frenk, C.~S.\ 2009, \mnras, 396, 1972

\bibitem[{{Peng} {et~al.}(2002){Peng}, {Ho}, {Impey}, \& {Rix}}]{Peng+2002AJ}
{Peng}, C.~Y., {Ho}, L.~C., {Impey}, C.~D., \& {Rix}, H.-W. 2002, \aj, 124, 266

\bibitem[{{Peng} {et~al.}(2010){Peng}, {Ho}, {Impey}, \& {Rix}}]{Peng+2010AJ}
{Peng}, C.~Y., {Ho}, L.~C., {Impey}, C.~D., \& {Rix}, H.-W. 2010, \aj, 139, 2097

\bibitem[{{Peng} {et~al.}(2006){Peng}, {Impey}, {Ho}, {Barton},
  \& {Rix}}]{Peng+2006aApJ}
{Peng}, C.~Y., {Impey}, C.~D., {Ho}, L.~C., {Barton}, E.~J., \& {Rix}, H.-W.
  2006, \apj, 640, 114

\bibitem[Petty et al. (2014)]{Petty2014} Petty, S. M., Armus, L., Charmandaris, V., et al. 2014, \aj, 148, 111

\bibitem[{{Popesso} {et~al.}(2019{\natexlab{a}}){Popesso}, {Concas},
  {Morselli}, {Schreiber}, {Rodighiero}, {Cresci}, {Belli}, {Erfanianfar},
  {Mancini}, {Inami}, {Dickinson}, {Ilbert}, {Pannella}, \&
  {Elbaz}}]{2019MNRAS.483.3213Popesso+}
{Popesso}, P., {Concas}, A., {Morselli}, L., {et~al.} 2019{\natexlab{a}},
  \mnras, 483, 3213

\bibitem[{{Popesso} {et~al.}(2019{\natexlab{b}}){Popesso}, {Morselli},
  {Concas}, {Schreiber}, {Rodighiero}, {Cresci}, {Belli}, {Ilbert},
  {Erfanianfar}, {Mancini}, {Inami}, {Dickinson}, {Pannella}, \&
  {Elbaz}}]{2019MNRAS.490.5285Popesso+}
{Popesso}, P., {Morselli}, L., {Concas}, A., {et~al.} 2019{\natexlab{b}},
  \mnras, 490, 5285

\bibitem[Qiu et al.(2020)]{Qiu+2020NatAs} Qiu, Y., Bogdanovi{\'c}, T., Li, Y., et al.\ 2020, Nature Astronomy, 4, 900


\bibitem[{{Rakshit} \& {Woo}(2018)}]{Rakshit&Woo2018ApJ}
{Rakshit}, S., \& {Woo}, J.-H. 2018, \apj, 865, 5

\bibitem[{{Renzini} \& {Peng}(2015)}]{Renzini&Peng2015ApJ}
{Renzini}, A., \& {Peng}, Y.-j. 2015, \apjl, 801, L29

\bibitem[{{Richards} {et~al.}(2006){Richards}, {Lacy}, {Storrie-Lombardi},
  {Hall}, {Gallagher}, {Hines}, {Fan}, {Papovich}, {Vanden Berk}, {Trammell},
  {Schneider}, {Vestergaard}, {York}, {Jester}, {Anderson}, {Budav{\'a}ri}, \&
  {Szalay}}]{Richards+2006ApJS}
{Richards}, G.~T., {Lacy}, M., {Storrie-Lombardi}, L.~J., {et~al.} 2006, \apjs,
  166, 470

\bibitem[{{Richstone} {et~al.}(1998){Richstone}, {Ajhar}, {Bender}, {Bower},
  {Dressler}, {Faber}, {Filippenko}, {Gebhardt}, {Green}, {Ho}, {Kormendy},
  {Lauer}, {Magorrian}, \& {Tremaine}}]{Richstone+1998Natur}
{Richstone}, D., {Ajhar}, E.~A., {Bender}, R., {et~al.} 1998, \nat, 395, A14

\bibitem[{{Robotham} {et~al.}(2017){Robotham}, {Taranu}, {Tobar}, {Moffett}, \&
  {Driver}}]{Robotham+2017MNRAS}
{Robotham}, A.~S.~G., {Taranu}, D.~S., {Tobar}, R., {Moffett}, A., \& {Driver},
  S.~P. 2017, \mnras, 466, 1513

\bibitem[{{Salim} {et~al.}(2007){Salim}, {Rich}, {Charlot}, {Brinchmann},
  {Johnson}, {Schiminovich}, {Seibert}, {Mallery}, {Heckman}, {Forster},
  {Friedman}, {Martin}, {Morrissey}, {Neff}, {Small}, {Wyder}, {Bianchi},
  {Donas}, {Lee}, {Madore}, {Milliard}, {Szalay}, {Welsh}, \&
  {Yi}}]{Salim+2007ApJS}
{Salim}, S., {Rich}, R.~M., {Charlot}, S., {et~al.} 2007, \apjs, 173, 267

\bibitem[Scannapieco et al. (2009)]{Scannapieco2009} Scannapieco, C., White, S. D. M., Springel, V., \& Tissera, P. B. 2009, \mnras, 396, 696

\bibitem[{{Scarlata} {et~al.}(2007){Scarlata}, {Carollo}, {Lilly}, {Sargent},
  {Feldmann}, {Kampczyk}, {Porciani}, {Koekemoer}, {Scoville}, {Kneib},
  {Leauthaud}, {Massey}, {Rhodes}, {Tasca}, {Capak}, {Maier}, {McCracken},
  {Mobasher}, {Renzini}, {Taniguchi}, {Thompson}, {Sheth}, {Ajiki}, {Aussel},
  {Murayama}, {Sanders}, {Sasaki}, {Shioya}, \&
  {Takahashi}}]{Scarlata+2007ApJS}
{Scarlata}, C., {Carollo}, C.~M., {Lilly}, S., {et~al.} 2007, \apjs, 172, 406

\bibitem[{{Schlafly} \&
  {Finkbeiner}(2011)}]{2011ApJ...737..103Schlafly&Finkbeiner}
{Schlafly}, E.~F., \& {Finkbeiner}, D.~P. 2011, \apj, 737, 103

\bibitem[{{Schlegel} {et~al.}(1998){Schlegel}, {Finkbeiner}, \&
  {Davis}}]{1998ApJ...500..525Schlegel+}
{Schlegel}, D.~J., {Finkbeiner}, D.~P., \& {Davis}, M. 1998, \apj, 500, 525

\bibitem[Schmidt \& Green(1983)]{Schmidt&Green1983ApJ} Schmidt, M., \& Green, R.~F.\ 1983, \apj, 269, 352

\bibitem[{{Scoville} {et~al.}(2007){Scoville}, {Aussel}, {Brusa}, {Capak},
  {Carollo}, {Elvis}, {Giavalisco}, {Guzzo}, {Hasinger}, {Impey}, {Kneib},
  {LeFevre}, {Lilly}, {Mobasher}, {Renzini}, {Rich}, {Sanders}, {Schinnerer},
  {Schminovich}, {Shopbell}, {Taniguchi}, \& {Tyson}}]{Scoville+2007ApJS}
{Scoville}, N., {Aussel}, H., {Brusa}, M., {et~al.} 2007, \apjs, 172, 1

\bibitem[{{S\'ersic} (1968)}]{Sersic}
S\'ersic, J. L. 1968, Atlas de Galaxias Australes (C\'ordoba: Obs. Astron., Univ. Nac. C\'ordoba)


\bibitem[Shangguan et~al.(2020a)]{Shangguan+2020aApJ} Shangguan, J., Ho, L. C., Bauer, F. E., Wang, R., \& Treister, E. 2020a, \apjs, 247, 15

\bibitem[Shangguan et~al.(2020b)]{Shangguan+2020bApJ} Shangguan, J., Ho, L. C., Bauer, F. E., Wang, R., \& Treister, E. 2020b, \apj, 899, 112

\bibitem[Shangguan et al. (2019)]{Shangguan2019} Shangguan, J., Ho, L. C., Li, R., et al. 2019, \apj, 870, 104

\bibitem[Shangguan et al. (2018)]{Shangguan2018} Shangguan, J., Ho, L. C., \& Xie, Y. 2018, \apj, 854, 15

\bibitem[{{Sharma} {et~al.}(2022){Sharma}, {Choi}, {Somerville}, {Snyder}, {Kocevski}, {Hirschmann}, {Moster}, {Naab}, {Narayanan}, {Ostriker}, \& {Rosario}}]{Sharma2022} {Sharma}, R.~S., {Choi}, E., {Somerville}, R.~S., {et~al.} 2022, \apj, submitted (arXiv:2101.01729)

\bibitem[{{Shimizu} {et~al.}(2015){Shimizu}, {Mushotzky}, {Mel{\'e}ndez},
  {Koss}, \& {Rosario}}]{Shimizu+2015MNRAS}
{Shimizu}, T.~T., {Mushotzky}, R.~F., {Mel{\'e}ndez}, M., {Koss}, M., \&
  {Rosario}, D.~J. 2015, \mnras, 452, 1841

\bibitem[Shin et al.(2019)]{Shin+2019ApJ} Shin, J., Woo, J.-H., Chung, A., et al.\ 2019, \apj, 881, 147

\bibitem[{{Silverman} {et~al.}(2019){Silverman}, {Treu}, {Ding}, {Jahnke},
  {Bennert}, {Birrer}, {Schramm}, {Schulze}, {Kartaltepe}, {Sanders}, \&
  {Cen}}]{Silverman+2019ApJ}
{Silverman}, J.~D., {Treu}, T., {Ding}, X., {et~al.} 2019, \apjl, 887, L5

\bibitem[{{Smirnova-Pinchukova} {et~al.}(2022){Smirnova-Pinchukova},
  {Husemann}, {Davis}, {Smith}, {Singha}, {Tremblay}, {Klessen}, {Powell},
  {Connor}, {Baum}, {Combes}, {Croom}, {Gaspari}, {Neumann}, {O'Dea},
  {P{\'e}rez-Torres}, {Rosario}, {Rose}, {Scharw{\"a}chter}, \&
  {Winkel}}]{Smirnova-Pinchukova+2021arXiv}
{Smirnova-Pinchukova}, I., {Husemann}, B., {Davis}, T.~A., {et~al.} 2022, A\&A, submitted (arXiv:2111.10419)

\bibitem[{{Somerville} \& {Dav{\'e}}(2015)}]{Somerville&Dave2015ARA&A}
{Somerville}, R.~S., \& {Dav{\'e}}, R. 2015, \araa, 53, 51

\bibitem[{{Speagle} {et~al.}(2014){Speagle}, {Steinhardt}, {Capak}, \&
  {Silverman}}]{Speagle+2014ApJS}
{Speagle}, J.~S., {Steinhardt}, C.~L., {Capak}, P.~L., \& {Silverman}, J.~D.
  2014, \apjs, 214, 15

\bibitem[{{Stacey} {et~al.}(2021){Stacey}, {McKean}, {Powell}, {Vegetti},
  {Rizzo}, {Spingola}, {Auger}, {Ivison}, \& {van der Werf}}]{Stacey+2021MNRAS}
{Stacey}, H.~R., {McKean}, J.~P., {Powell}, D.~M., {et~al.} 2021, \mnras, 500,
  3667

\bibitem[Stalevski et al.(2019)]{Stalevski+2019MNRAS} Stalevski, M., Tristram, K.~R.~W., \& Asmus, D.\ 2019, \mnras, 484, 3334

\bibitem[{{Steinborn} {et~al.}(2018){Steinborn}, {Hirschmann}, {Dolag},
  {Shankar}, {Juneau}, {Krumpe}, {Remus}, \& {Teklu}}]{Steinborn+2018MNRAS}
{Steinborn}, L.~K., {Hirschmann}, M., {Dolag}, K., {et~al.} 2018, \mnras, 481,
  341

\bibitem[{{Stemo} {et~al.}(2020){Stemo}, {Comerford}, {Barrows}, {Stern},
  {Assef}, \& {Griffith}}]{Stemo+2020ApJ}
{Stemo}, A., {Comerford}, J.~M., {Barrows}, R.~S., {et~al.} 2020, \apj, 888, 78

\bibitem[{{Suh} {et~al.}(2019){Suh}, {Civano}, {Hasinger}, {Lusso}, {Marchesi},
  {Schulze}, {Onodera}, {Rosario}, \& {Sanders}}]{Suh+2019ApJ}
{Suh}, H., {Civano}, F., {Hasinger}, G., {et~al.} 2019, \apj, 872, 168

\bibitem[Thomas et al.(2018)]{Thomas+2018ApJ} Thomas, A.~D., Dopita, M.~A., Kewley, L.~J., et al.\ 2018, \apj, 856, 89

\bibitem[Toomre \& Toomre(1972)]{Toomre&Toomre1972ApJ} Toomre, A., \& Toomre, J.\ 1972, \apj, 178, 623

\bibitem[Torbaniuk et al.(2021)]{Torbaniuk+2021MNRAS} Torbaniuk, O., Paolillo, M., Carrera, F., et al.\ 2021, \mnras, 506, 2619

\bibitem[{{Vanden Berk} {et~al.}(2006){Vanden Berk}, {Shen}, {Yip}, {Schneider}, {Connolly}, {Burton}, {Jester}, {Hall}, {Szalay}, \& {Brinkmann}}]{Vanden_Berk+2006AJ} {Vanden Berk}, D.~E., {Shen}, J., {Yip}, C.-W., {et~al.} 2006, \aj, 131, 84

\bibitem[{{van der Wel} {et~al.}(2012){van der Wel}, {Bell}, {H{\"a}ussler},
  {McGrath}, {Chang}, {Guo}, {McIntosh}, {Rix}, {Barden}, {Cheung}, {Faber},
  {Ferguson}, {Galametz}, {Grogin}, {Hartley}, {Kartaltepe}, {Kocevski},
  {Koekemoer}, {Lotz}, {Mozena}, {Peth}, \& {Peng}}]{van_der_Wel+2012ApJS}
{van der Wel}, A., {Bell}, E.~F., {H{\"a}ussler}, B., {et~al.} 2012, \apjs,
  203, 24

\bibitem[{{van der Wel} {et~al.}(2014){van der Wel}, {Franx}, {van Dokkum},
  {Skelton}, {Momcheva}, {Whitaker}, {Brammer}, {Bell}, {Rix}, {Wuyts},
  {Ferguson}, {Holden}, {Barro}, {Koekemoer}, {Chang}, {McGrath},
  {H{\"a}ussler}, {Dekel}, {Behroozi}, {Fumagalli}, {Leja}, {Lundgren},
  {Maseda}, {Nelson}, {Wake}, {Patel}, {Labb{\'e}}, {Faber}, {Grogin}, \&
  {Kocevski}}]{van_der_Wel2014ApJ}
{van der Wel}, A., {Franx}, M., {van Dokkum}, P.~G., {et~al.} 2014, \apj, 788,
  28

\bibitem[{{Vika} {et~al.}(2013){Vika}, {Bamford}, {H{\"a}u{\ss}ler}, {Rojas},
  {Borch}, \& {Nichol}}]{Vika+2013MNRAS}
{Vika}, M., {Bamford}, S.~P., {H{\"a}u{\ss}ler}, B., {et~al.} 2013, \mnras,
  435, 623

\bibitem[{{Villforth} {et~al.}(2017){Villforth}, {Hamilton}, {Pawlik},
  {Hewlett}, {Rowlands}, {Herbst}, {Shankar}, {Fontana}, {Hamann}, {Koekemoer},
  {Pforr}, {Trump}, \& {Wuyts}}]{Villforth+2017MNRAS}
{Villforth}, C., {Hamilton}, T., {Pawlik}, M.~M., {et~al.} 2017, \mnras, 466,
  812

\bibitem[{{Waters} {et~al.}(2020){Waters}, {Magnier}, {Price}, {Chambers},
  {Burgett}, {Draper}, {Flewelling}, {Hodapp}, {Huber}, {Jedicke}, {Kaiser},
  {Kudritzki}, {Lupton}, {Metcalfe}, {Rest}, {Sweeney}, {Tonry}, {Wainscoat},
  \& {Wood-Vasey}}]{Waters+2020ApJS}
{Waters}, C.~Z., {Magnier}, E.~A., {Price}, P.~A., {et~al.} 2020, \apjs, 251, 4

\bibitem[{{Wild} {et~al.}(2007){Wild}, {Kauffmann}, {Heckman}, {Charlot},
  {Lemson}, {Brinchmann}, {Reichard}, \& {Pasquali}}]{Wild2007MNRAS}
{Wild}, V., {Kauffmann}, G., {Heckman}, T., {et~al.} 2007, \mnras, 381, 543

\bibitem[{{Woo} {et~al.}(2017){Woo}, {Son}, \& {Bae}}]{Woo+2017ApJ}
{Woo}, J.-H., {Son}, D., \& {Bae}, H.-J. 2017, \apj, 839, 120

\bibitem[{{Woo} {et~al.}(2020){Woo}, {Son}, \& {Rakshit}}]{Woo+2020ApJ}
{Woo}, J.-H., {Son}, D., \& {Rakshit}, S. 2020, \apj, 901, 66

\bibitem[{{Xie} {et~al.}(2021){Xie}, {Ho}, {Zhuang}, \& {Shangguan}}]{Xie+2021ApJ} {Xie}, Y., {Ho}, L.~C., {Zhuang}, M.-Y., \& {Shangguan}, J. 2021, \apj, 910, 124

\bibitem[Yang et al.(2020)]{Yang+2020MNRAS} Yang, G., Boquien, M., Buat, V., et al.\ 2020, \mnras, 491, 740

\bibitem[{{Yesuf} \& {Ho}(2019)}]{Yesuf2019} Yesuf, H. M., \& Ho, L. C. 2019, \apj, 884, 177

\bibitem[{{Yesuf} \& {Ho}(2020)}]{Yesuf2020} Yesuf, H. M., \& Ho, L. C. 2020, \apj, 901, 42

\bibitem[{{York} {et~al.}(2000){York}, {Adelman}, {Anderson}, {Anderson},
  {Annis}, {Bahcall}, {Bakken}, {Barkhouser}, {Bastian}, {Berman}, {Boroski},
  {Bracker}, {Briegel}, {Briggs}, {Brinkmann}, {Brunner}, {Burles}, {Carey},
  {Carr}, {Castander}, {Chen}, {Colestock}, {Connolly}, {Crocker}, {Csabai},
  {Czarapata}, {Davis}, {Doi}, {Dombeck}, {Eisenstein}, {Ellman}, {Elms},
  {Evans}, {Fan}, {Federwitz}, {Fiscelli}, {Friedman}, {Frieman}, {Fukugita},
  {Gillespie}, {Gunn}, {Gurbani}, {de Haas}, {Haldeman}, {Harris}, {Hayes},
  {Heckman}, {Hennessy}, {Hindsley}, {Holm}, {Holmgren}, {Huang}, {Hull},
  {Husby}, {Ichikawa}, {Ichikawa}, {Ivezi{\'c}}, {Kent}, {Kim}, {Kinney},
  {Klaene}, {Kleinman}, {Kleinman}, {Knapp}, {Korienek}, {Kron}, {Kunszt},
  {Lamb}, {Lee}, {Leger}, {Limmongkol}, {Lindenmeyer}, {Long}, {Loomis},
  {Loveday}, {Lucinio}, {Lupton}, {MacKinnon}, {Mannery}, {Mantsch}, {Margon},
  {McGehee}, {McKay}, {Meiksin}, {Merelli}, {Monet}, {Munn}, {Narayanan},
  {Nash}, {Neilsen}, {Neswold}, {Newberg}, {Nichol}, {Nicinski}, {Nonino},
  {Okada}, {Okamura}, {Ostriker}, {Owen}, {Pauls}, {Peoples}, {Peterson},
  {Petravick}, {Pier}, {Pope}, {Pordes}, {Prosapio}, {Rechenmacher}, {Quinn},
  {Richards}, {Richmond}, {Rivetta}, {Rockosi}, {Ruthmansdorfer}, {Sandford},
  {Schlegel}, {Schneider}, {Sekiguchi}, {Sergey}, {Shimasaku}, {Siegmund},
  {Smee}, {Smith}, {Snedden}, {Stone}, {Stoughton}, {Strauss}, {Stubbs},
  {SubbaRao}, {Szalay}, {Szapudi}, {Szokoly}, {Thakar}, {Tremonti}, {Tucker},
  {Uomoto}, {Vanden Berk}, {Vogeley}, {Waddell}, {Wang}, {Watanabe},
  {Weinberg}, {Yanny}, {Yasuda}, \& {SDSS Collaboration}}]{York+2000AJ}
{York}, D.~G., {Adelman}, J., {Anderson}, Jr., J.~E., {et~al.} 2000, \aj, 120,
  1579

\bibitem[{{Yue} {et~al.}(2018){Yue}, {Jiang}, {Shen}, {Hall}, {Yu},
  {Schneider}, {Ho}, {Horne}, {Petitjean}, \& {Trump}}]{Yue+2018ApJ}
{Yue}, M., {Jiang}, L., {Shen}, Y., {et~al.} 2018, \apj, 863, 21

\bibitem[{{Zhao} {et~al.}(2019){Zhao}, {Ho}, {Zhao}, {Shangguan}, \& {Kim}}]{Zhao+2019ApJ} {Zhao}, D., {Ho}, L.~C., {Zhao}, Y., {Shangguan}, J., \& {Kim}, M. 2019, \apj, 877, 52

\bibitem[{{Zhao} {et~al.}(2021){Zhao}, {Ho}, {Shangguan}, {Kim}, {Zhao}, \& {Gao}}]{Zhao+2021ApJ} {Zhao}, Y., {Ho}, L.~C., {Shangguan}, J., {et~al.} 2021, \apj, 911, 94

\bibitem[{{Zhao} {et~al.}(2022)}]{Zhao2022} 
Zhao, Y., Li, Y. A., Shangguan, J., Zhuang, M.-Y., \& Ho, L. C. 2022, \apj, 925, 70

\bibitem[{{Zhuang} \& {Ho}(2019)}]{Zhuang&Ho2019ApJ}
{Zhuang}, M.-Y., \& {Ho}, L.~C. 2019, \apj, 882, 89

\bibitem[{{Zhuang} \& {Ho}(2020)}]{Zhuang&Ho2020ApJ}
{Zhuang}, M.-Y., \& {Ho}, L.~C. 2020, \apj, 896, 108

\bibitem[Zhuang et al.(2018)]{Zhuang+2018ApJ} Zhuang, M.-Y., Ho, L.~C., \& Shangguan, J.\ 2018, \apj, 862, 118

\bibitem[Zhuang et al.(2019)]{Zhuang+2019ApJ} Zhuang, M.-Y., Ho, L.~C., \& Shangguan, J.\ 2019, \apj, 873, 103

\bibitem[{{Zhuang} {et~al.}(2021){Zhuang}, {Ho}, \&
  {Shangguan}}]{Zhuang+2021ApJ}
{Zhuang}, M.-Y., {Ho}, L.~C., \& {Shangguan}, J. 2021, \apj, 906, 38

\bibitem[Zubovas et al.(2013)]{Zubovas+2013MNRAS} Zubovas, K., Nayakshin, S., King, A., et al.\ 2013, \mnras, 433, 3079

\end{thebibliography}
\end{document}